\documentclass[fleqn,usenatbib]{mnras}

\usepackage{newtxtext,newtxmath}

\usepackage[T1]{fontenc}
\usepackage{ae,aecompl}

\usepackage{graphicx}	
\usepackage{subcaption}
\graphicspath{{./age/}{./stars/}{./young/}{./SFH/}{./gas/}{./Transfer_report/}{./}}
\usepackage{amsmath}	
\usepackage{amssymb}	
\usepackage{float}
\usepackage{tabu}
\usepackage{array}
\usepackage{color}

\title[Stars and gas in barred galaxies]{Redistribution of Stars and Gas in the Star Formation Deserts of Barred Galaxies}

\author[C. E. Donohoe-Keyes]{
C. E. Donohoe-Keyes$^{1}$\thanks{E-mail: C.E.DonohoeKeyes@2013.ljmu.ac.uk},
M. Martig$^{1}$,
P. A. James$^{1}$,
K. Kraljic$^{2}$
\\
$^{1}$Astrophysics Research Institute, Liverpool John Moores University, IC2, Brownlow Hill, Liverpool, Merseyside L3 5RF, UK\\
$^{2}$Institute for Astronomy
University of Edinburgh
Royal Observatory
Blackford Hill
Edinburgh EH9 3HJ
U.K.}

\date{Accepted XXX. Received YYY; in original form ZZZ}

\pubyear{2019}

\begin{document}
\label{firstpage}
\pagerange{\pageref{firstpage}--\pageref{lastpage}}
\maketitle

\begin{abstract}
Bars strongly influence the distribution of gas and stars within the central regions of their host galaxies. This is particularly pronounced in the star formation desert (SFD) which is defined as two symmetrical regions either side of the bar that show a deficit in young stars. Previous studies proposed that, if star formation is truncated because of the influence of the bar, then the age distribution of stars within the SFD could be used to determine the epoch of bar formation. To test this, we study the properties of SFDs in 6 galaxies from zoom-in cosmological re-simulations. Age maps reveal old regions on both sides of the bars, with a lack of stars younger than 10 Myr, confirming the SFD phenomenon. Local star formation is truncated in the SFDs because after the bar forms, gas in these regions is removed on 1 Gyr timescales. However, the overall age distribution of stars in the SFD does not show a sharp truncation after bar formation but rather a gradual downturn in comparison to that of the bar. This more subtle signature may still give information on bar formation epochs in observed galaxies, but the interpretation will be more difficult than originally hoped. The gradual drop in the SFD age distribution, instead of a truncation, is due to radial migration of stars born in the disk. The SFD is thus one of the only regions where an uncontaminated sample of stars only affected by radial migration can be studied.

\end{abstract}




\section{Introduction}\label{1}  

Strong bars are clearly seen in the optical morphologies of ~30\% of disc galaxies \citep{Knapen2000,Marinova2007,Nair2010}, with 60\% displaying bars or bar-like features in the near infrared. Bars exert strong torques on the gas and stars in their host galaxy and are one of the major drivers of secular evolution \citep{LyndenBell1979,Combes2008,Cheung2013,Sellwood2014}. They can have a strong influence on the redistribution of gas thus impacting star formation activity and the overall stellar populations and structure in the centre of galaxies.\par

The torques which bars induce drive gas towards the leading edges of the bar. This gas becomes compressed, loses angular momentum and energy, and falls towards the centre of the galaxy \citep{Athanassoula1992,Heller1994,Knapen1995,Sheth2005}, which explains the higher central gas concentrations observed in barred galaxies compared to unbarred spirals \citep{Sakamoto1999,Sheth2005}.\par
 
By redistributing the gas, bars can influence the star formation of their galaxies; however, the details of these effects are still unclear. While many numerical \citep{Shlosman1989,Berentzen1998,Combes2001,Kim2011,Kim2012,Seo2013,Shin2017} and observational \citep{Heckman1980,Hawarden1986,Devereux1987,Hummel1990,Laurikainen2004,Jogee2005,Regan2006,Ellison2011,Wang2012,Lin2017} studies have found that bars are associated with an increase in central star formation rates (SFR), others find that there is no association between bars and an increase in star formation \citep{Pompea1990,Martinet1997,Chapelon1999,Cheung2013,Willett2015}.\par 

Similarly, \cite{Vera2016} find there is an increase in the metallicity in the centre of barred galaxies compared to non-barred galaxies, while others find no changes in metallicities \citep{Henry1999,Considere2000,Cacho2014}.\par

The disagreement between the effects of bars on the SFRs and metallicities of their hosts could be related to morphology \citep{Ho1997,Oh2012} where early-type barred galaxies show enhanced SFRs, while late-types show no difference between the SFRs seen in barred and unbarred spirals. This might also be explained through the different lengths and strengths of bars \citep{Martin1995,Martinet1997,Kim2017}. Indeed, numerical studies find that bar strength and 3D structure can impact the efficiency of gas inflow and hence the SFR and metallicity of the host \citep{Athanassoula2003,Buta2005,Nair2010,Hoyle2011,fragkoudi2016}.\par  

Given the influence that bars can have over the properties of their host, determining their age becomes an important step in understanding the evolution of their galaxies. What actually matters is not only when bars form but also if they are long lived features.\par

While some numerical studies in idealized galaxies find that bars can be  destroyed by central mass concentrations, gravity torques and supermassive black holes \citep{Bournaud2002,Bournaud2005,Hozumi2005,Hozumi2012}, in cosmological simulations most of the bars are long lived features \citep{KK2012,Fiacconi2015}. Observationally, barred galaxies show central gas concentrations which are not seen in unbarred galaxies \citep{Sakamoto1999}, with some bars containing old nuclear disks \citep{Gadotti2015} which would be difficult to explain if bars were short-lived features.\par

If bars are long lived then determining the age of the bar can help uncover when galactic disks begin to settle \citep{Gadotti2001}. However, when talking about bar ages we must use some caution since the age of the stellar population within the bar is not necessarily related with the bar formation epoch \citep{Wozniak2007}. Observationally, there have been several methods proposed to provide bar ages.  \cite{Perez2009} and \cite{Perez2011} used optical spectroscopy to analyse the properties of the stellar populations in bars, finding a wide range of bar ages. \cite{Gadotti2005} used the vertical velocity dispersion of the bar and found that older bars are vertically thick when compared to recently formed bars. By comparing gas mass with accretion rate \cite{Elmegreen2009} determined a lower limit on the age of the bar in their study. \cite{Kim2014} determined that as bars evolve their light profiles move from exponential and disk-like to flat. From this, they associate the flattening of the profile with bar age (i.e. older bars have flatter profiles while younger bars have more exponential, disk-like profiles). \cite{Gadotti2015} used the age of stars in the nuclear ring to define a lower limit of the epoch of bar formation. \cite{Carles2016} proposed that it might be possible to determine when the bar forms from changes in the star formation histories of the central regions of barred galaxies. \par

\cite{James2016,James2018} used a feature first noticed by \cite{James2009}, which they named the `star formation desert' (SFD), to determine the ages of the bars. They define the SFD as a region lying within the inner ring, either side of the bar in the area the bar sweeps out that shows little to no H$\alpha$ emission. These regions also display a deficit in surface stellar density \citep{Gadotti2003,Gadotti2008,Kim2016}. \cite{James2016,James2018} assumed a truncated star formation model and found that SFD regions can be very old. If the truncation of star formation is caused by the bar, this feature can be used to determine the epoch of bar formation. This leads to some interesting questions:
\begin{itemize}
    \item Is the SFD region observable in simulations? Can the mechanism behind this cessation of star formation be determined?
    \item Is it a result of gas being dynamically heated against star formation, or is the gas being removed by the formation of the bar? If the gas is removed then where does it go? 
    \item Can the properties of the SFD be used as a method for determining the formation epoch of the bar? 
    \item Are the SFD stars only born before the formation of the bar and, if they are not, where do the later-forming stars come from? 
    \item Is the cessation of star formation in the SFDs related to a global downturn in star formation?
\end{itemize}\par

In the following paper we attempt to answer these questions by presenting a numerical analysis of a sample of simulated galaxies selected from \cite{Martig2012}. The structure of the paper is as follows: Section \ref{section:simulation} contains a description of the simulation techniques used to produce our sample, a description of the sample itself and the method used to obtain the properties of the bars. Section \ref{section:results} contains our results and analysis of stars within the SFD region in comparison with the bar and global galaxy properties. Section \ref{section:discussion} contains our discussion of the main results in terms of determining the epoch of bar formation and the analysis of the stars within the SFD region. Our main conclusions are presented in Section \ref{section:conclusion} along with our plans for future analysis of this region.\par

\begin{figure*}
	\begin{tabular}{lcr}
	\includegraphics[trim={.5cm .5cm .5cm 1.5cm},clip,width=3.1cm]{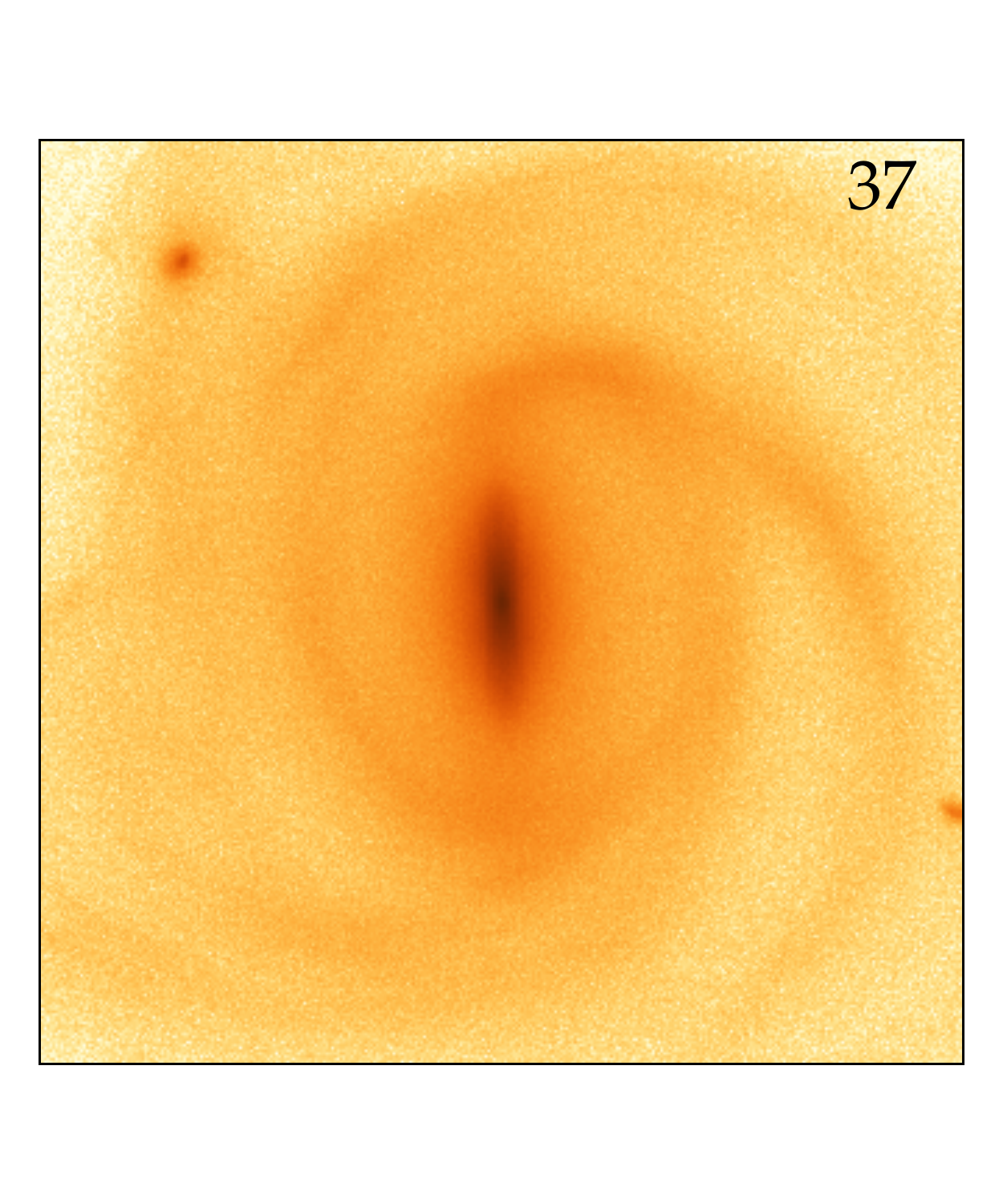}&
	\includegraphics[trim={.5cm .5cm .5cm 1.5cm},clip,width=3.1cm]{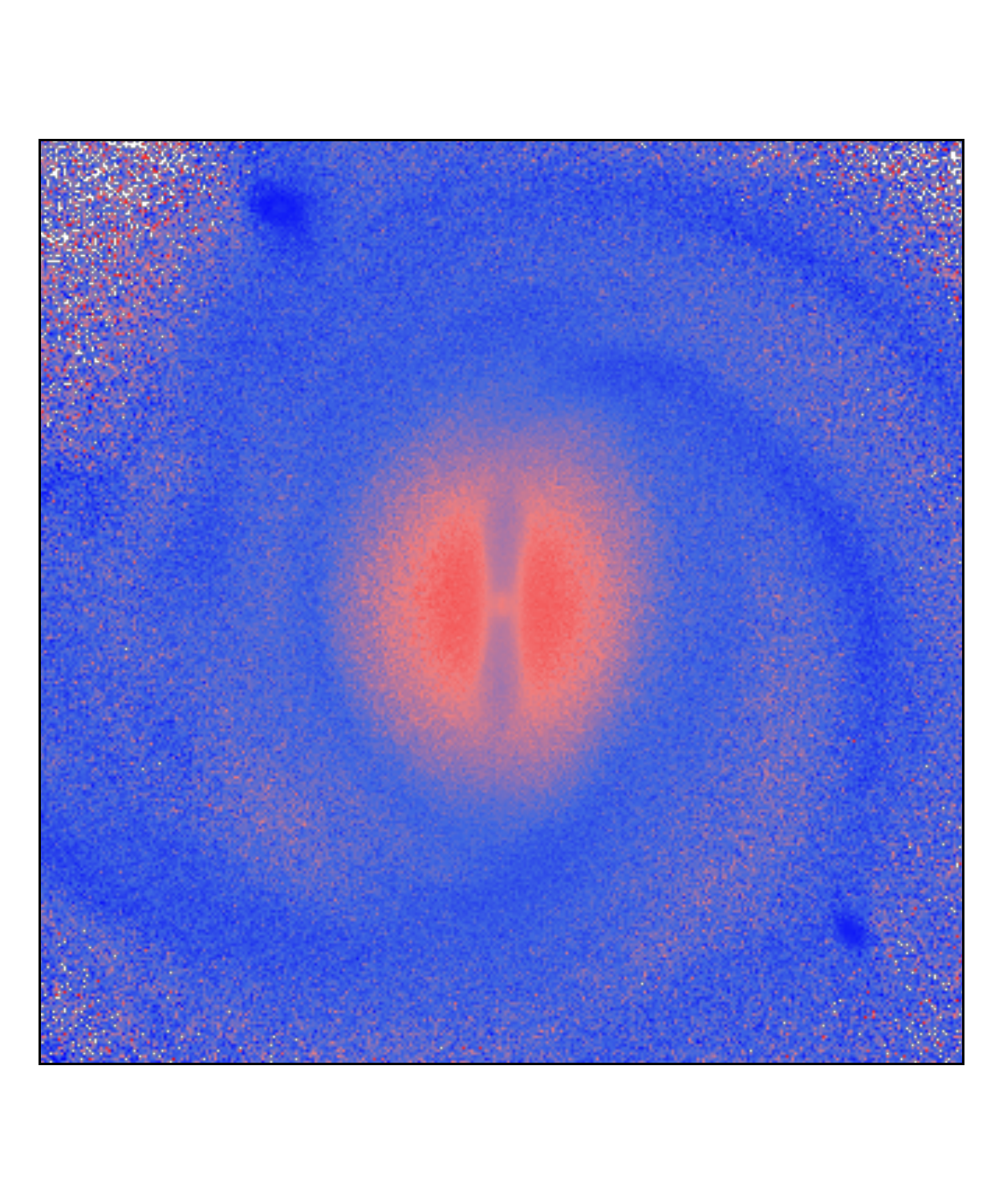}&
	\includegraphics[trim={.5cm .5cm .5cm 1.5cm},clip,width=3.1cm]{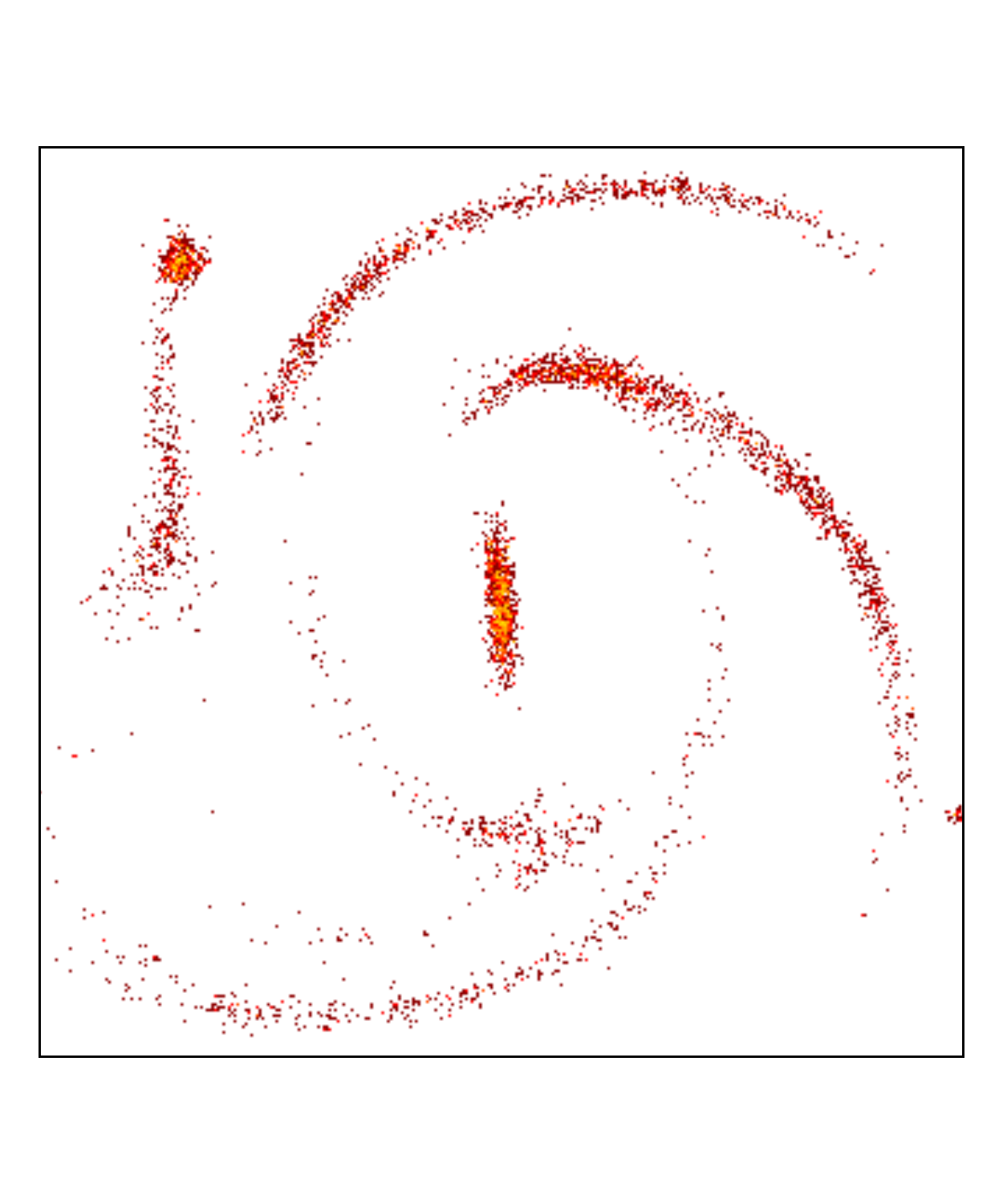}\\
	\includegraphics[trim={.5cm .5cm .5cm 1.5cm},clip,width=3.1cm]{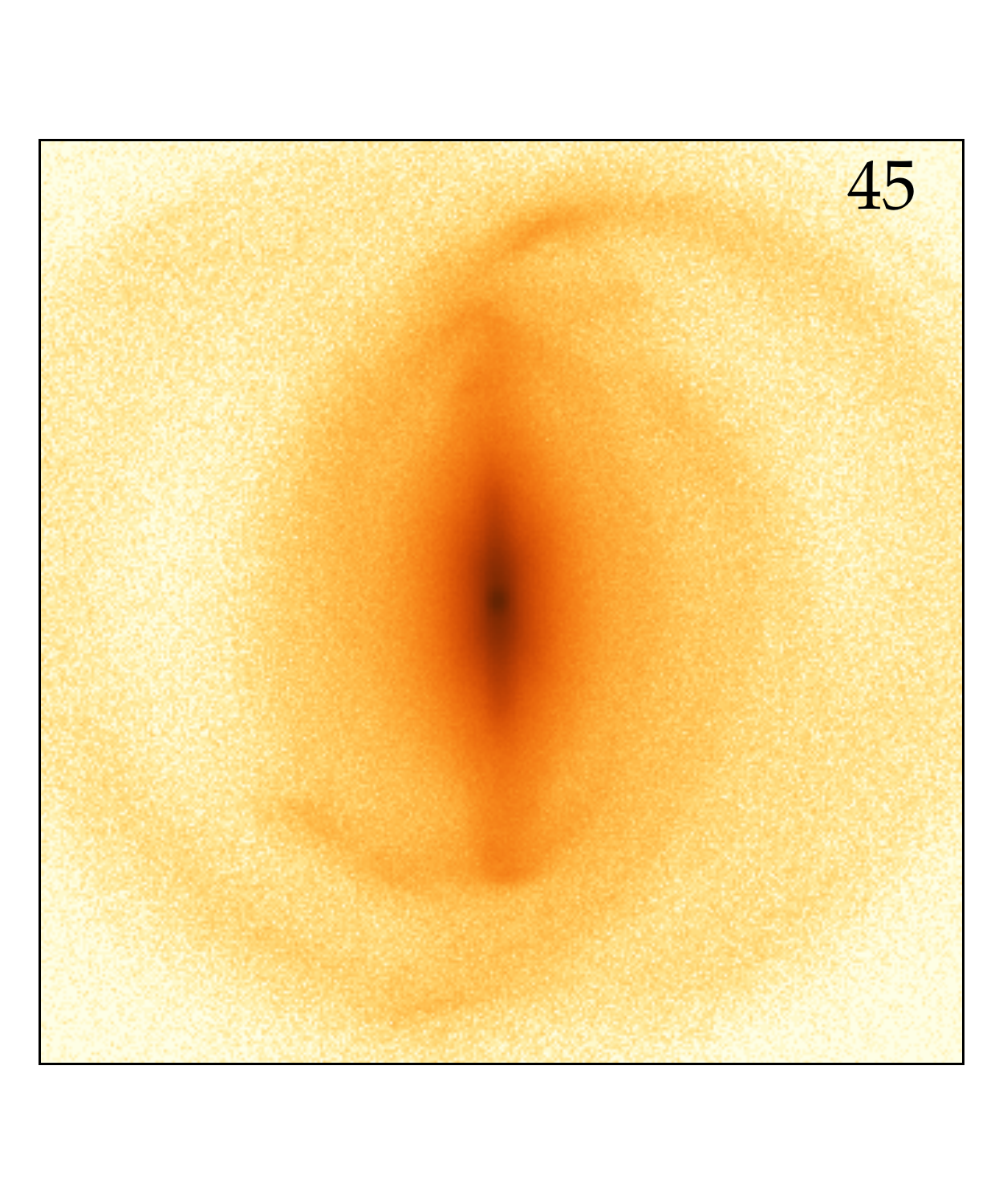}&
	\includegraphics[trim={.5cm .5cm .5cm 1.5cm},clip,width=3.1cm]{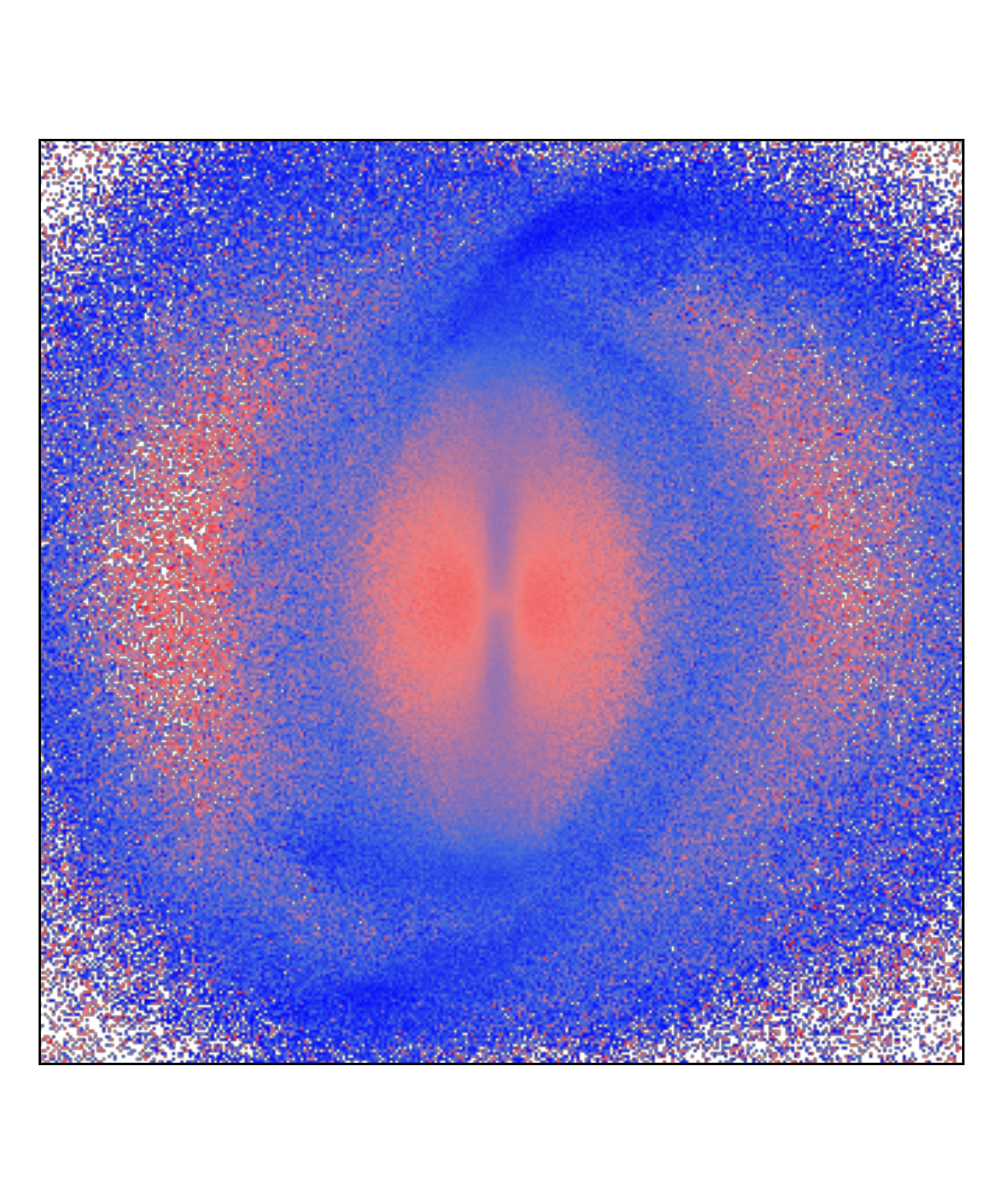}&
	\includegraphics[trim={.5cm .5cm .5cm 1.5cm},clip,width=3.1cm]{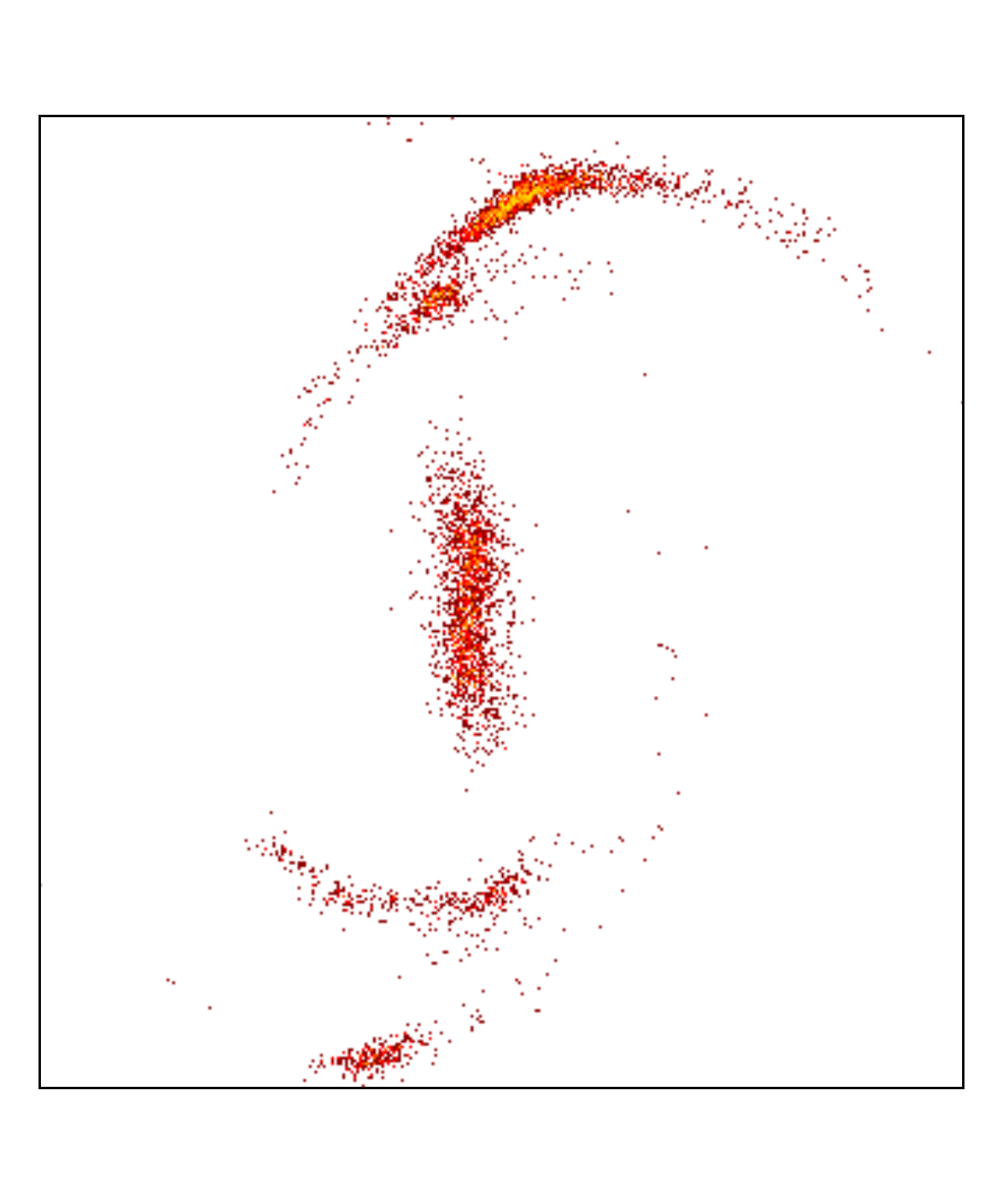}\\
	\includegraphics[trim={.5cm .5cm .5cm 1.5cm},clip,width=3.1cm]{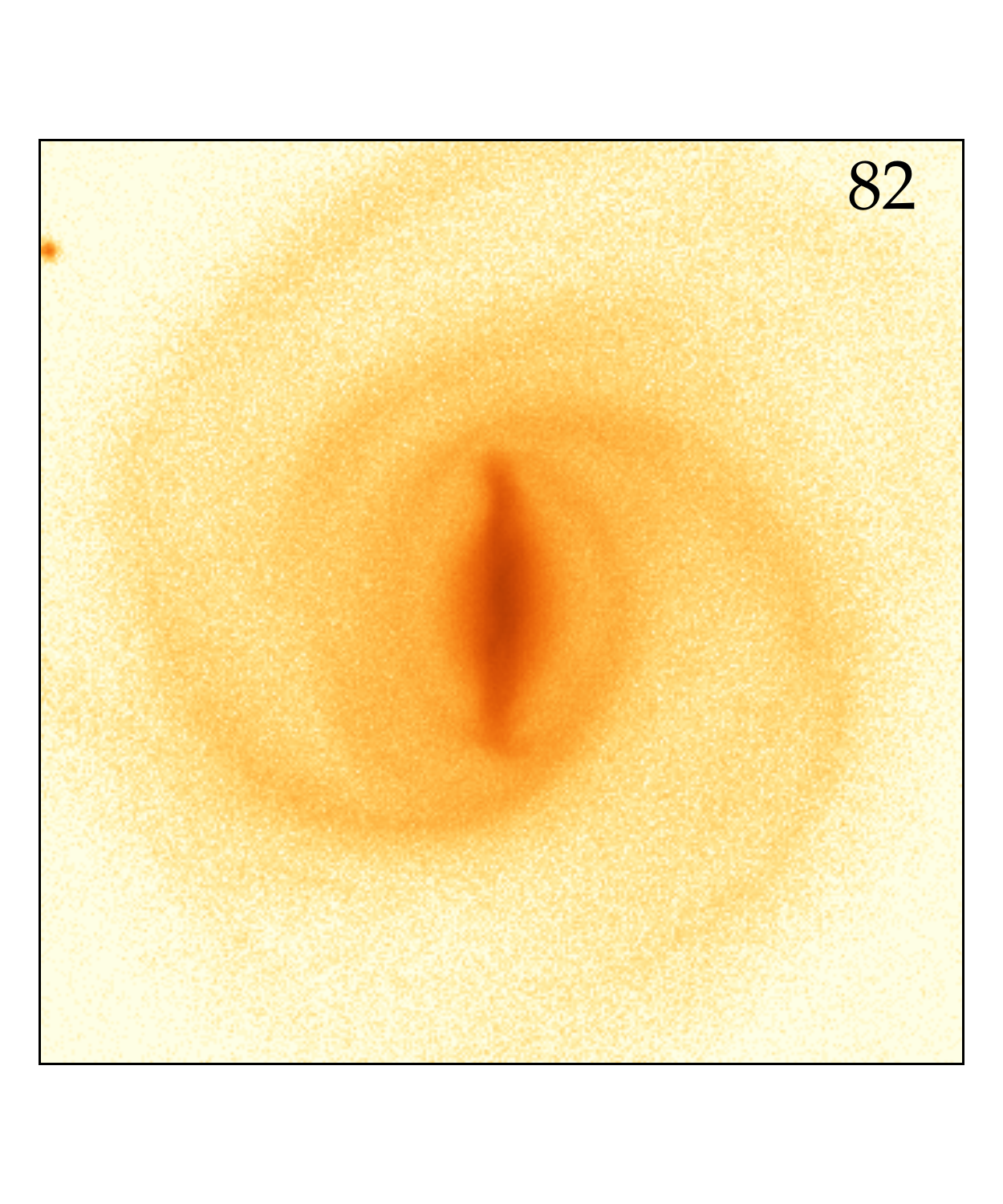}&
	\includegraphics[trim={.5cm .5cm .5cm 1.5cm},clip,width=3.1cm]{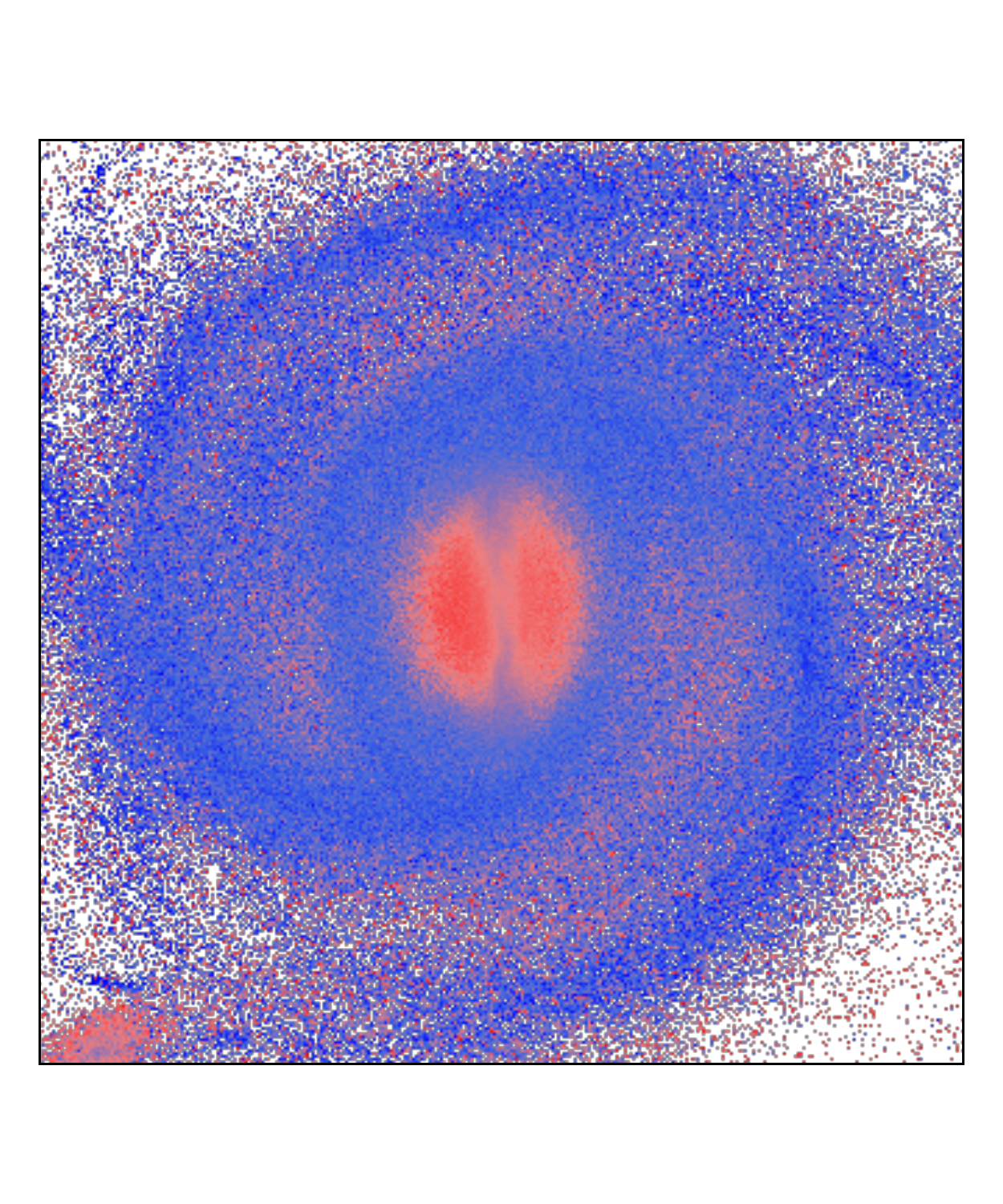}&
	\includegraphics[trim={.5cm .5cm .5cm 1.5cm},clip,width=3.1cm]{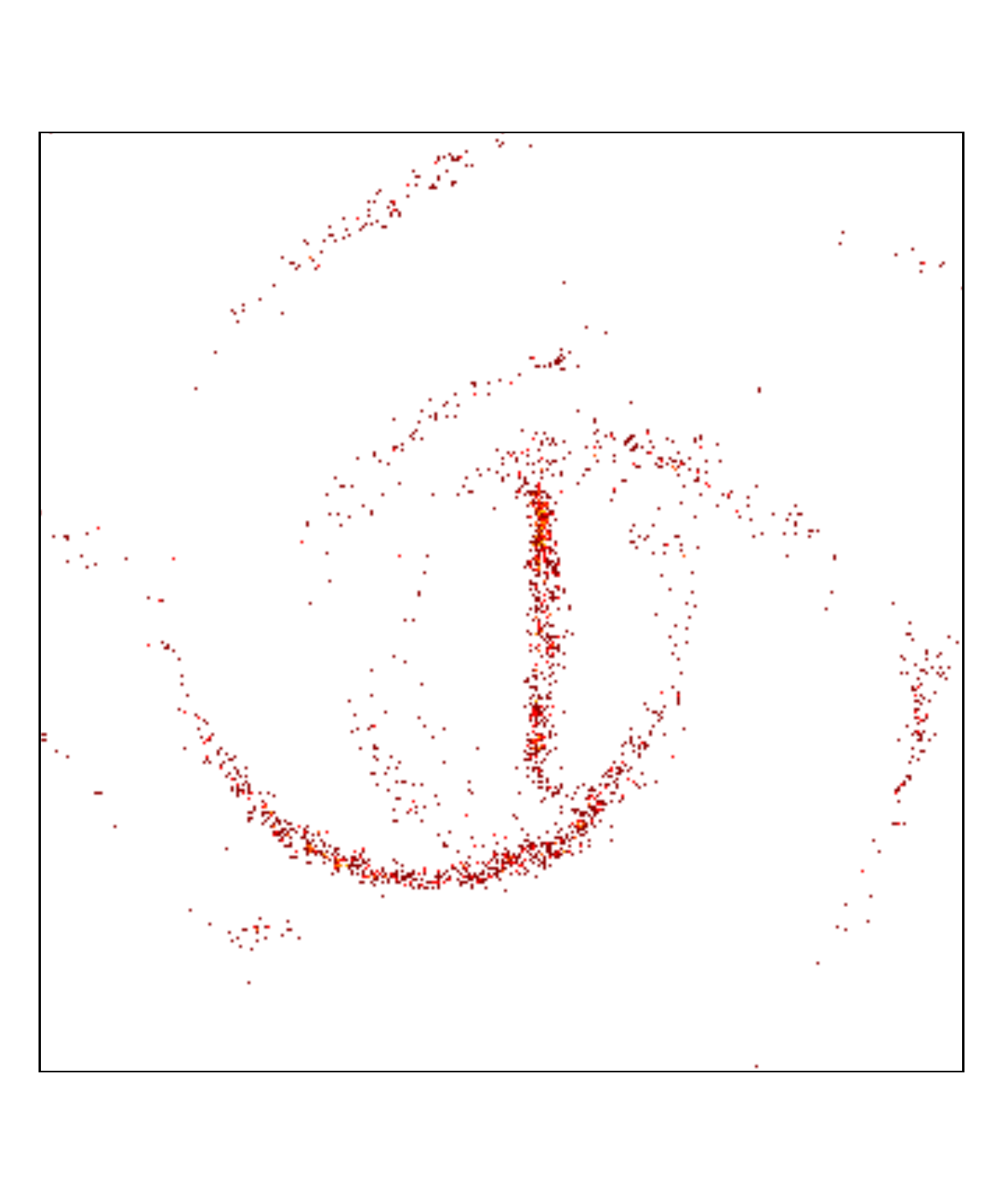}\\
	\includegraphics[trim={.5cm .5cm .5cm 1.5cm},clip,width=3.1cm]{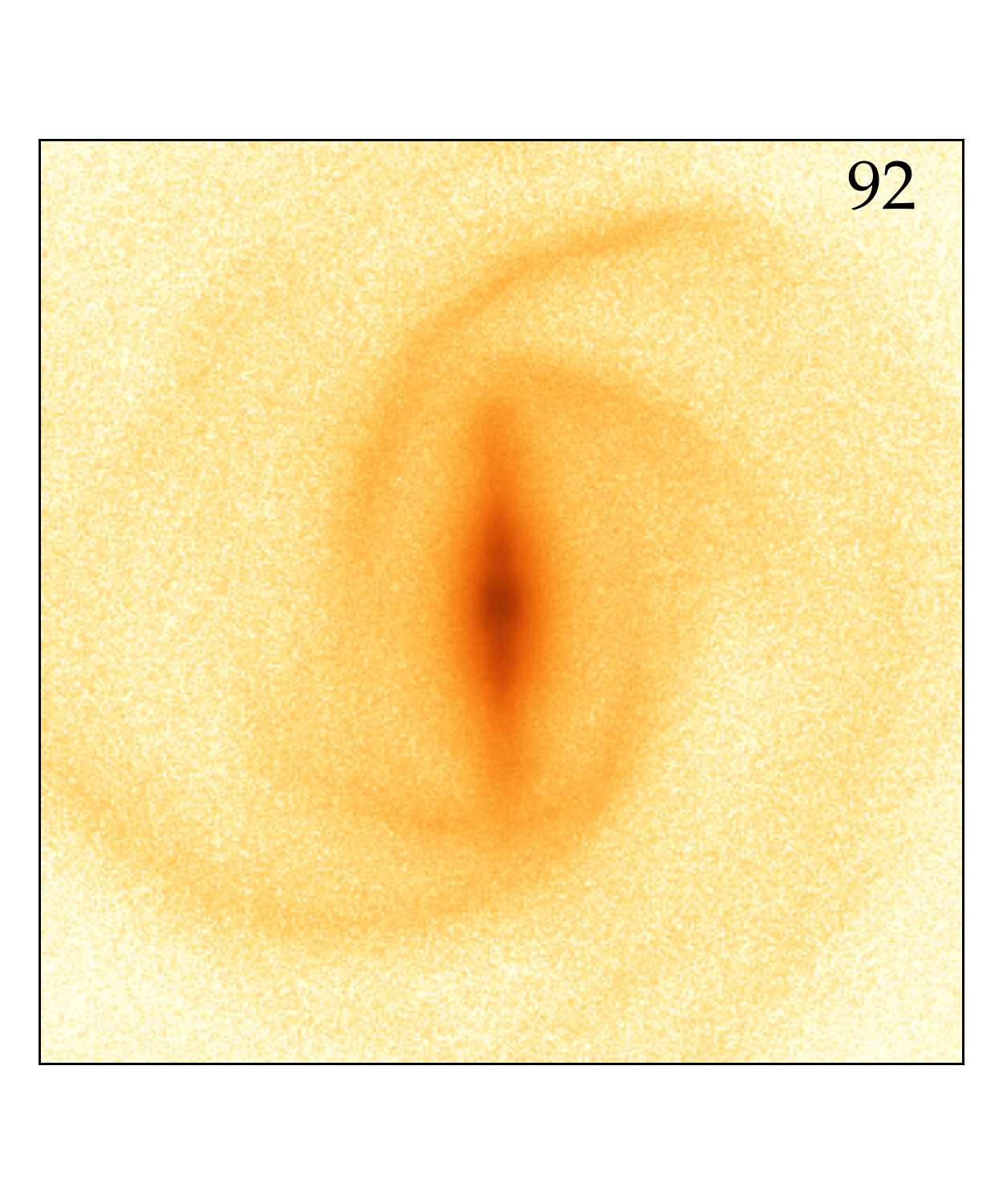}&
	\includegraphics[trim={.5cm .5cm .5cm 1.5cm},clip,width=3.1cm]{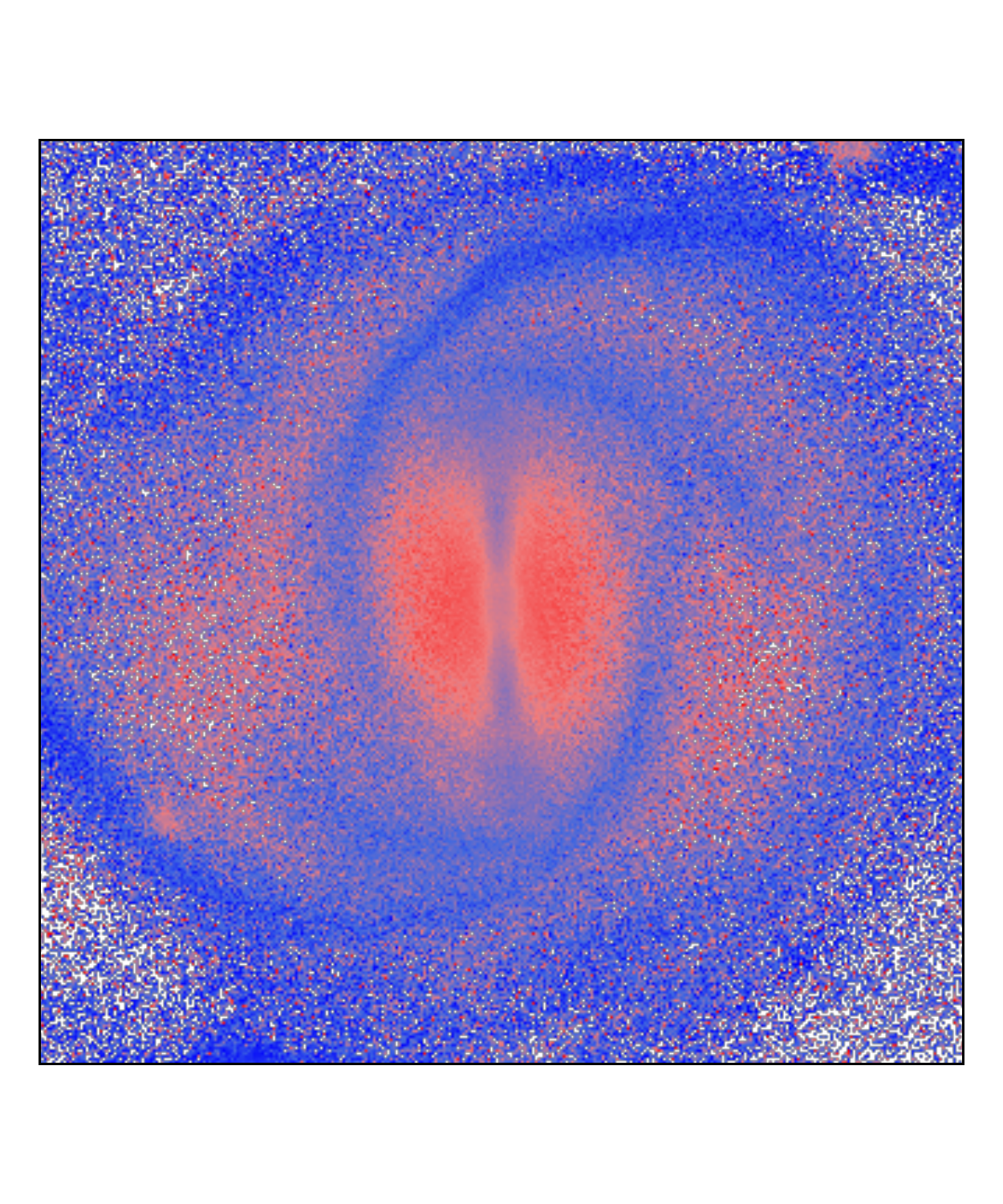}&
	\includegraphics[trim={.5cm .5cm .5cm 1.5cm},clip,width=3.1cm]{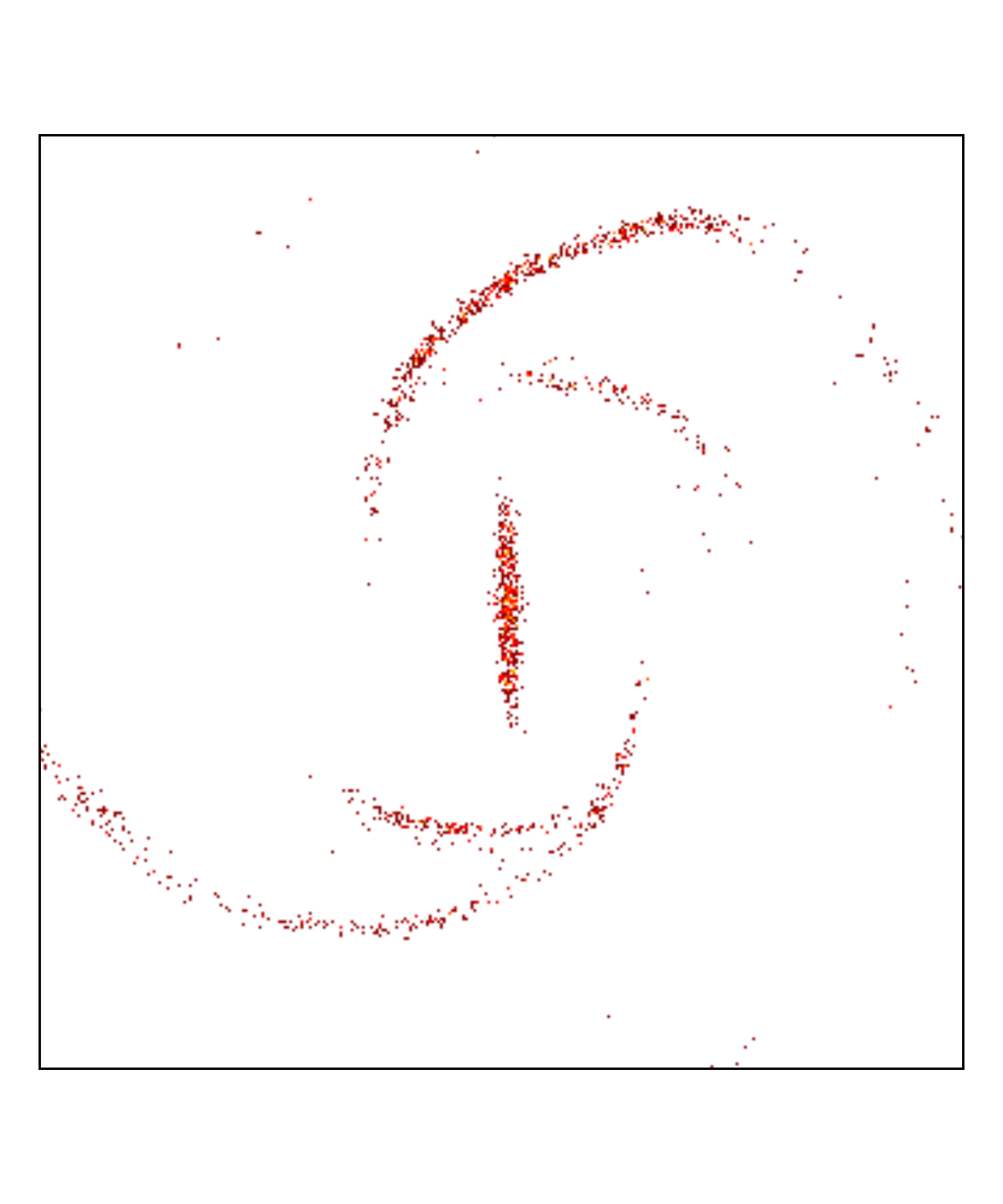}\\
	\includegraphics[trim={.5cm .5cm .5cm 1.5cm},clip,width=3.1cm]{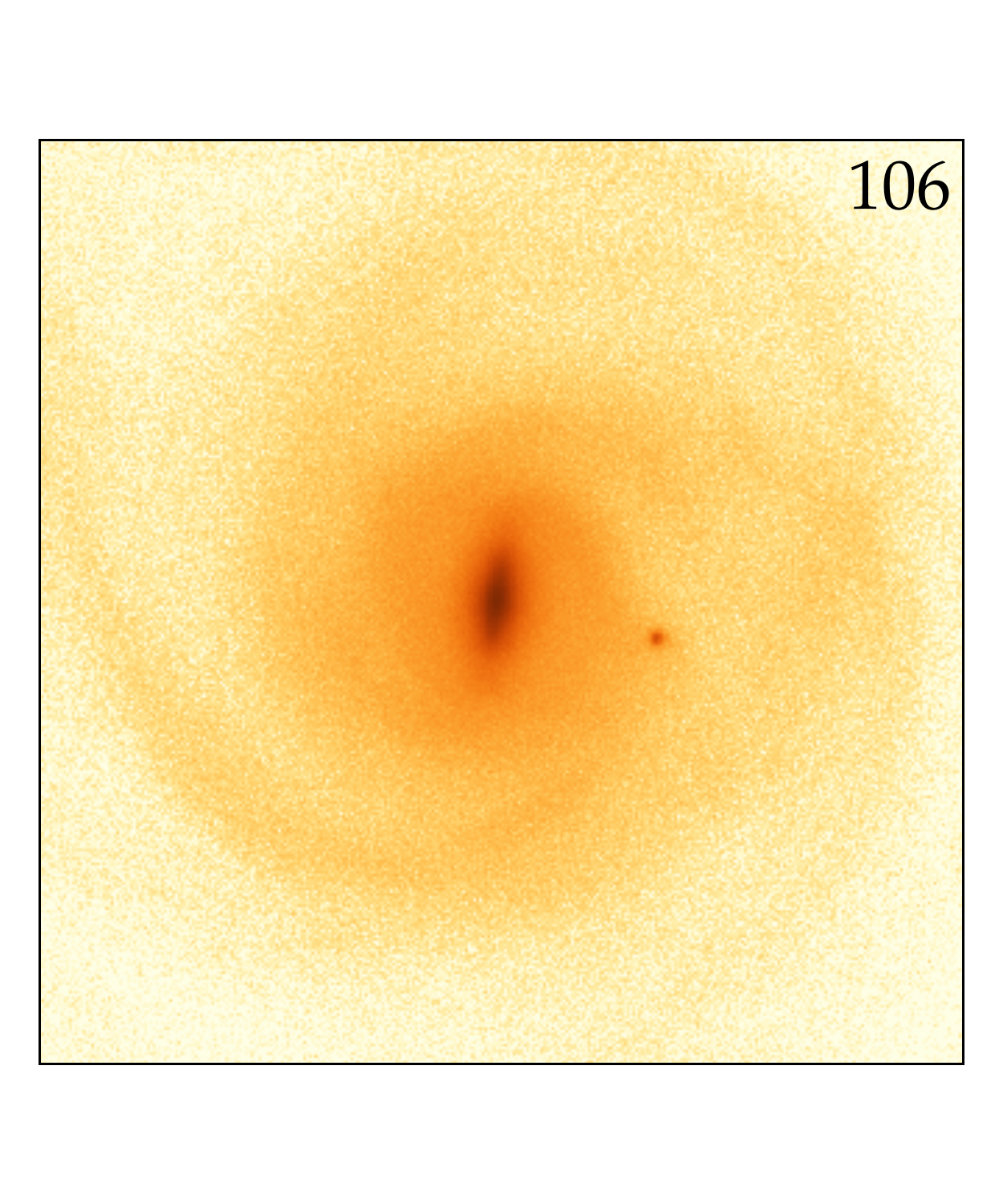}&
	\includegraphics[trim={.5cm .5cm .5cm 1.5cm},clip,width=3.1cm]{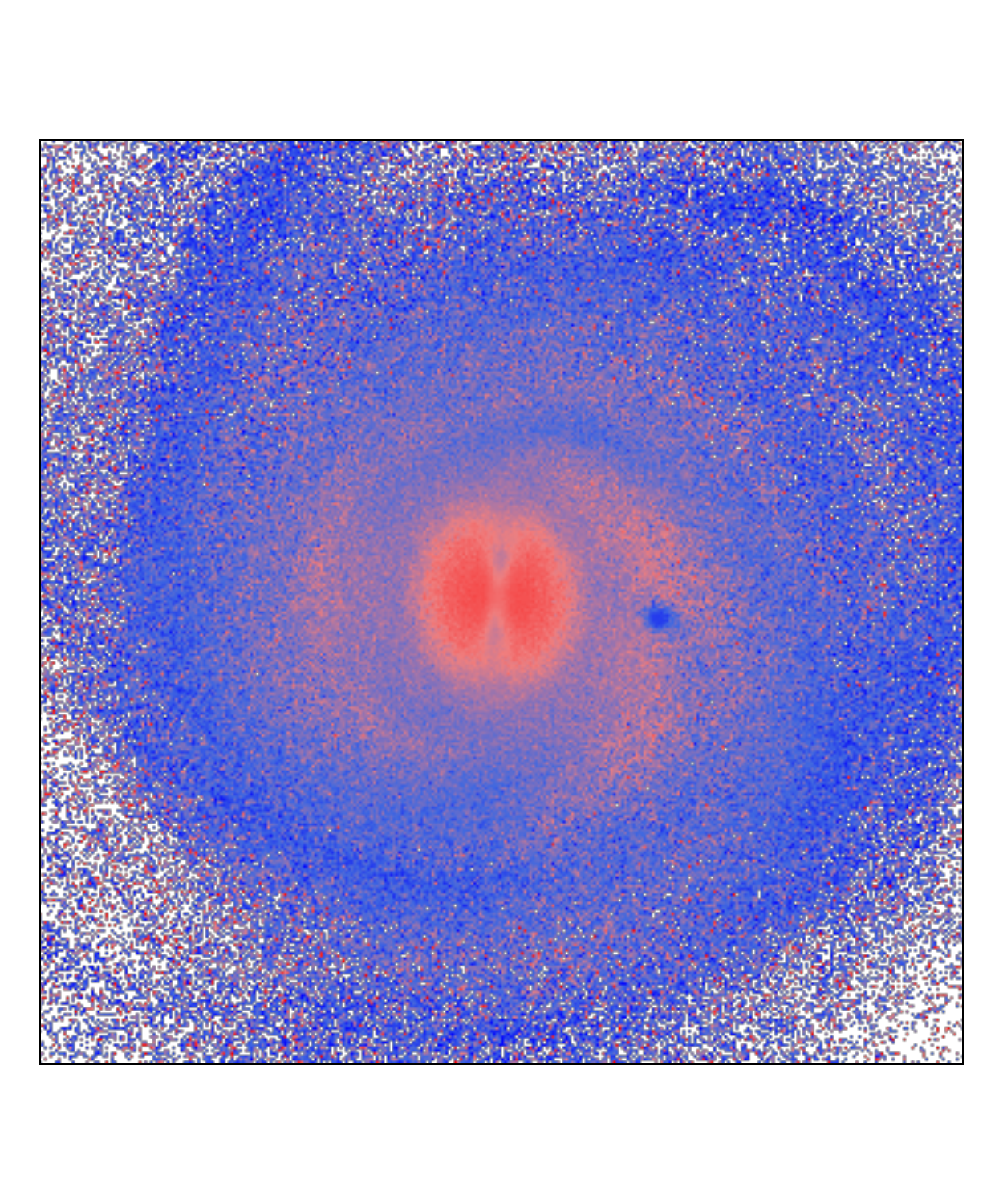}&
	\includegraphics[trim={.5cm .5cm .5cm 1.5cm},clip,width=3.1cm]{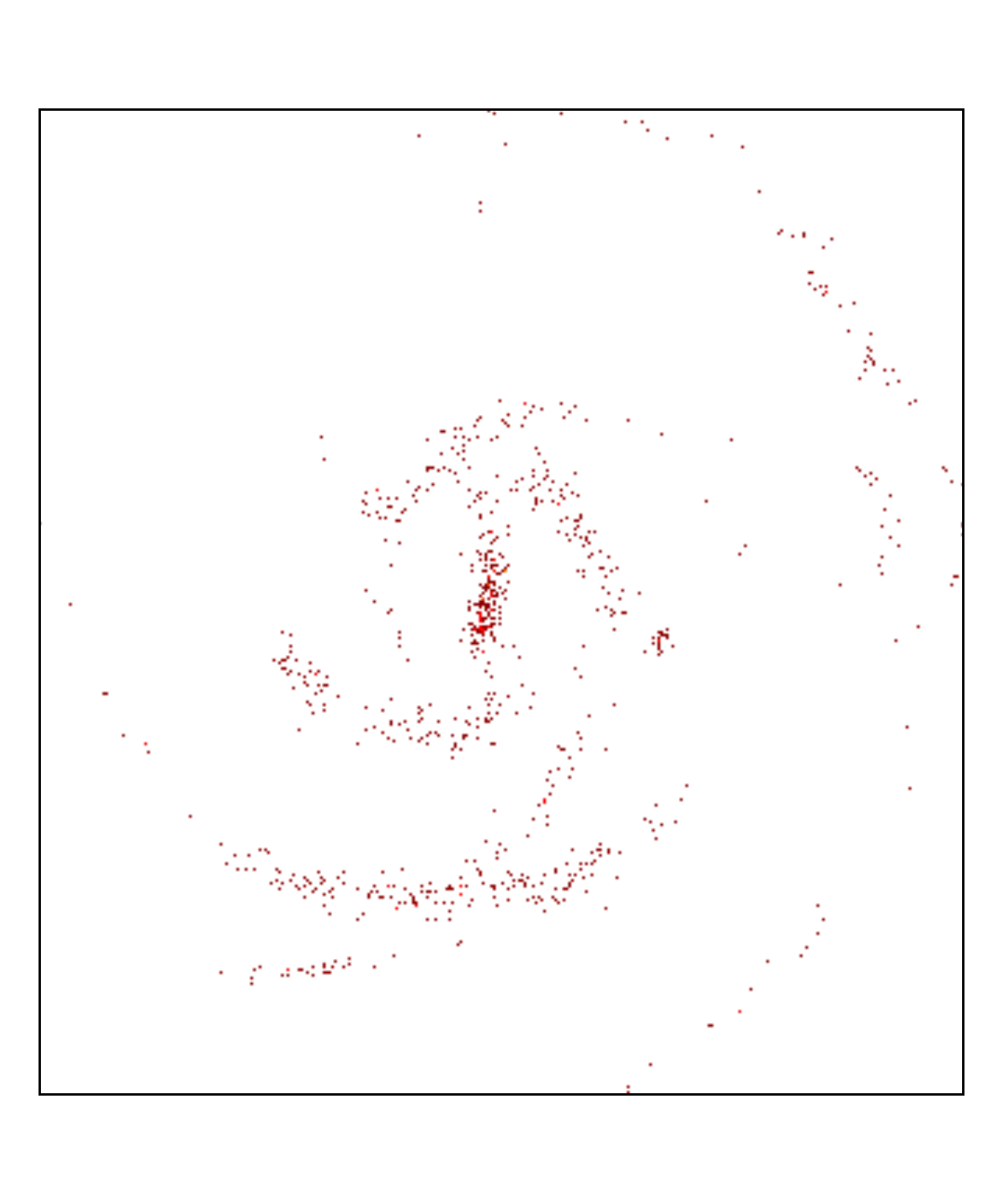}\\
	\includegraphics[trim={.5cm .5cm .5cm 0cm},clip,width=3.4cm]{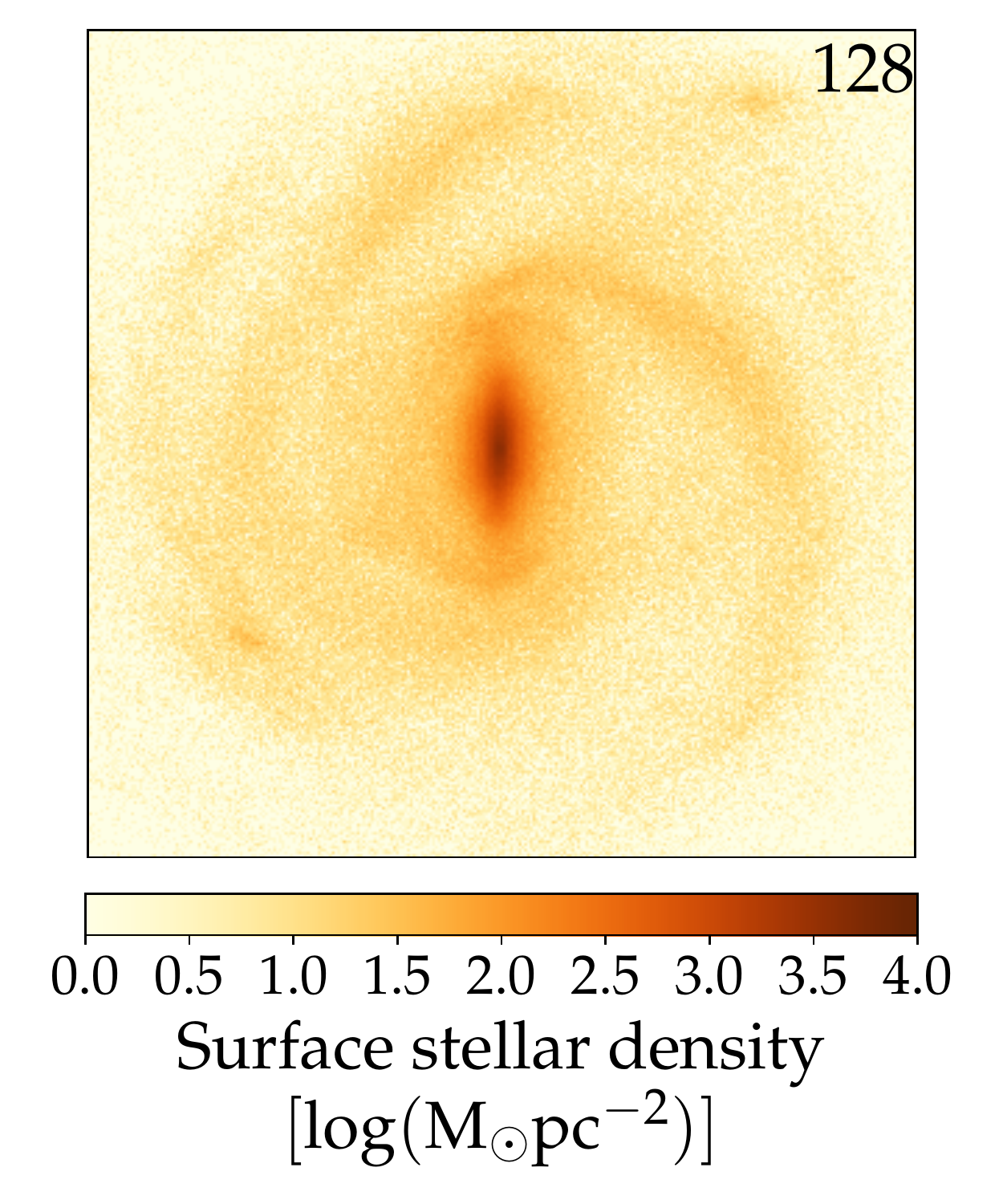}&
	\includegraphics[trim={0cm 0cm 1cm 0cm},clip,width=3.3cm]{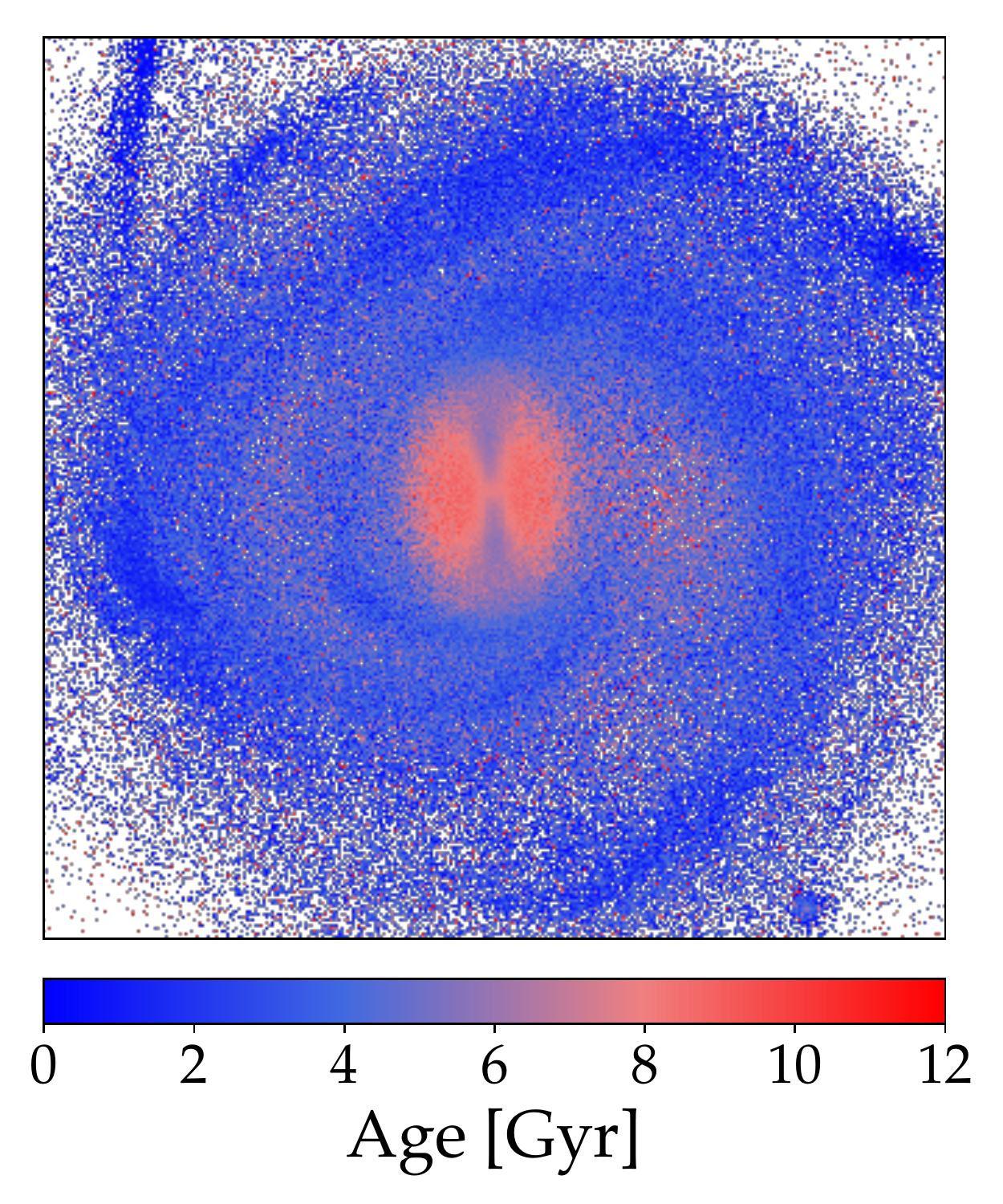}&
	\includegraphics[trim={.5cm .5cm .5cm 0cm},clip,width=3.4cm]{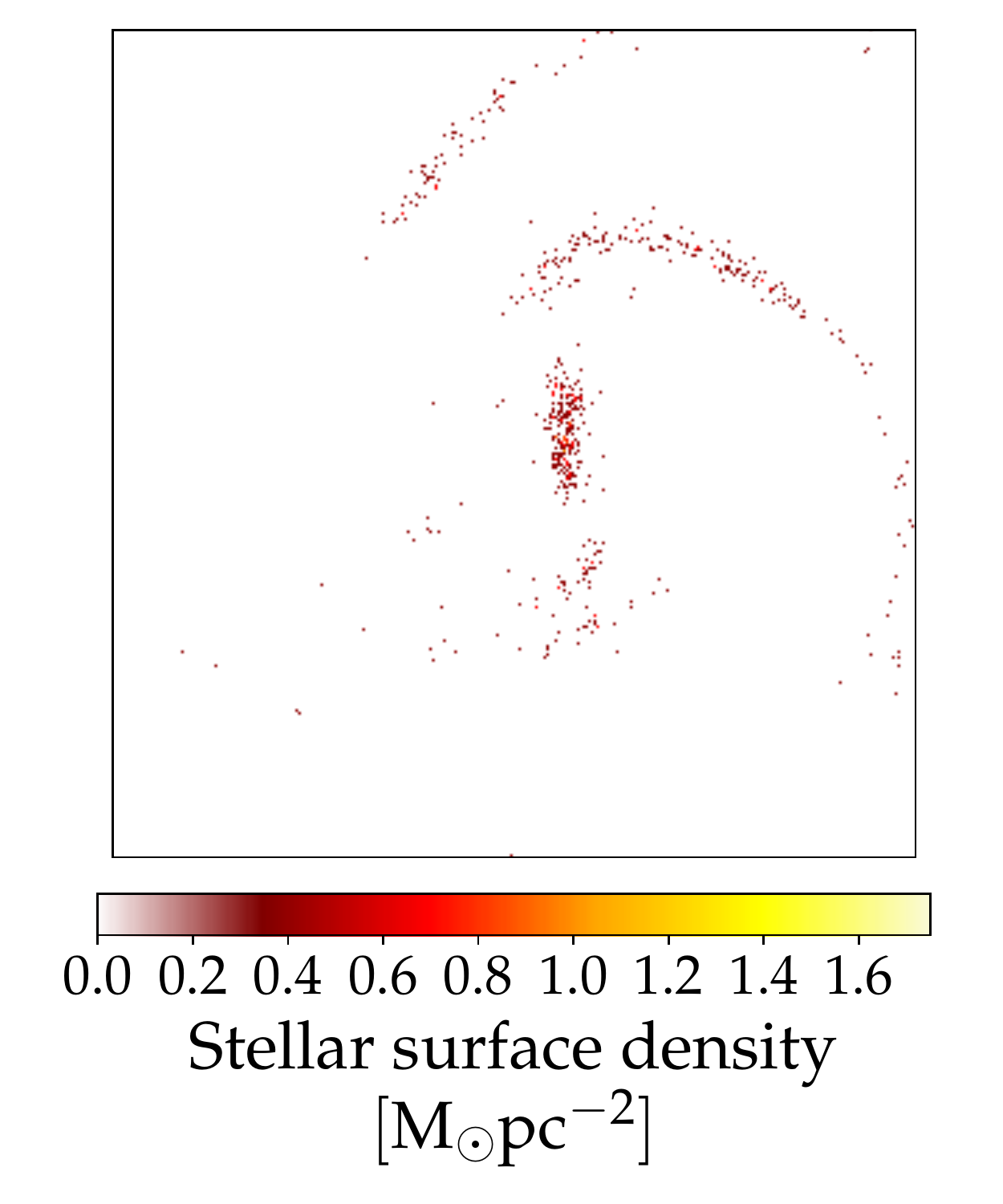}
	\end{tabular}
	\caption{Each plot represents a 40$\times$40$\times$40\,kpc box with the galaxy centred within the box. Left: Face-on surface stellar density maps with the total halo mass decreasing down the column. Middle: Average age maps displaying strong signals for the SFD desert feature. Right: Surface stellar density maps for the young stars, <10 Myrs, also displaying the SFD feature with SF mainly located within the bar region and along the spiral arms of the galaxies.}
\label{Figure:AGE}   
\end{figure*}

\section{Simulations}\label{section:simulation}

In this paper we analyse 6 simulated barred spiral galaxies with a range of star formation histories and bar formation epochs selected from the 33 galaxies presented in \cite{Martig2012}. In this section we give a brief overview of the simulation technique: the motivations behind our sample, the algorithm used to determine the strengths, lengths and formation epoch of bars, and the properties of the galaxies in our sample.\par

\subsection{Simulation Technique}

The simulation technique requires two parts. The first involves a dark matter-only cosmological simulation with the adaptive mesh refinement code \textsc{ramses} \citep{teyssier2002}. From this simulation, the merger and accretion histories for halos within isolated environments at z=0, with masses between $\mathrm{{2.7} \times {10^{11}}}$ and $\mathrm{{2} \times {10^{12}} M_{\odot}}$, are extracted.\par 

The target halos are then re-simulated at higher resolution. The re-simulations begin at z=5 with a seed galaxy containing stars, gas and dark matter. As shown in Appendix A.5 of \cite{Martig2009} the initial conditions do not affect the subsequent evolution of the simulated galaxy due to the very small mass of the initial seed galaxy. This galaxy's evolution is followed down to z=0 with mergers, as well as dark matter and gas accretion prescribed by the cosmological simulation; we refer the reader to \cite{Martig2012} for details on the properties of the incoming galaxies. The re-simulation has a spatial resolution of 150 pc, mass resolution of $\mathrm{{1.5}\times{{10}^4} M_{\odot}}$ for gas particles, of $\mathrm{{7.5}\times{{10}^4} M_{\odot}}$ for star particles (or $\mathrm{{1.5}\times{{10}^4} M_{\odot}}$ for star particles formed during the simulation from the gas) and $\mathrm{{3}\times{{10}^5} M_{\odot}}$ for dark matter particles in a box of 800 kpc using the particle mesh-code described in \cite{Bournaud2002,Bournaud2003}. Gas dynamics are modelled using a  sticky particle algorithm.\par

Star formation is modeled using a Kennicutt-Schmidt relation \citep{1998kennicutt} with a 1.5 exponent and a star formation threshold of $\mathrm{0.03M_{\odot}pc^{-3}}$. Kinetic feedback from supernovae is included such that 20 percent of supernova energy is redistributed to the gas particles, and stellar mass loss is also taken into account \citep{Martig2012}.\par

\begin{center}
\begin{table}
\centering
\begin{tabu} to 0.45\textwidth { | X[c] | X[c] | X[c] | X[c] | X[c] | }
\hline
	\textbf{Halo} & \textbf{ $\mathrm{M_*}$ [$\mathrm{10^{10}M_{\odot}}$]} &\textbf{$\mathrm{L_{bar}}$ [kpc]} &\textbf{$\mathrm{S_{bar}}$} & \textbf{$\mathrm{T_{bar}}$ [Gyr]} \\
    \hline
    37 & $\mathrm{12.0}$ & 6.0 & 0.70 & 8.5\\
    
    45 & $\mathrm{10.2}$ & 6.6 & 0.76& 6.8\\
    
    82 & $\mathrm{3.81}$ & 4.4 & 0.38 & 2.0\\
    
    92 & $\mathrm{4.38}$ & 5.6 & 0.71 & 6.8\\
    
    106 & $\mathrm{4.29}$ & 3.1 & 0.45 & 6.6\\
    
    128 & $\mathrm{2.69}$ & 3.3 & 0.74 & 4.7\\
    \hline
\end{tabu}
\caption{Properties of the model galaxies taken from z=0. For each halo we provide the halo index number, the stellar mass ($\mathrm{M_*}$) calculated by summing star particles to the R${}_{25}$ limit, the bar length ($\mathrm{L_{bar}}$), and the bar strength ($\mathrm{S_{bar}}$). The final column gives the bar formation epoch of the galaxy in lookback time.}
\end{table}
\end{center}
\subsection{Sample Selection}

From the sample of 33 simulated galaxies described in \cite{Martig2012} we select 6 that display a wide range of star formation histories, masses, and bar lengths, strengths, and formation epochs. By selecting this limited sample we can do a more detailed analysis but still explore the diversity of the larger sample.\par

Column 1 of Figure 1 shows the surface stellar density maps of the galaxies face-on at z=0, ranked in order of largest halo mass (top) to lowest (bottom). The main properties are highlighted in Table 1.\par

All of the galaxies begin with a merger-intense phase which contributes to the build up of a hot stellar component for ages greater than 9 Gyr. After this the disk builds with features such as spiral arms, and, more pivotal to the focus of this paper, the bars and star formation desert regions. Halo 106 differs from this scenario by having three epochs of bar formation with the first two being destroyed by mergers. For this case we list properties relevant to the final bar, for which the bar formation epoch is given in the final column of Table 1.\par

\subsection{Bar detection}
The sample chosen has a range of bar formation epochs as listed in Table 1. Bars can be identified visually but, for a more systematic study, we identify bars through an automatic detection method. Using this method also allows for the computation of bar strengths and lengths, as explained in more detail in \cite{KK2012}. This method of bar detection is founded on the azimuthal spectral analysis of surface density profiles of face-on galaxies. \par

Bars are identified in this method with even-mode phase signatures, m=2 being the most prominent, within the `bar detection region'. The `bar detection region' we define as starting between 900 and 1500 pc. We do not begin detecting bars within 900 pc because small variations in $\mathrm{\Phi_2}$ are produced by off-centering (a result of the resolution limits) and central asymmetries cause the mis-identification of barred or non-barred systems. Once a $\mathrm{\Phi_2}$ phase is detected it must be constant for at least 1500 pc for the galaxy to be classified as barred. After a bar has been found its length (determined by the extent of the constant phase $\mathrm{\Phi_{2}}$) and strength are measured. To calculate strength we use the definition proposed by \cite{AJAL1998}:\par

\begin{equation}
\begin{array}{c@{\qquad}}
	\mathrm{S \equiv r^{-1}_{bar} \int^{r_{bar}}_{0} \frac{A_2}{A_0} dr}
\end{array}
\end{equation}

\noindent where the radial limit of the bar is defined by $\mathrm{r_{bar}}$ and A${}_{2}$ and A${}_{0}$ represent the Fourier amplitudes for the 0${}^{\mathrm{th}}$ and 2${}^{\mathrm{nd}}$ modes.\par 

Bars observable to high redshift have a strength $\mathrm{S\geqslant 0.2}$ \citep{Sheth2008}. At this strength, bars can still be confused with flattened early-type galaxies. To reduce this effect we identify true bars by using the constraint that the strengths of the m=2 mode must be greater than, or equivalent to, 0.3 in two orthogonal edge-on projections. \par

\subsection{Defining the SFD}\label{method:sfd_definition}
\begin{figure}
\centering
     \includegraphics[width=0.4\textwidth]{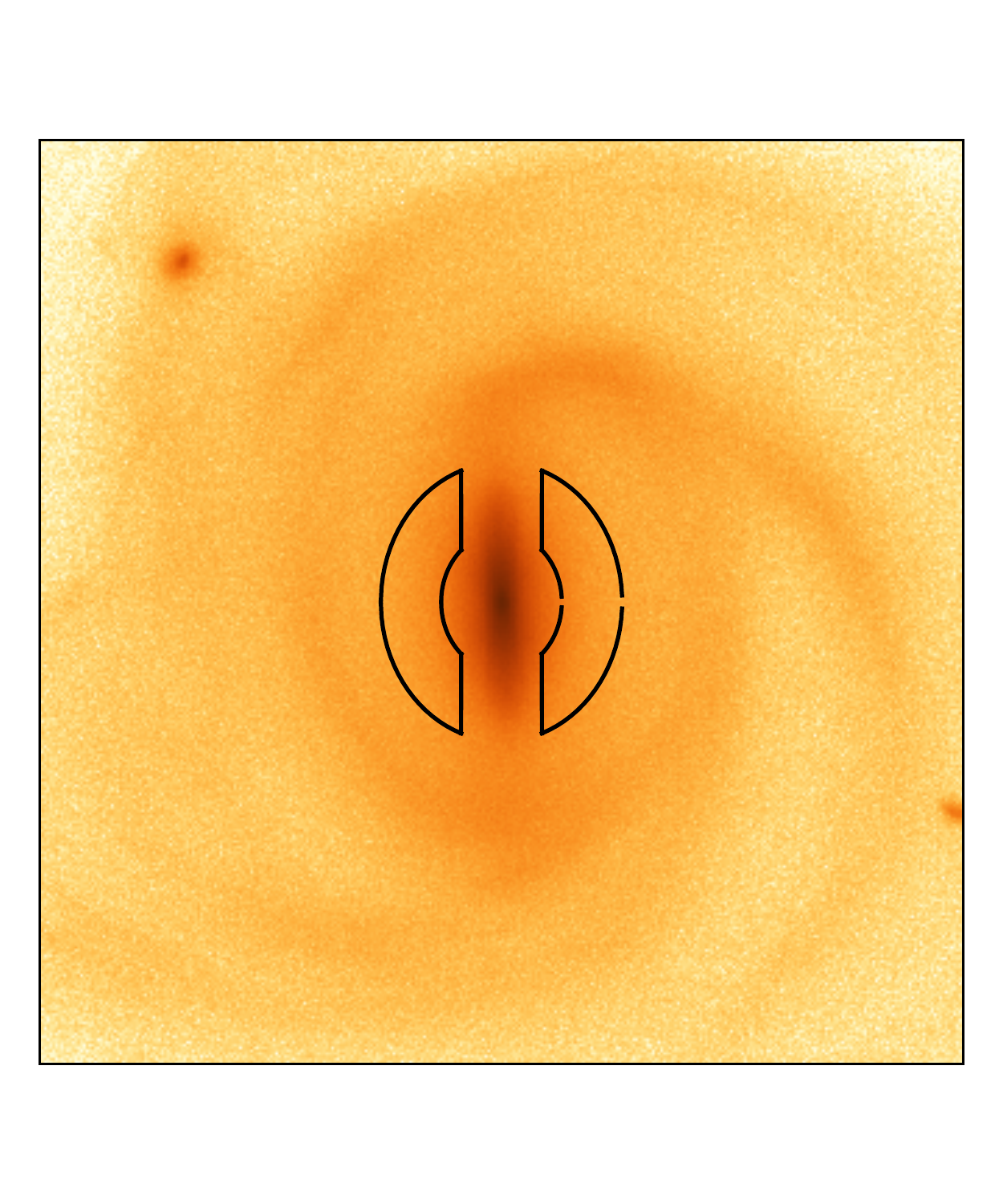}\\
    \caption{The two `C'-shaped regions we define as the SFD.}
    \label{figure:SFD_region}
\end{figure}

Figure \ref{Figure:AGE}, column 2 shows the mean age maps for our sample of simulated galaxies. The blue colour highlights younger stellar populations while the red shows older populations. In all of the galaxies in the sample there is a region either side of the bar, within the region the bar sweeps out, displaying consistently older populations. This coincides with the SFD region seen observationally in  \cite{James2015}. The size of the SFD is closely associated with bar length and it never extends further than the radius of the bar. The SFD region is bordered by the inner ring which contains a younger population. In all the cases the bar appears to be a younger feature than the SFD but, in these simulated galaxies, older than the ring and disk.\par

We define the SFD as the region encompassed in a ring excluding the bar and the bulge. We fit the shape of the ring as an ellipse using the bar-length as the major axis and take the width of the bar as 1 kpc. Additionally, we remove stars which are associated with the bulge from the SFD by removing an inner ellipse shaped region and then removing the bar itself. This results in two `C'-shaped regions shown in Figure \ref{figure:SFD_region}.\par
Finally we remove `interloper' stars. These are stars which are only passing through the SFD region at the point of selection. To remove them from the SFD sample we define a z-axis (perpendicular to the plane of the galaxy) limit of 2 kpc either side of the central plane on a snapshot 0.075 Gyr from the selection snapshot and compare the stellar IDs to those in the selection snapshot, only keeping the stars which appear in both snapshots.\par

\section{Results}\label{section:results}
\subsection{Age Maps}

To determine whether the SFD region in the simulated galaxy sample is a result of a lack of star formation we refer to the young star maps shown in Figure \ref{Figure:AGE}, column 3. Here we present the surface stellar density of stars less than 10 Myr old, at z=0. High concentrations of young stars are seen within the bar, the spiral arms, and along the inner ring. Some of the rings are populated fully with young stars, while others exhibit broken profiles. For those that do show broken inner rings, the stars are more concentrated at the regions connecting to the ends of the bar. Very few, if any, young stars are seen in the SFD regions. When making side-by-side comparisons between the age and young star maps it is clear that they both highlight the SFD region, the age maps through the older mean age populations and the young star maps through a lack of young stars.\par

However, the figures presented in this section only show the mean age population and do not tell us about the distribution in ages within the SFD region in comparison to the bar and global populations. To understand how the age distributions differ between regions we need to investigate how the age distributions change with respect to lookback time. \par

\subsection{Star Formation Histories} \label{sec:SFH}

From the mean stellar age maps in Figure \ref{Figure:AGE} centre column there is a clear difference between the mean ages of stellar populations within the SFDs, bars, and inner rings of the galaxies.\par 

\begin{figure*}
\centering
	\begin{tabular}{lr}
	\includegraphics[width=0.48\textwidth]{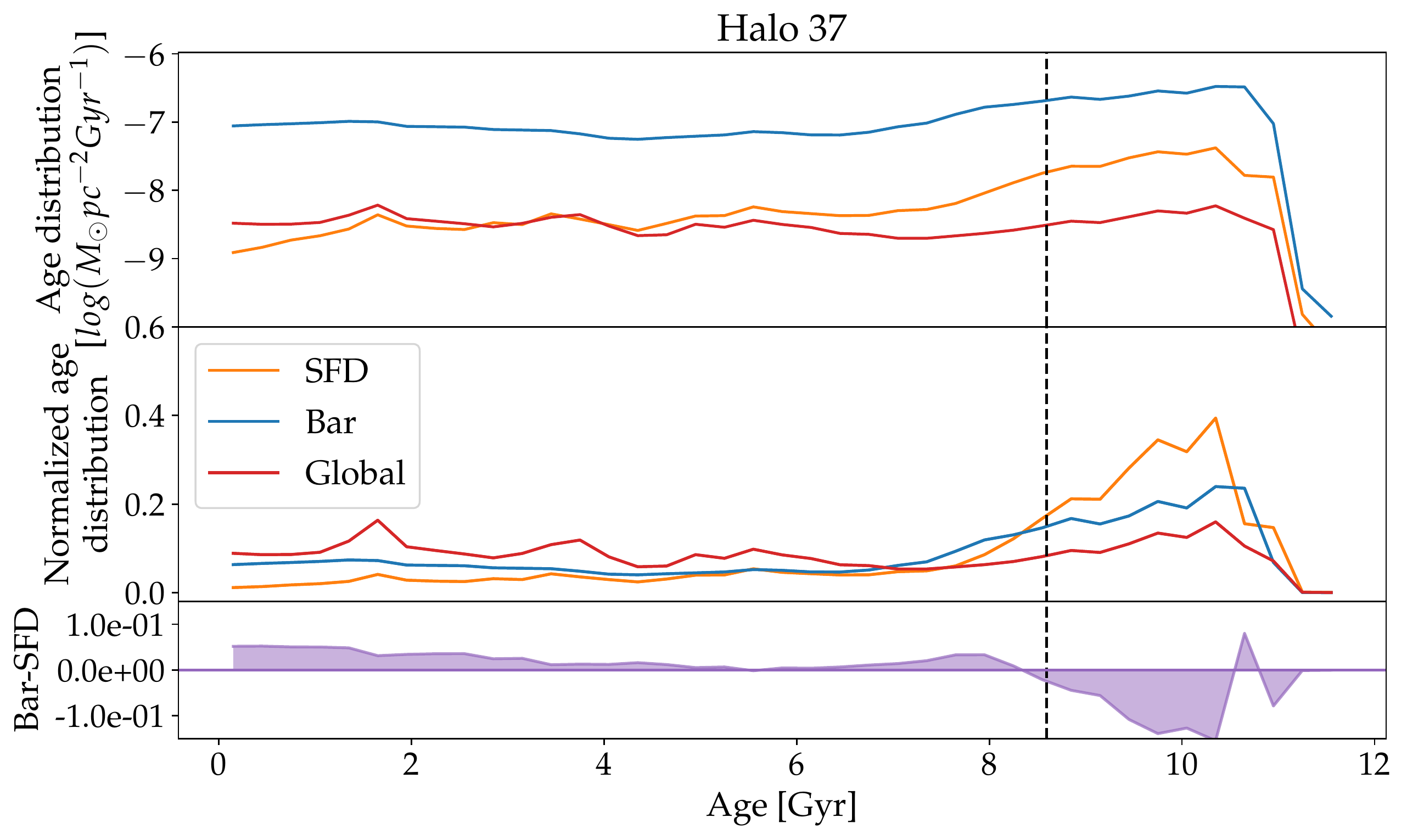}&
	\includegraphics[width=0.48\textwidth]{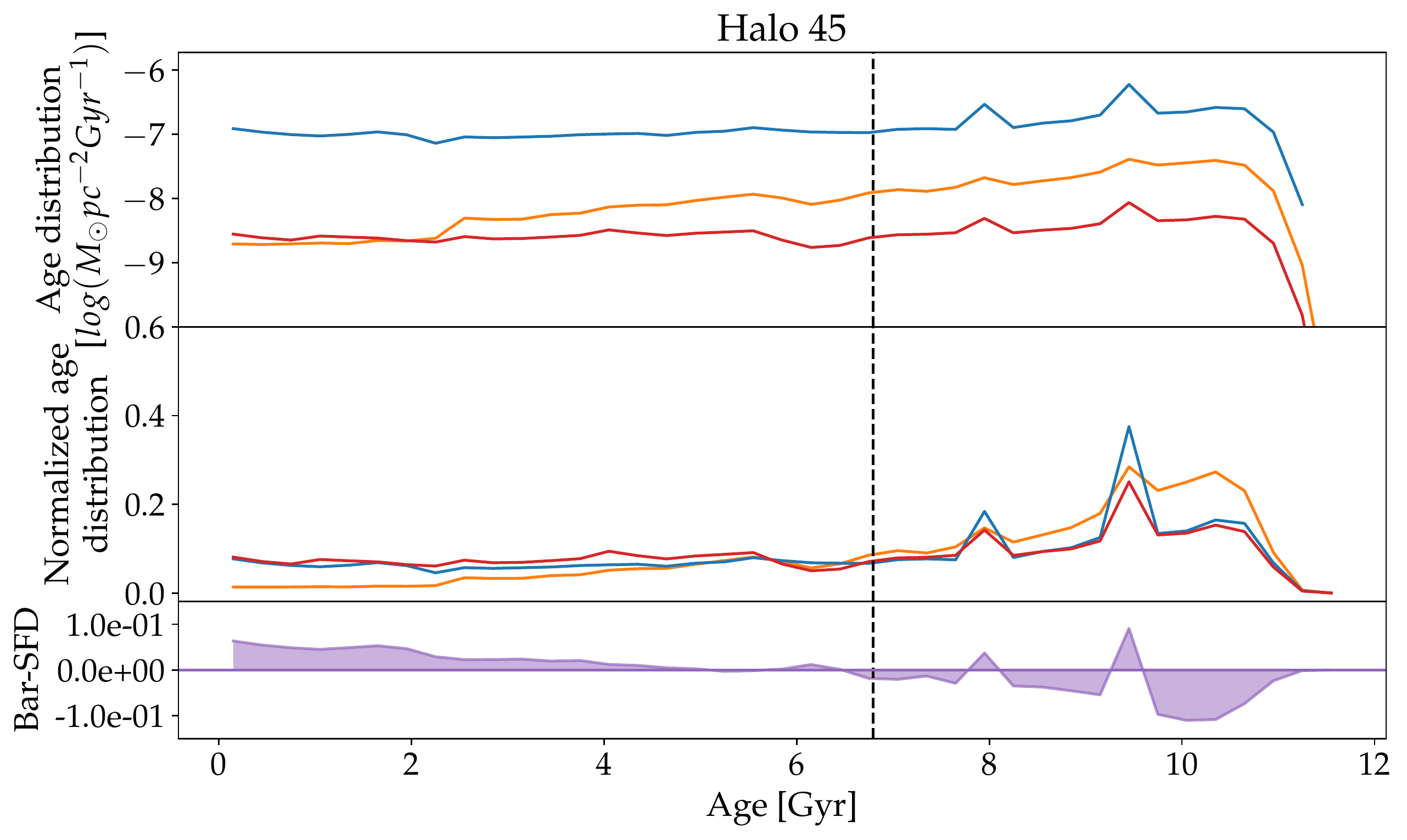}\\
	\includegraphics[width=0.48\textwidth]{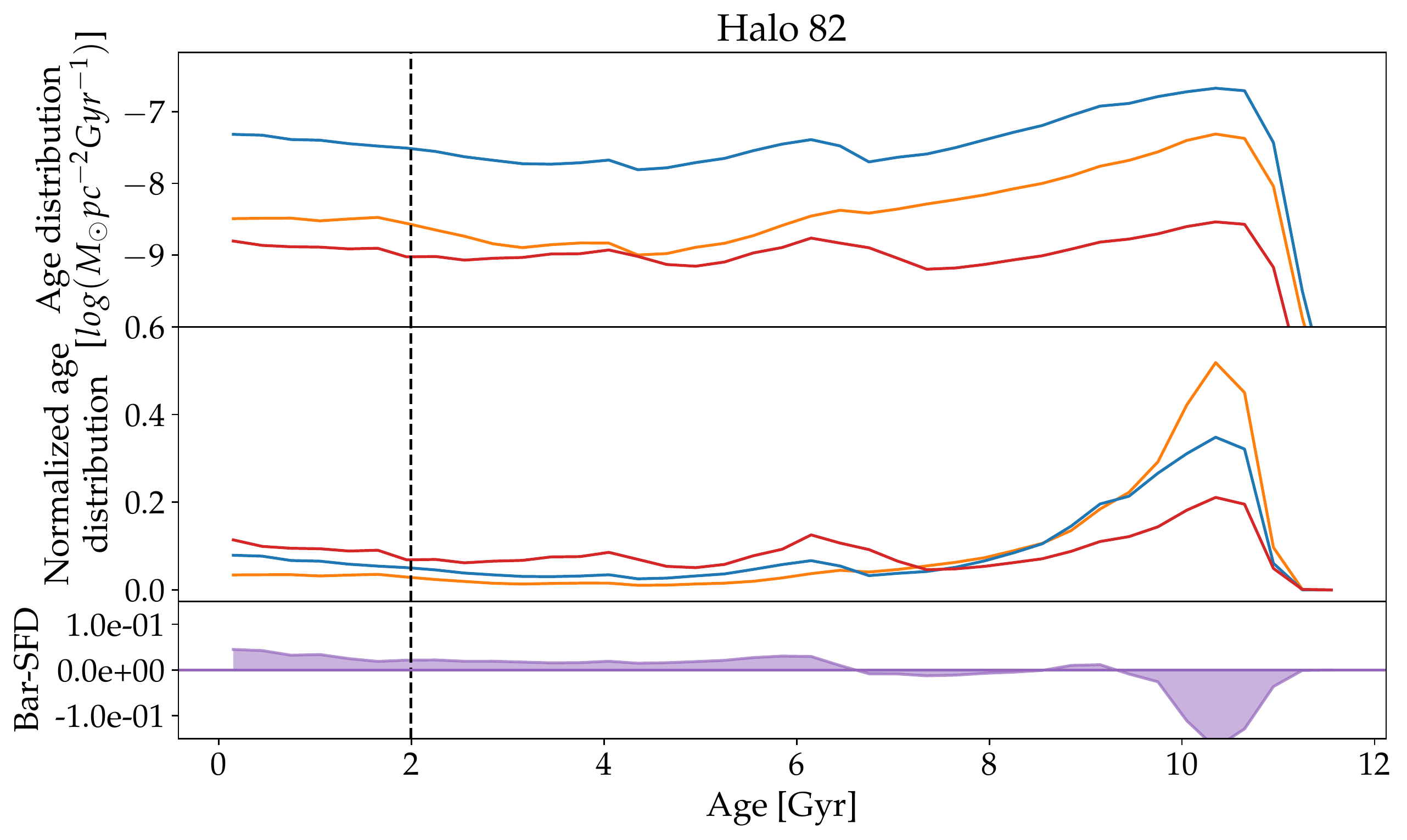}&
	\includegraphics[width=0.48\textwidth]{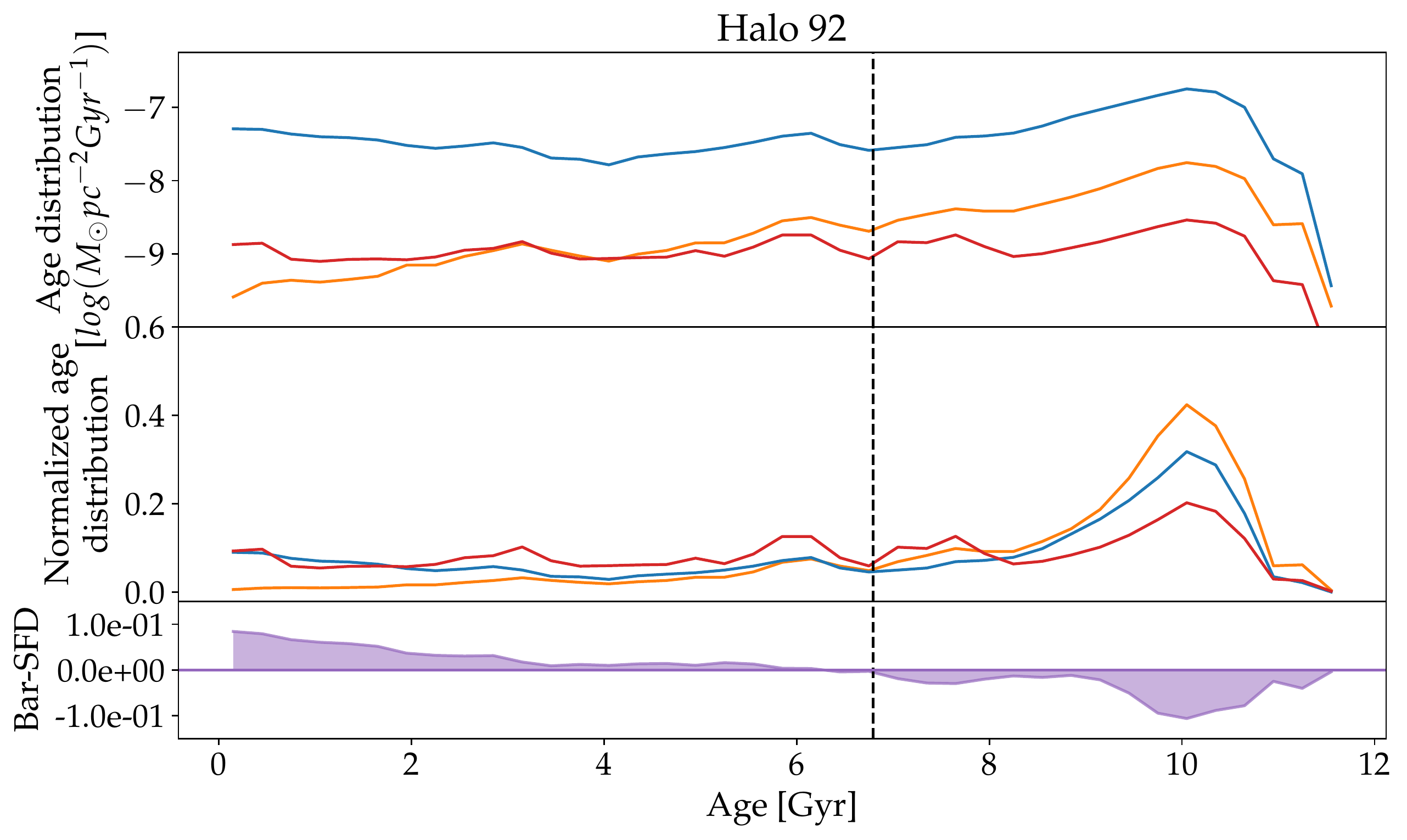}\\
	\includegraphics[width=0.48\textwidth]{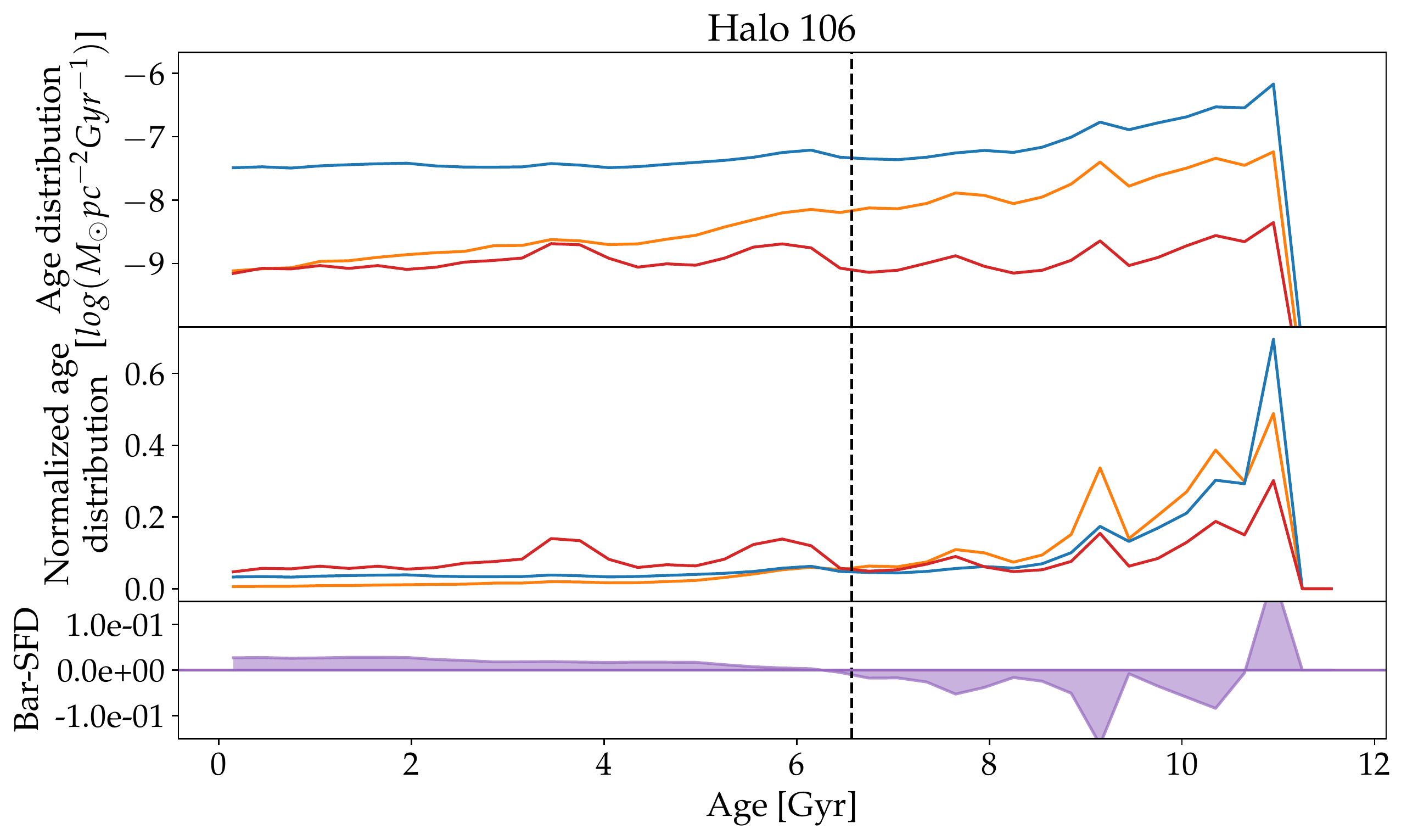}&
	\includegraphics[width=0.48\textwidth]{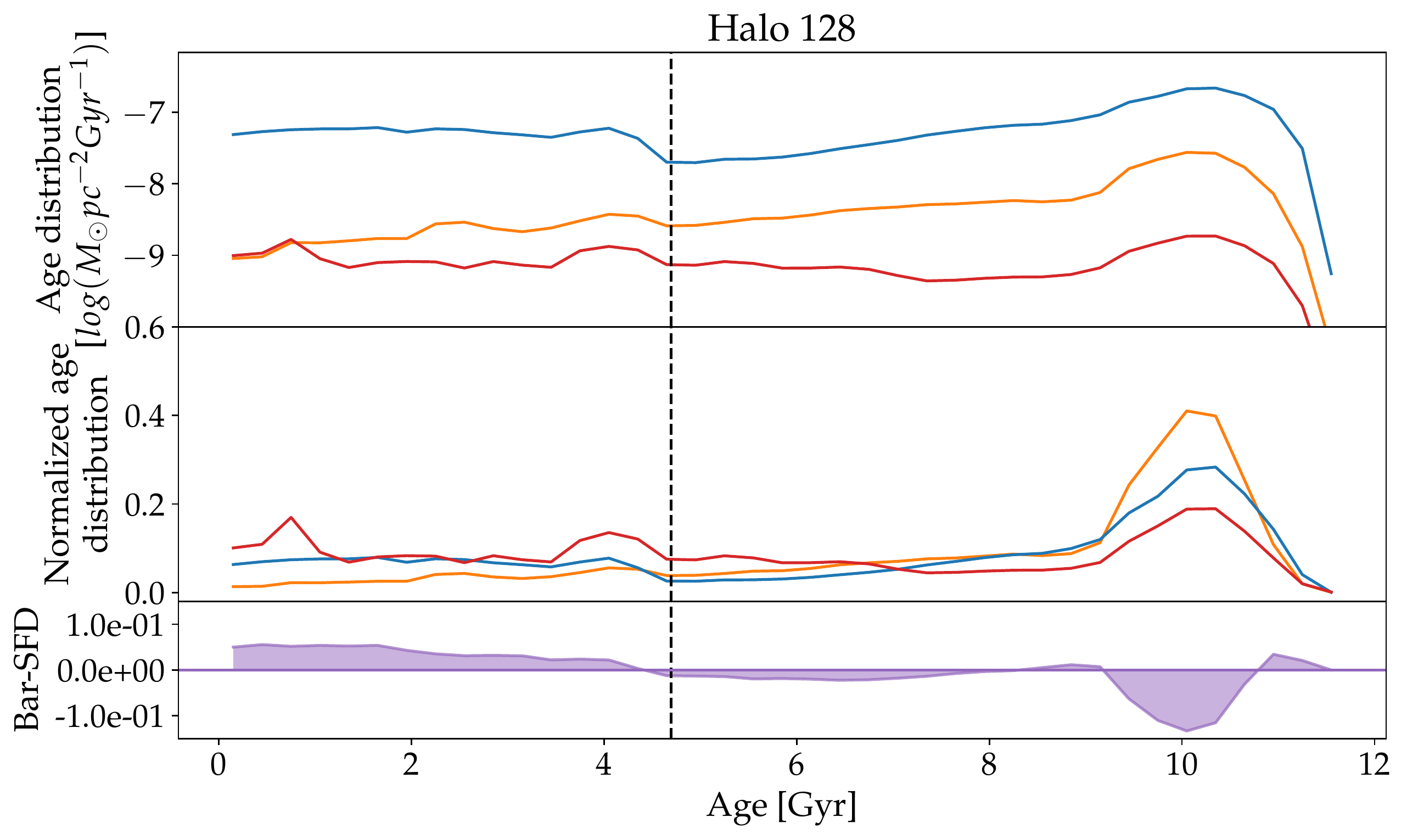}\\
	\end{tabular}
        \caption{For each of the simulated galaxies in our sample we present the age distributions taken from the SFD region, the bar, and the total galaxy at z=0. In each plot we display the age distribution normalised to the surface area of the corresponding regions, the age distribution normalised to an area of 1, and the residual (the bar minus the SFD age distribution). Marked on each plot by the vertical dashed line is the time of bar formation. This line coincides with the downturn in the age distribution of the SFD and, in most cases (see section \ref{sec:SFH}), the change of the residual from negative to positive.}
\label{Figure:SFH}
\end{figure*}

In Figure \ref{Figure:SFH} we plot the age distribution of stars found in the bar and SFD regions, together with the age distribution of all stars found within a 20$\times$20 kpc $\mathrm{{}^{2}}$ box with a height of 4 kpc. The top section of each plot shows the bar, SFD, and global age distributions normalised by area. The onset of the bar is marked with a black dashed line. The bar always shows a $\sim$10 times higher surface density in the age distribution when compared to the SFD and global galaxy, reflecting the higher mass surface density in the bar. The shape of the age distributions for the bar and global galaxy are actually very similar, and the formation of the bar does not seem to have any impact on star formation globally in the galaxy. By contrast, the age distribution of the SFD shows a relative lack of young stars after the formation of the bar.\par

For galaxies 37, 45, 92, 106, and 128 the drop in the age distribution of the SFD coincides with the onset of the bar. However, in galaxy 82, the drop happens long before the formation of the bar (see section 4.2.1 for more details).\par

To better compare the shapes of the different age distributions, we normalize them to 1 and plot them in the middle panels of each plot. In all cases the global and bar age distributions follow similar shapes, while the SFD gradually drops relative to that of the bar after bar formation. We highlight this effect by showing the difference between the age distributions of the bar and SFD in the bottom panels. For the majority of cases this difference moves from negative to positive after bar formation (corresponding to a change to a lower value for the SFD after bar formation). As the galaxy continues to evolve the residual difference between the bar and SFD tends to increase which we associate with a suppression in the star formation of the SFD region.\par   

Again, galaxy 82 remains an outlier. The transfer of the residual from negative to positive occurs $\sim$5 Gyr before the onset of the bar. While this is not associated with the formation of the bar, there is a ring-like feature which does form during this period.\par

In all galaxies the age distribution of the stars in the SFD does not show a sudden drop at the time of bar formation, contrary to what could have been expected from the mean age maps which show a striking contrast between the mean ages of the SFD and the bar regions. For almost all of the galaxies we see a more gradual decrease in the age distribution of the SFD. If this is a true representation of the star formation histories in observed galaxies, this will make using the SFDs to time the formation of the bar harder than expected. However, there is information in the shape of the difference between the SFD and bar age distributions. Once the bar has formed, for almost all the galaxies, we see a change from negative to positive in the difference between the SFD and the bar. This difference is subtle but, it does imply that there is a suppression of star formation within the SFD after the formation of the bar.\par

\begin{figure*}
	\begin{tabular}{ll}
    \includegraphics[trim={0.25cm 0 2.8cm 0},clip,width=0.29\textwidth]{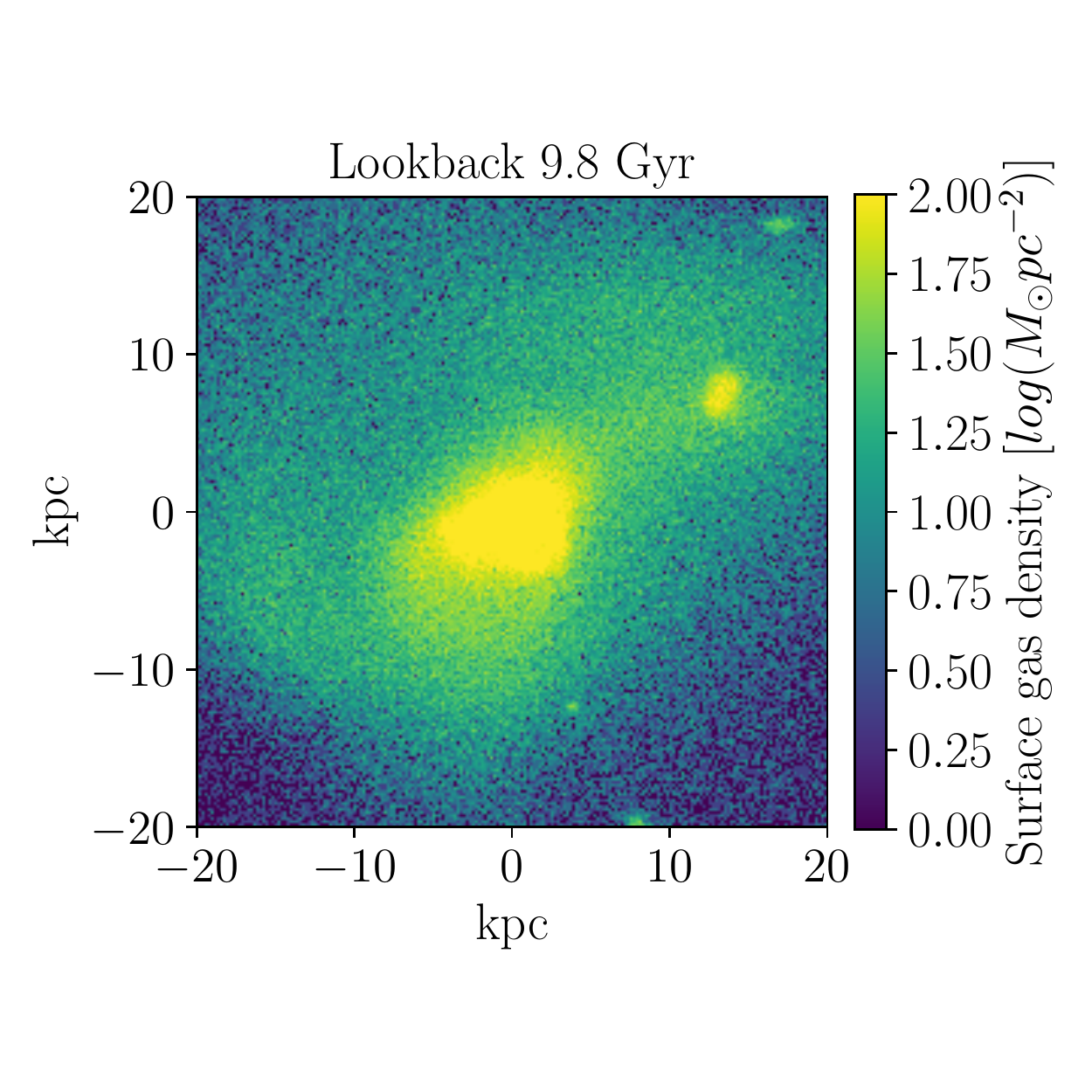}&\includegraphics[trim={0.25cm 0 2.8cm 0},clip,width=0.29\textwidth]{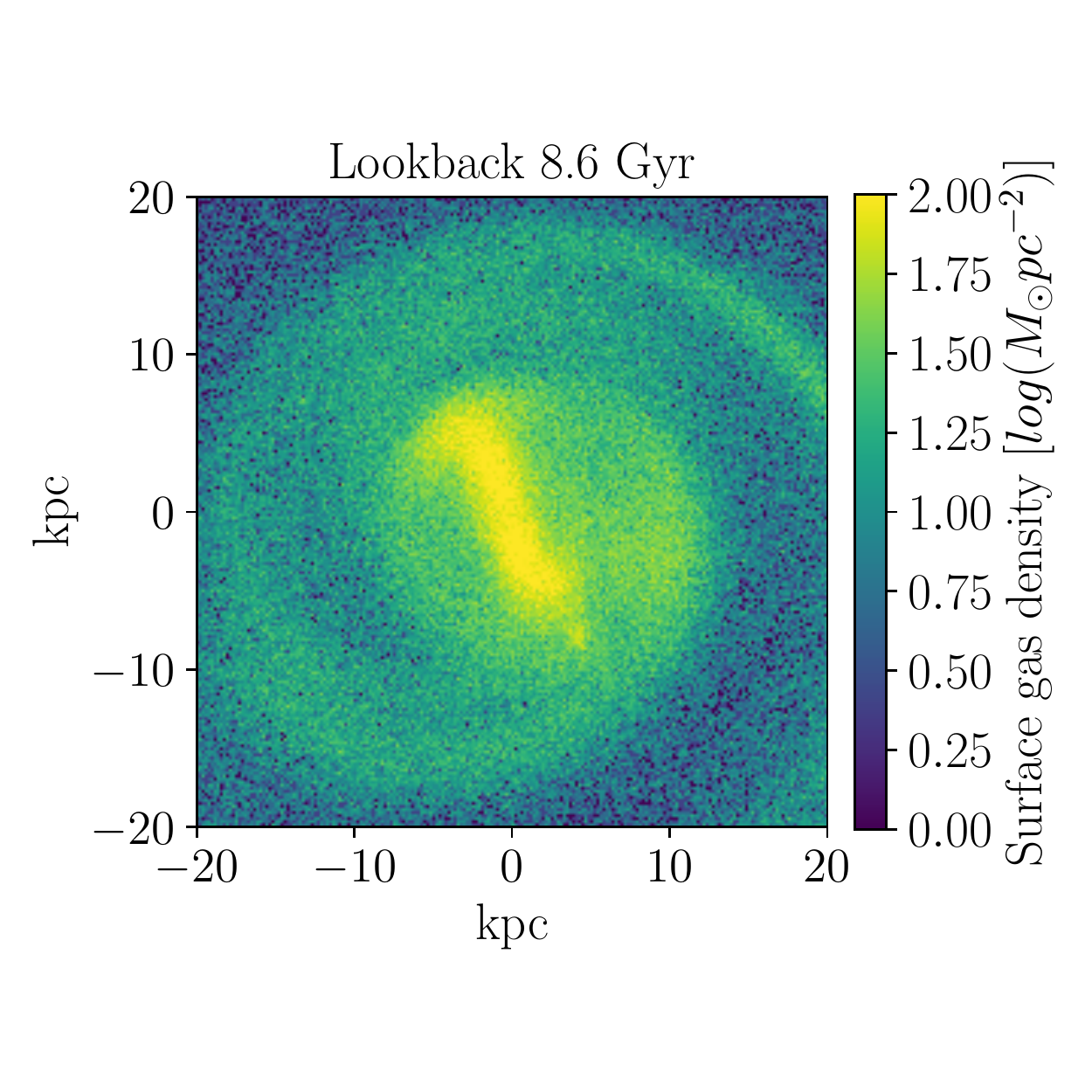}\\
    \includegraphics[trim={0.25cm 0 2.8cm 0},clip,width=0.29\textwidth]{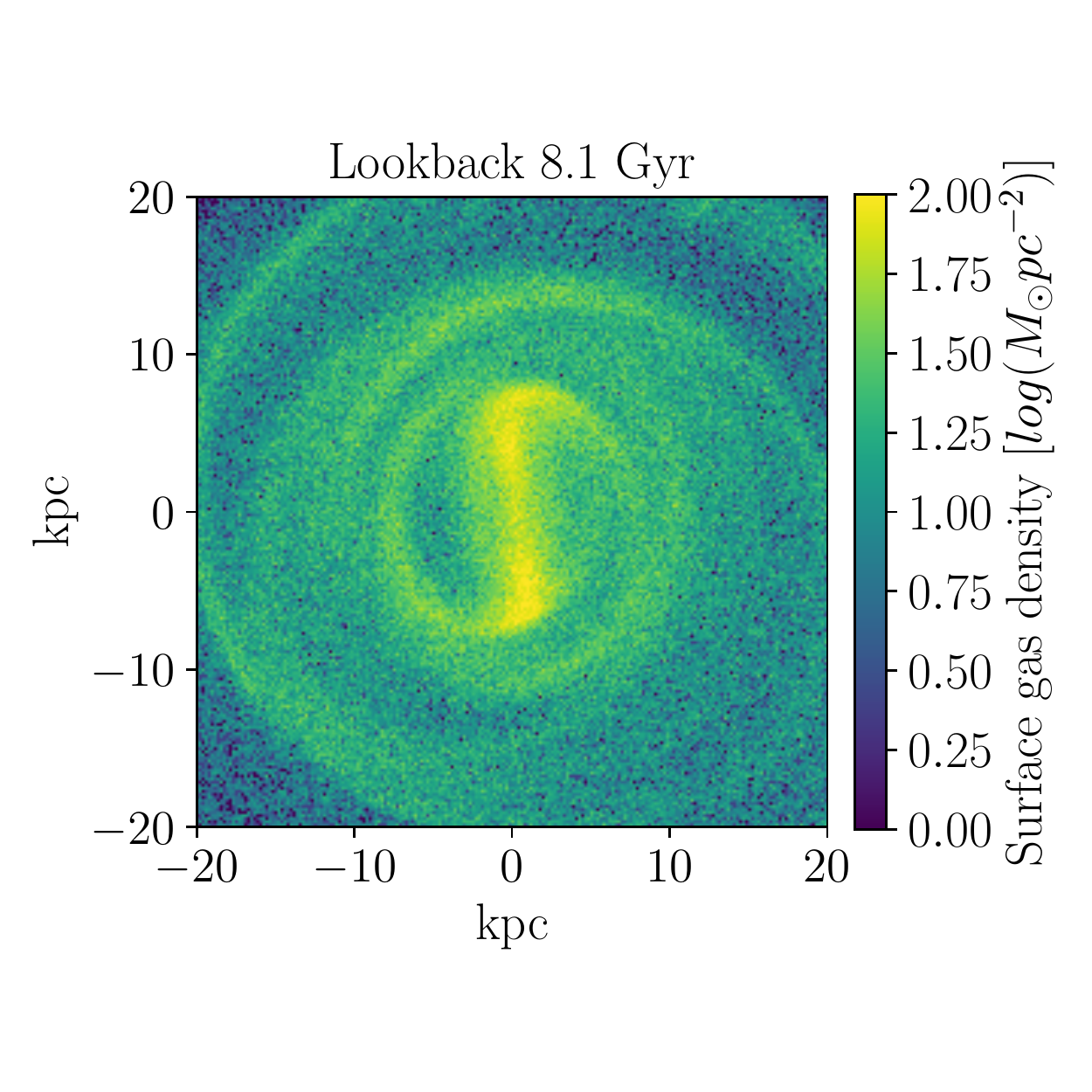}&\includegraphics[trim={0.25cm 0 0 0},clip,width=0.365\textwidth]{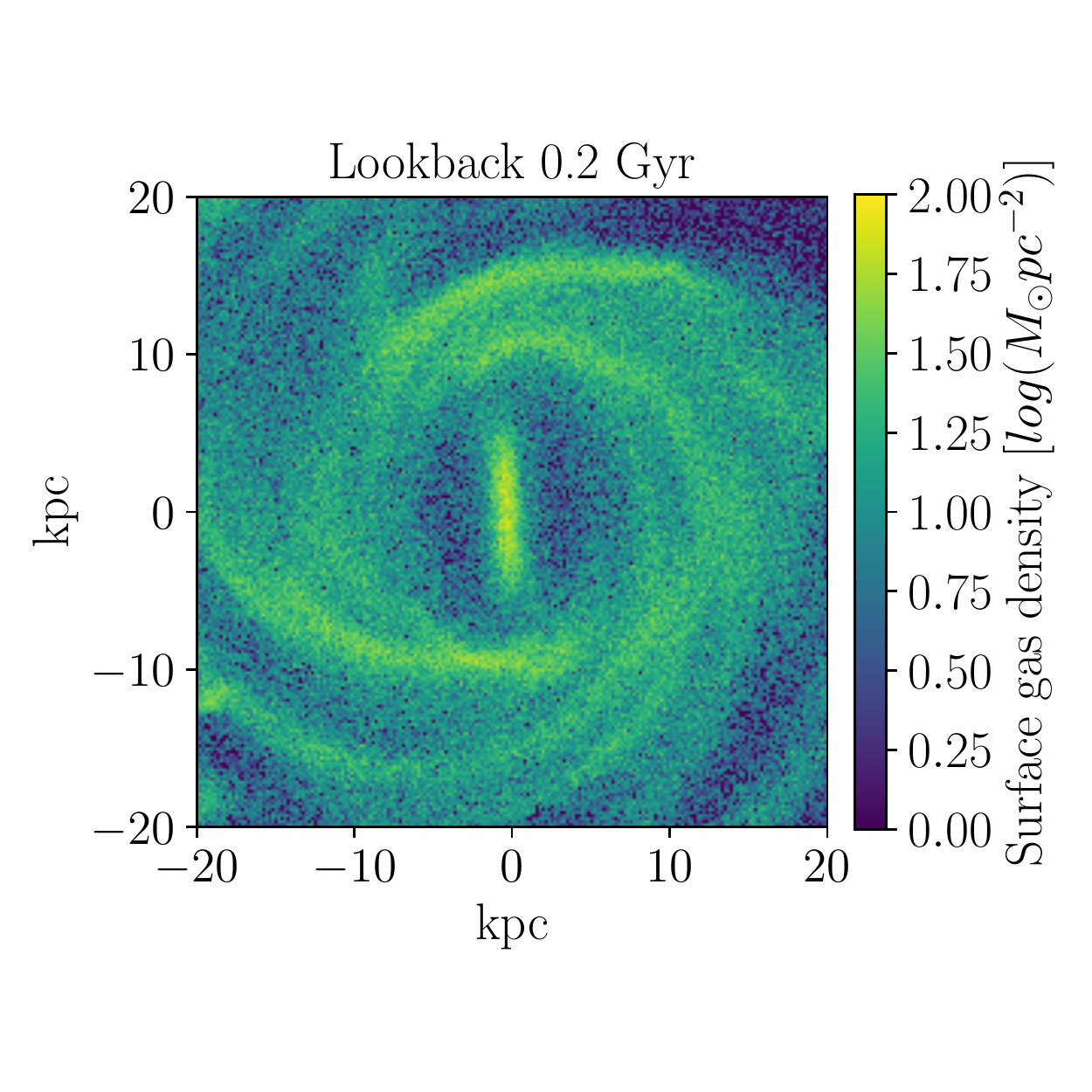}
	\end{tabular}
	\caption{Here we show the evacuation of gas from the SFD regions. Initially, the gas is diffuse before spiral arms begin to appear. When the bar forms, the central gas concentration elongates along the major axis of the bar, and the spiral arms strengthen. Once the bar is established the gas is removed from the SFD region progressively over 1-2 Gyr. Over time the size of the SFD changes corresponding to variations in the length of the bar.}
\label{Figure:gas}
\end{figure*}

\subsection{Gas Removal}	
To understand the drop in star formation in the SFD after the bar forms, we now explore how the gas disk responds to bar formation.
As an example, in Figure \ref{Figure:gas} we present the time evolution of the gas in galaxy 37. \par

Before the bar forms (top left panel, lookback time of 9.8 Gyr), the gas density is peaked in the center and does not show any other overdensities. The slight lopsidedness is due to tidal effects following a fly-by. As the gas disk grows and cools, it first develops spiral arms.
A bar then starts to form at a lookback time of 8.6 Gyr (top right panel). At first, the gas density contrast between the bar and its surroundings is small, but after $\sim$1 Gyr the gas within the bar region starts to be collected by the bar. After 500 Myr (bottom left panel) the bar has strengthened  and it becomes clear that there is a deficit of gas within the SFD region, with the bar surrounded by a ring connected to clear spiral arms.
By z=0, there is very little gas remaining inside the SFD region (bottom right panel). \par

In all six galaxies, the gas in the central regions follows a similar evolution, although the bars form at different times.
The removal of gas from the SFD region is a relatively fast process, taking between 1-2 Gyr. This also means that star formation within the SFD is quickly suppressed after the bar forms. However, the star formation histories in Figure \ref{Figure:SFH} (discussed in Section \ref{sec:SFH}) do not show a sharp decline around the time of bar formation and instead imply a more gradual decline in the age distribution of SFD region. With no gas to continue forming young stars in the SFD after the bar formed, the younger population found in that region must be coming from elsewhere in the galaxy.\par

\begin{figure}
\centering
	\begin{tabular}{cc}
     \includegraphics[width=0.4\textwidth]{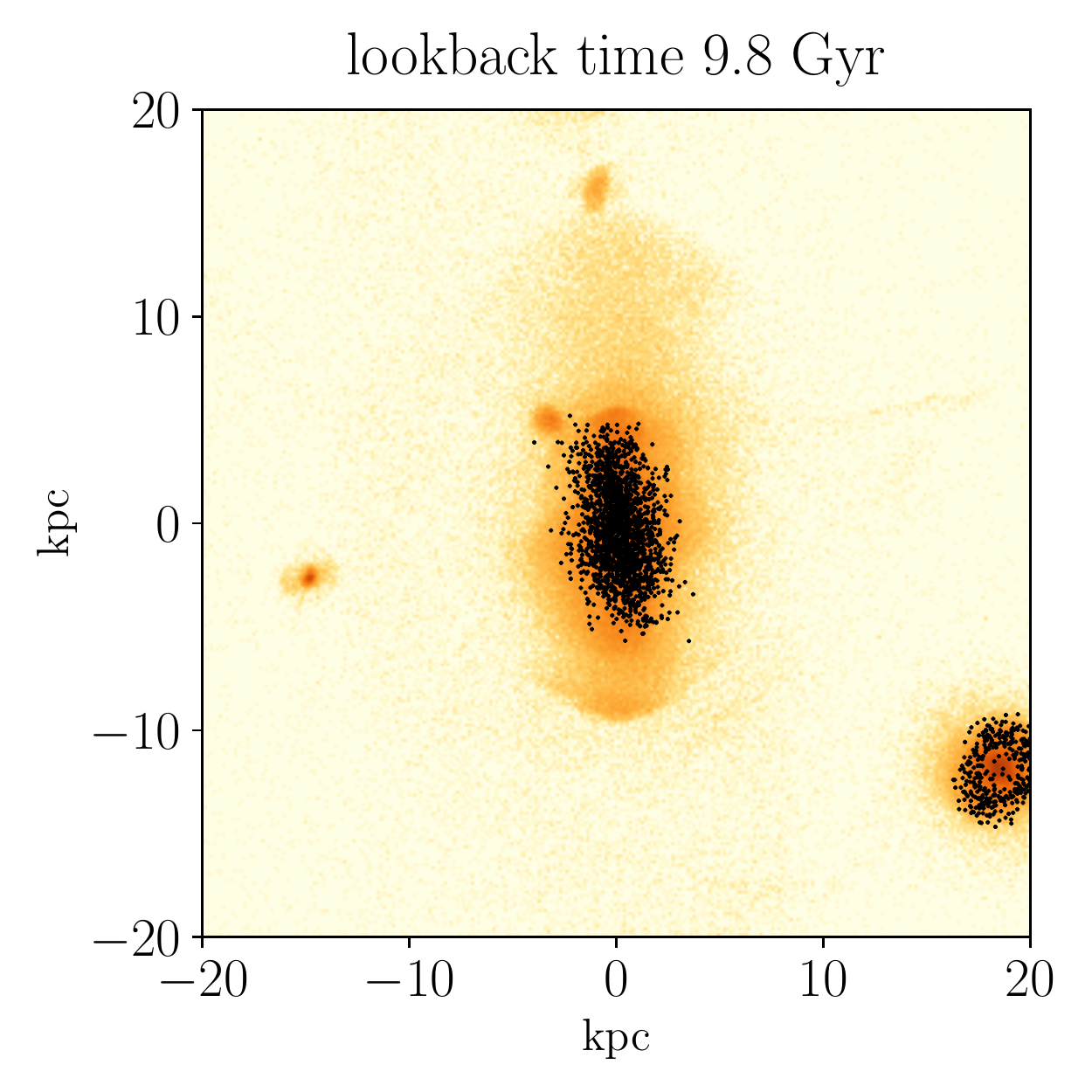}\\
     \includegraphics[width=0.4\textwidth]{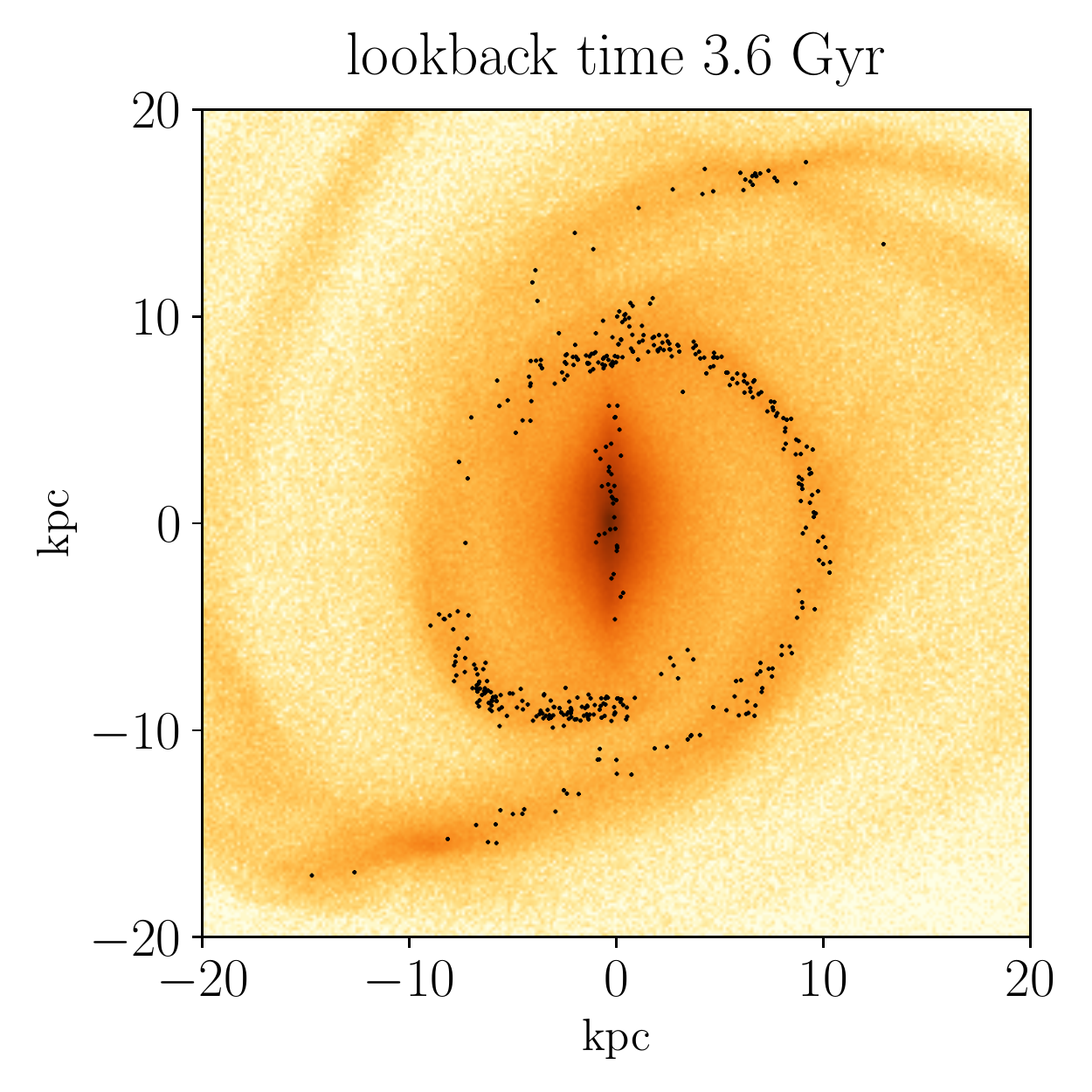}\\	
	\end{tabular}
    \caption{The birth positions of SFD stars before and after the formation of the bar overlaid on the surface stellar density maps for galaxy 37. Upper: Birth positions of SFD stars before bar formation. Lower: Birth positions of SFD stars after the formation of the bar.}
    \label{figure:birth_pos}
\end{figure}

\begin{figure}
\centering
\begin{tabular}{cc}
\includegraphics[width=0.36\textwidth]{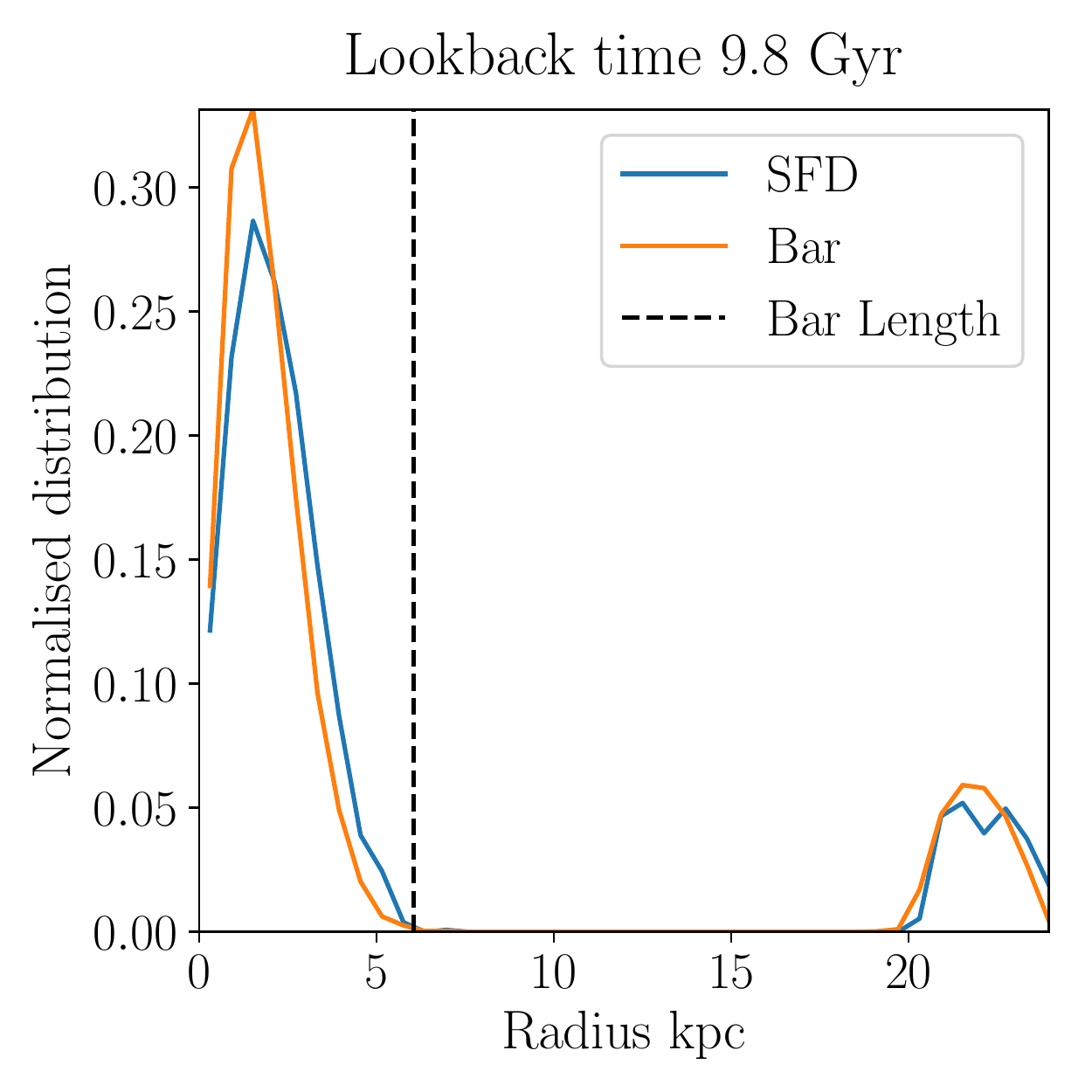}\\
\includegraphics[width=0.36\textwidth]{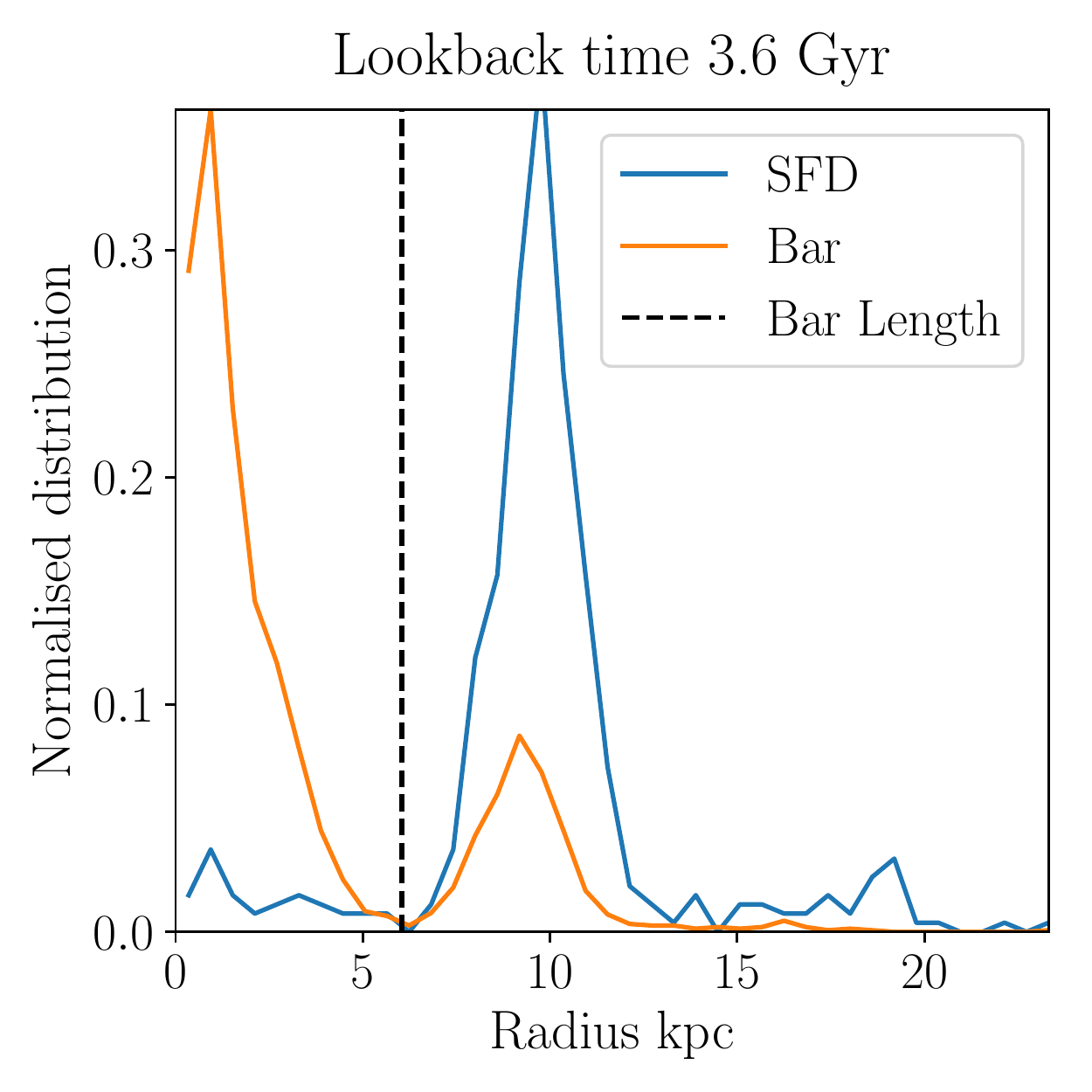}\\
\end{tabular}
\caption{Top: the radial distribution of birth positions for stars born before the formation of the bar. The blue line shows the radial distribution for the SFD stars and orange the radial distribution for bar stars. Before the formation of the bar the stars are mainly born in the same region, within 6\,kpc. Some stars are born in merging satellite galaxies, beyond 20\,kpc. Bottom: the radial distribution of stars born after the formation of the bar, with blue representing the SFD and orange the bar. Bar stars are mainly born in the central regions while SFD stars are mainly born outside the radius of the bar.}
\label{figure:birth_rad}
\end{figure}

\begin{figure}
\centering
\begin{tabular}{c}
\includegraphics[width=0.5\textwidth]{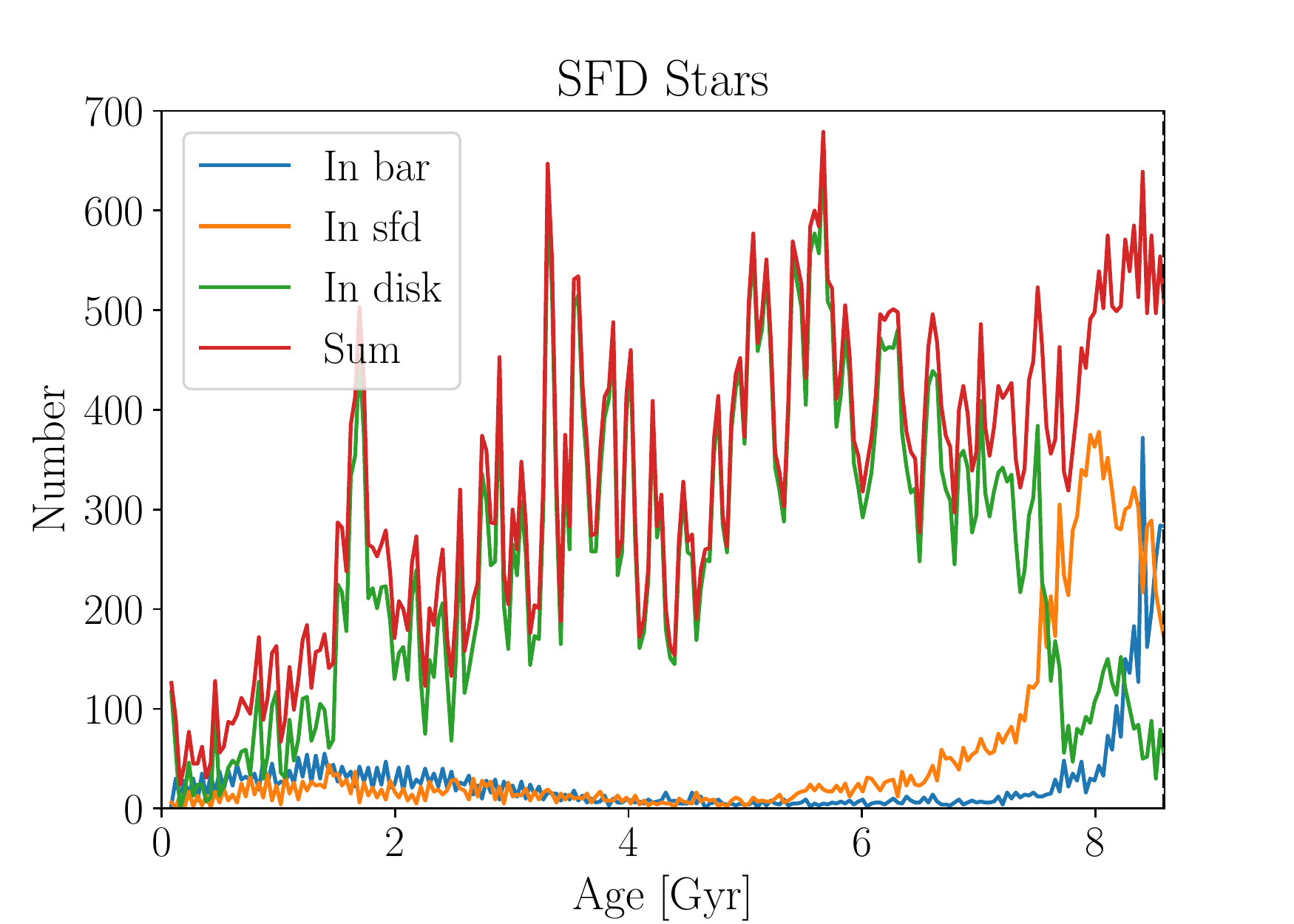}\\
\includegraphics[width=0.5\textwidth]{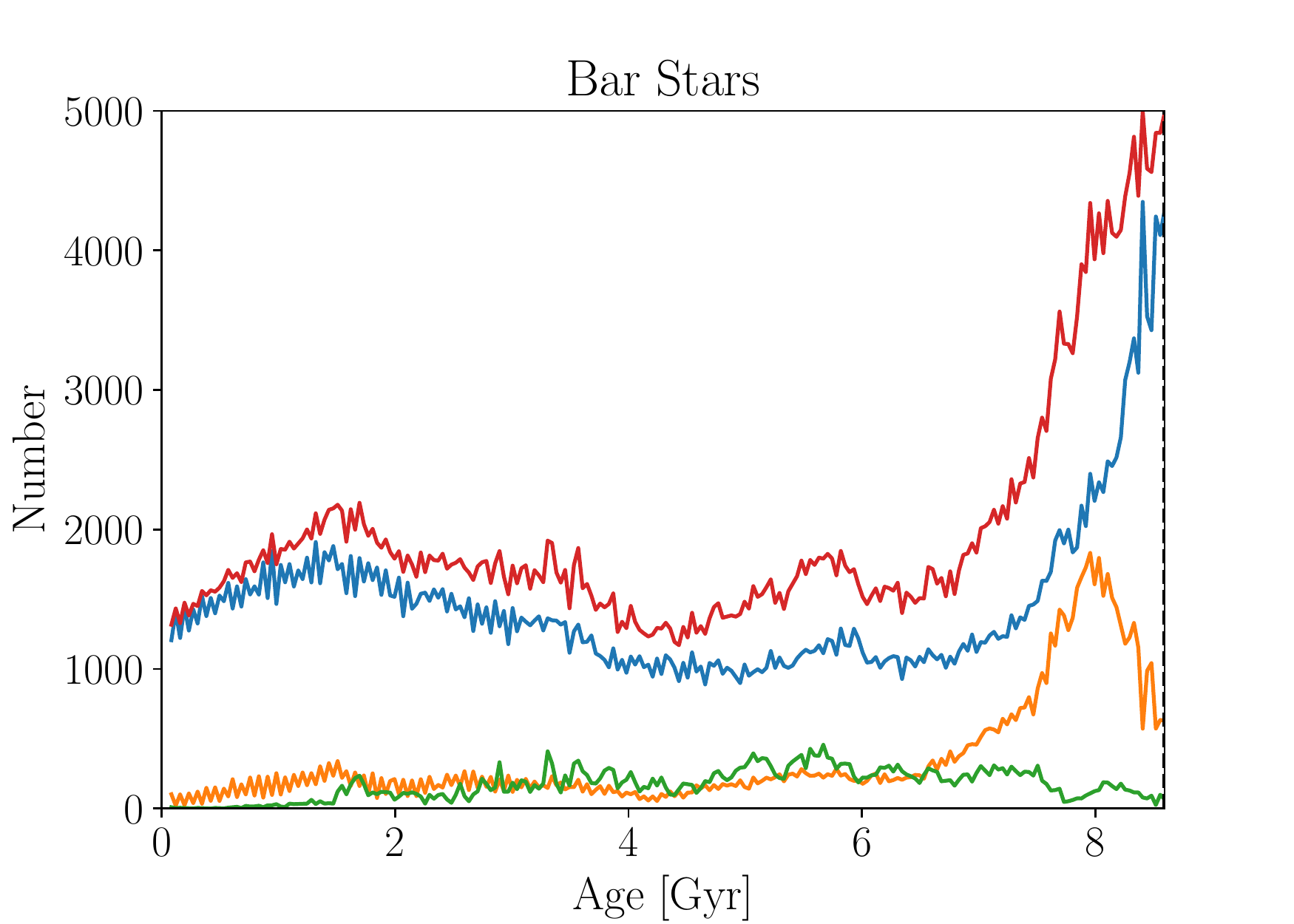}\\
\end{tabular}
\caption{Top: the fraction of stars born after the formation of the bar in the SFD, bar and disk selected to be SFD stars at z=0 for galaxy 37. Red represents the total SFD stars born at that time, green the number of SFD stars born in the disk, blue the number of SFD stars born in the bar, and orange the number of SFD stars born inside the SFD region. The majority of the stars ending up in the SFD after the bar is formed come from the disk. Very few stars come from the SFD region. Bottom: the fraction of stars selected to be bar stars at z=0 born in the SFD, bar and disk.}
\label{figure:birth_frac}
\end{figure}

\subsection{Birth positions of SFD stars before \& after bar formation} \label{section:birth_positions}

From Figure \ref{Figure:SFH} it is clear that there is no truncation in the age distribution associated with the onset of the bar: instead it is a gradual process with the number of young stars in the SFD decreasing after the formation of the bar. However, when looking at the evolution of the gas density within the SFD after bar formation we see a distinct lack of gas in the SFD within about 1 Gyr. This is a relatively fast process and does not match up with what we inferred from the age distribution plots, which imply a gradual down turn in the age distribution. This implies that the SFD region, after the formation of the bar, is being supplemented with young stars from elsewhere in the galaxy.\par

Figure \ref{figure:birth_pos} shows the birth positions of stars found in the SFD at z=0 and born before and after the formation of the bar, for galaxy 37. Before the formation of the bar, the stars are born throughout the galaxy. After the formation of the bar there is a distinct difference; the SFD stars are born mainly in the inner ring surrounding the bar with some along the spiral arms. \par

No stars are born within the defined SFD regions. This explains the disparity between Figures \ref{Figure:SFH} and \ref{Figure:gas}. There are no stars forming within the SFD region but younger stars are coming into the SFD from the inner ring and spiral arms, which explains the gradual drop of the SFD age distribution.\par

Figure \ref{figure:birth_rad} shows the distribution of birth radii of SFD and bar stars born before (upper) and after (lower) the formation of the bar for galaxy 37 at the same ages as Figure \ref{figure:birth_pos}. This further supports the conclusion that the SFD is being supplemented with young stars from outside the inner ring and that in the SFD star formation is suppressed. This is a trend that can be seen in all of the galaxies in our sample. For all cases, before bar formation the SFD and bar stars are coming from the same regions. However, stars ending up in the bar and SFD that form after the onset of the bar come from two different regions. SFD stars come mainly from outside the bar radius (mainly from the inner ring and the spiral arms), while bar stars are mainly born inside the bar radius with a portion coming from the spiral arms.\par

Figure \ref{figure:birth_frac} shows the number of stars being born in the disk, SFD and bar for galaxy 37. The top plot in Figure \ref{figure:birth_frac} shows that almost all (75.2\%) of the SFD stars born after the formation of the bar are coming from the region we define as the disk, with only a small fraction (8.1\%) coming from the SFD. The bar also contributes a minor fraction (16.6\%) of SFD stars which may represent some of the bar stars we were not able to remove from the SFD sample selection. At $\sim$1.5 Gyr there is a drop in the age distribution which coincides with a drop in the contribution of SFD stars from the disk. This could be accounted for by the time it takes stars from the disk to migrate to the SFD region. In that case, when we take our SFD sample from the final snapshot (z=0) we are missing out on disk stars which would become SFD stars after this time.\par

The lower half of Figure \ref{figure:birth_frac} shows the number of bar stars being born in the same region defined for the top plot of the same figure. The majority (73.8\%) of bar stars are born within the bar, with a small contribution (17.7\%) from the disk and a negligible amount (8.6\%) coming from the SFD. At late times, less than 1 Gyr, there is no contribution from the disk.\par

By looking at the three plots discussed in this section in conjunction with Figure \ref{Figure:SFH} we find that before the formation of the bar the population in the SFD and bar regions come from the same regions, which is supported by the similarities of the SFD and bar age distributions. However, after the formation of the bar there is a disparity in the regions in which bar and SFD stars are born. The star formation in the SFD region is truncated quickly as gas is removed from the SFD, but young stars are being born in the disk which migrate into the SFD. To determine how the stars from the disk and ring migrate into the SFD we need to track their progression from their birth positions to the SFD region.\par

\subsection{Collective dynamics}

\begin{figure*}
    \centering
    \begin{tabular}{c c}
    \includegraphics[width=0.4\textwidth]{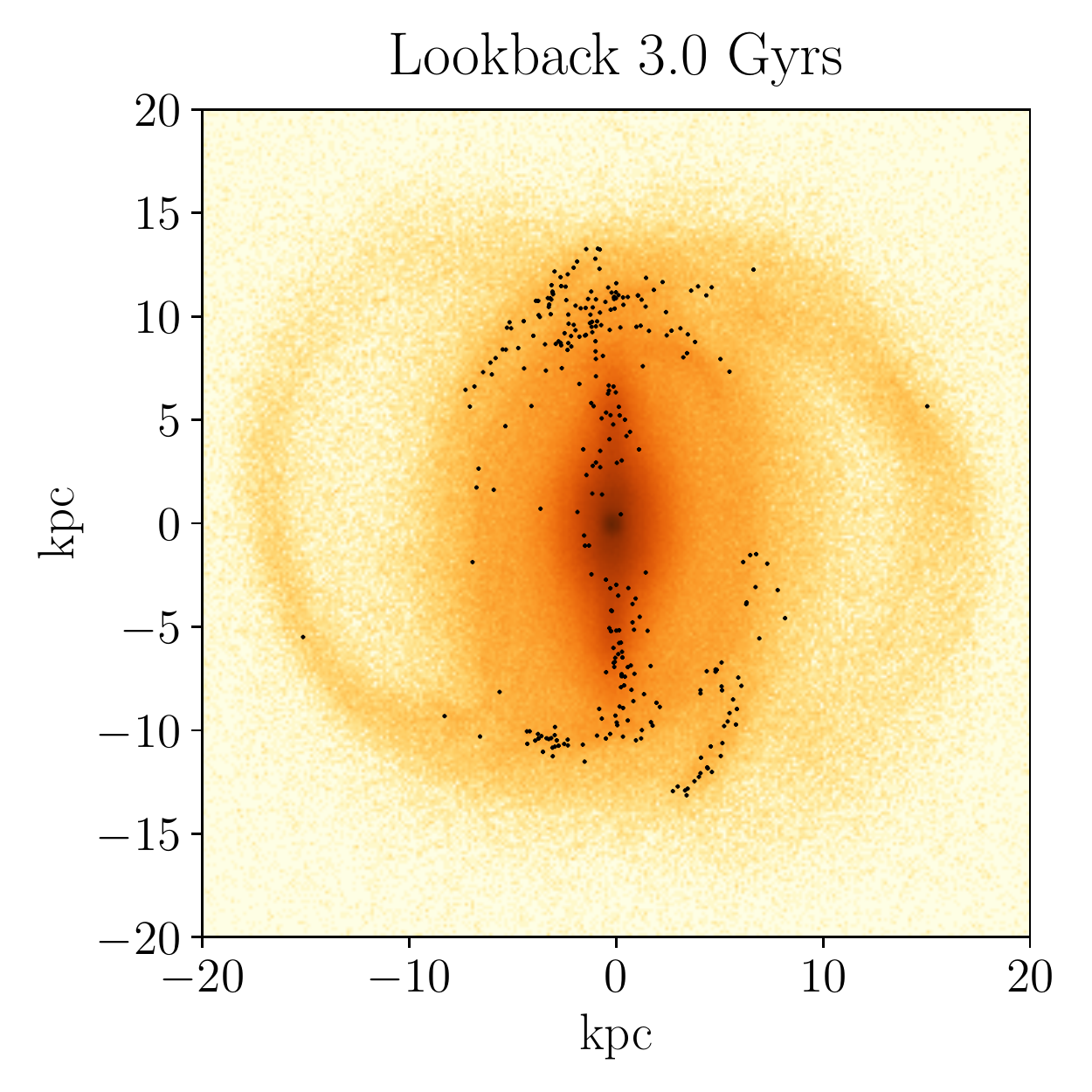} & \includegraphics[width=0.4\textwidth]{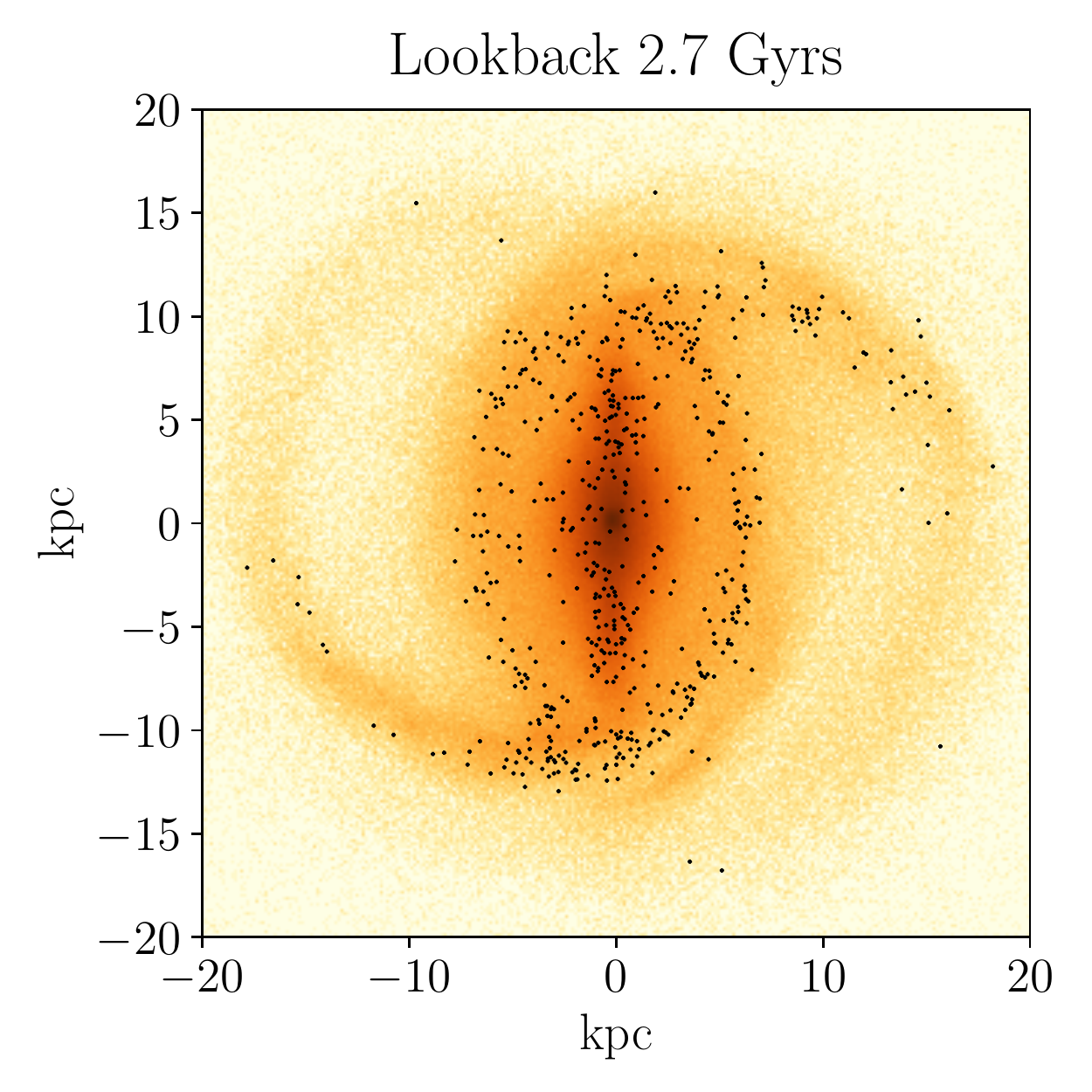}\\  
    \includegraphics[width=0.4\textwidth]{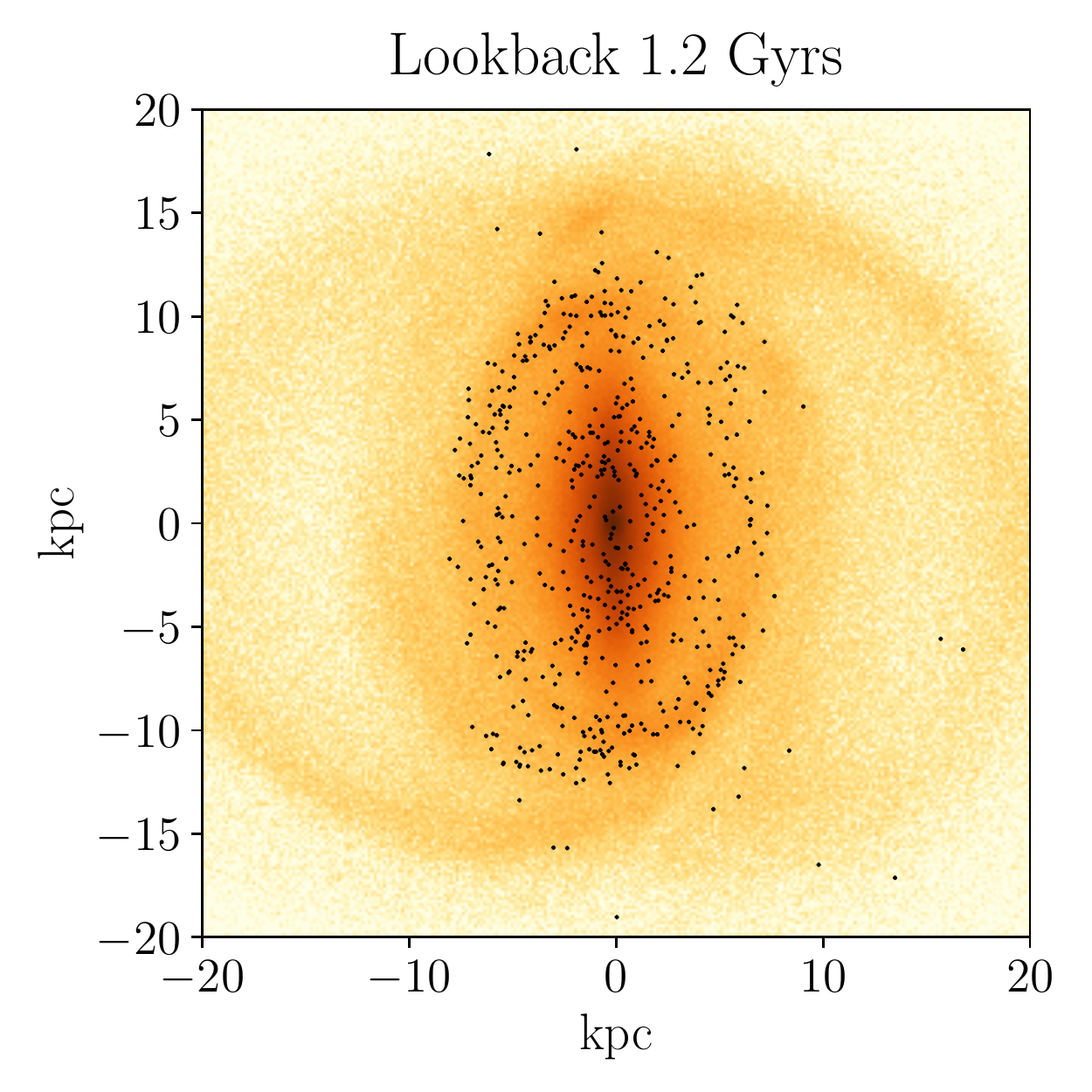} & \includegraphics[width=0.4\textwidth]{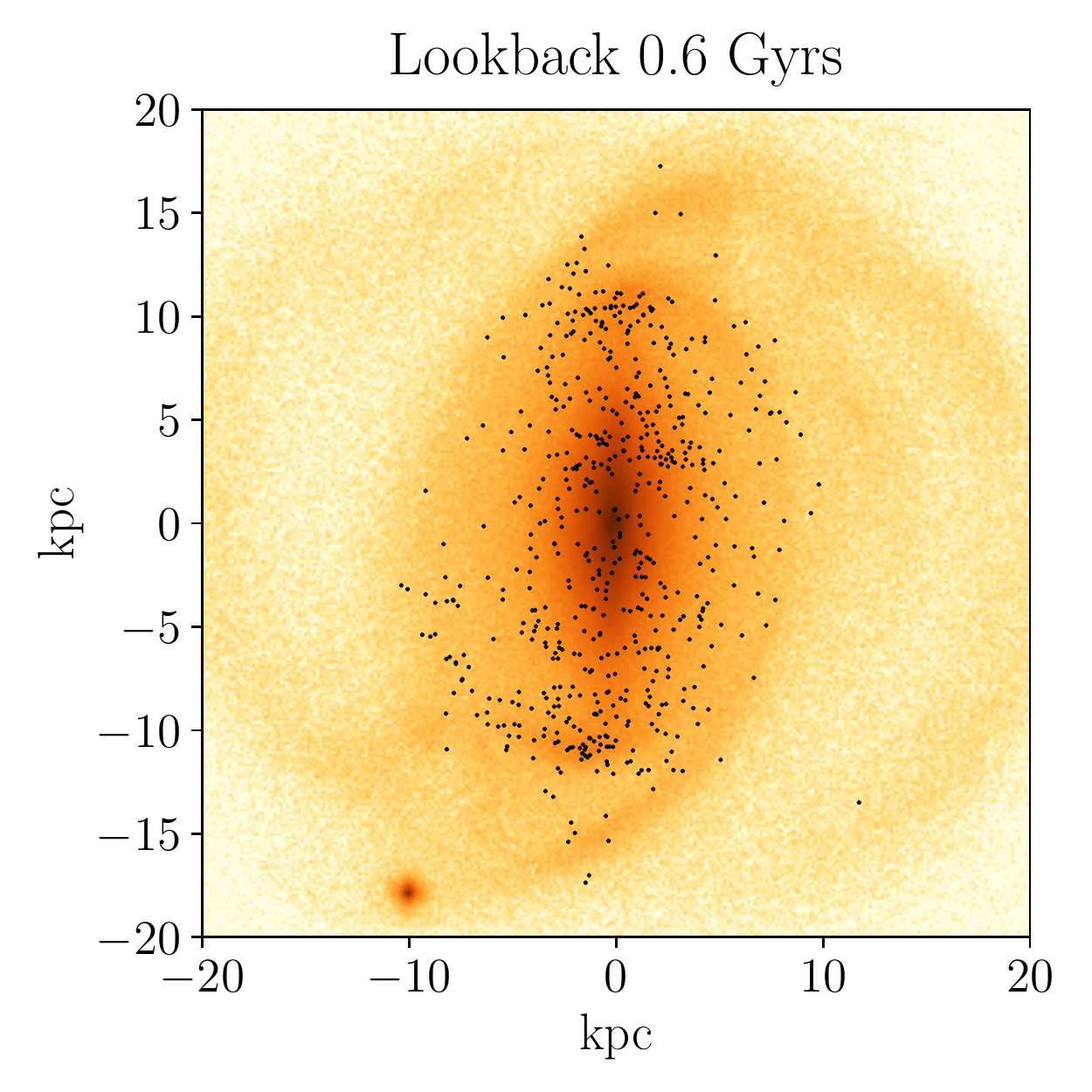}\\
    \includegraphics[width=0.4\textwidth]{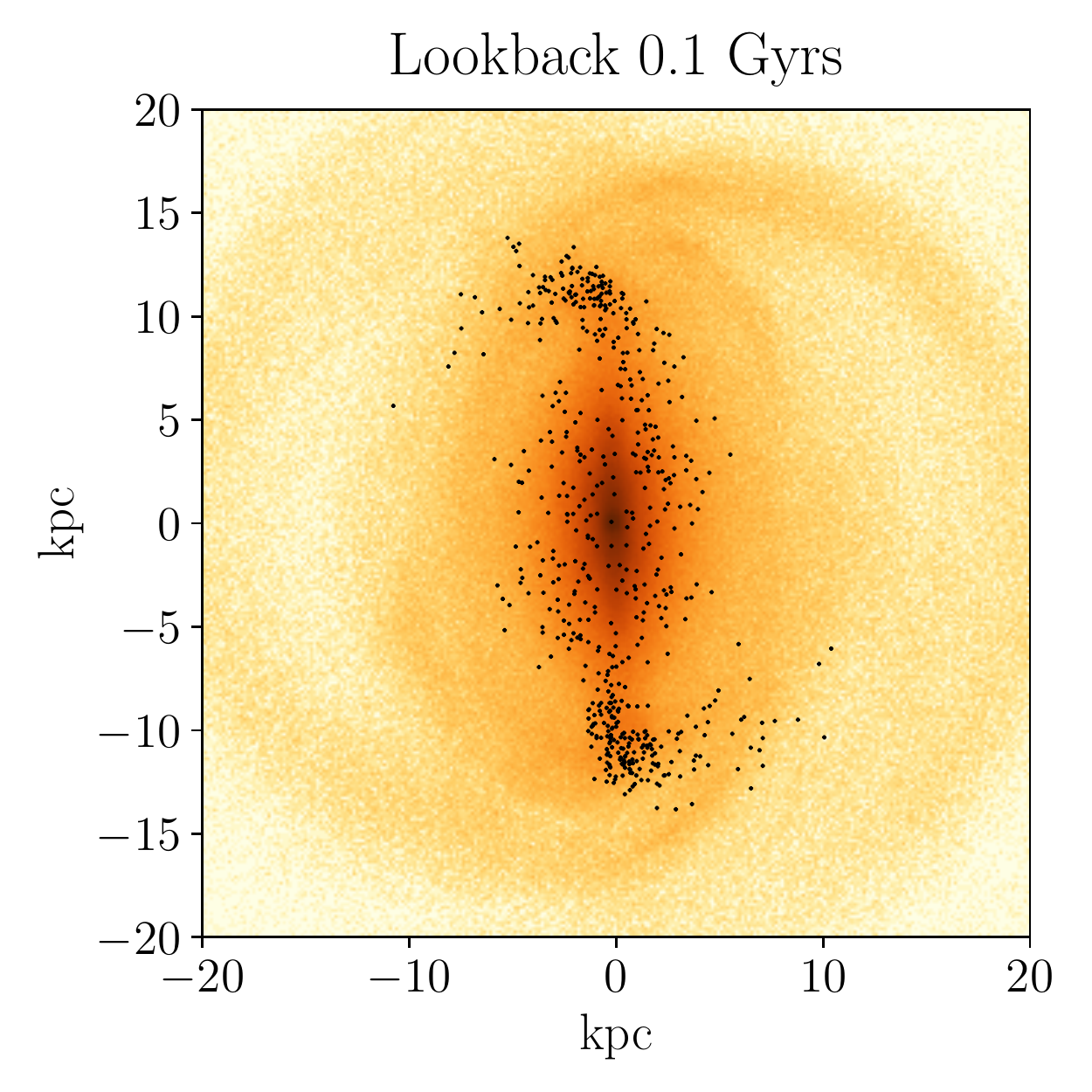} & \includegraphics[width=0.4\textwidth]{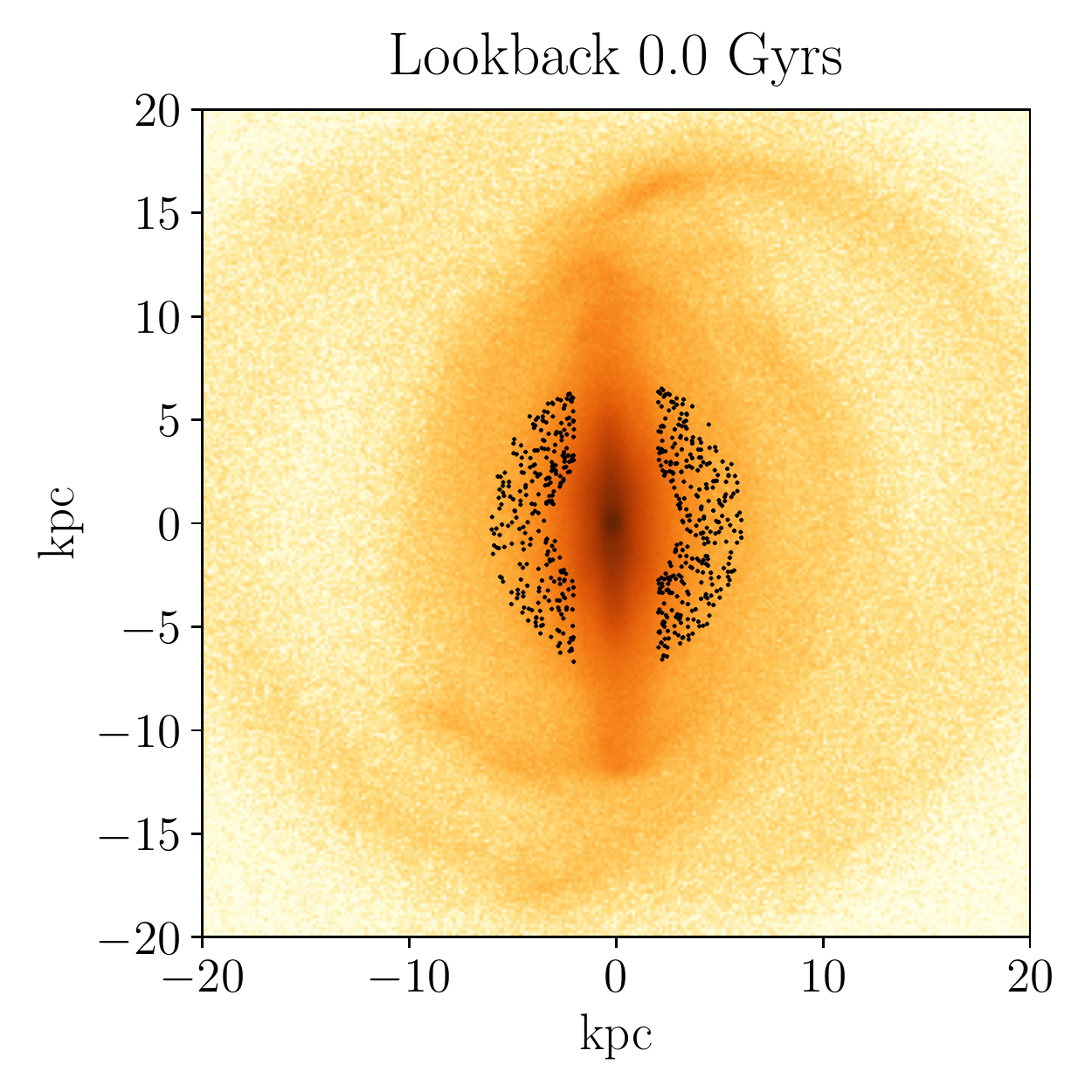}\\
    \end{tabular}
    \caption{Tracking of SFD stars from their birth positions to z=0. Initially stars are born in the inner ring near the ends of the bar and along the spiral arms. They then move along the spiral arms and around the inner ring. Slowly stars begin to spiral from the inner ring into the SFD region. Finally the stars collect near the ends of the bar before circling back into the SFD selection region at z=0.}
    \label{figure:tracking}
\end{figure*}

After the formation of the bar the SFD region is supplemented with young stars which are born along the inner ring and spiral arms. To determine how these stars end up in the SFD we track the progression of stars born at a lookback time of 3 Gyr to z=0 in Figure \ref{figure:tracking}. The plot at 3 Gyr shows the birth positions of the SFD stars. Correlating with the results from Section \ref{section:birth_positions}, the stars are born mainly along the inner ring and spiral arms with very few being born in the bar and SFD. Within 300 Myr the stars begin to move along the spiral arms and inner ring. By 1.2 Gyr almost all of the stars are moving along the inner ring and are beginning to fall towards the SFD region by 600 Myr. At 100 Myr the stars are collected near the ends of the bar before they reach their selection point in the SFD regions at 0 Gyr. This implies that it takes approximately 2.4 Gyr before ring stars begin to reach the SFD region, which supports our conclusion that the reduction in SFD stars being born in the disk for the final 1.5 Gyr seen in Figure \ref{figure:birth_frac} could be a result of the time taken for disk stars to migrate to the SFD.\par

\section{Discussion}\label{section:discussion}

\subsection{Limitations of the simulations}
A number of previous papers have explored the properties of simulated disks in the \cite{Martig2012} sample, and have found those disks to be realistic overall, when compared to a range of observational data. 
Most importantly for this paper, \cite{KK2012} showed that the fraction of barred galaxies  in the simulated sample ($\sim$70\%) is consistent with observations in the local universe, and that the time evolution of the fraction of barred galaxies matches observations by \cite{Sheth2008,Simmons2014}. Additionally, in our simulations, bars, on average, form later in low mass galaxies, which agrees with \cite{Sheth2008}. 
\cite{Martig2014a,Martig2014b} have further shown that the vertical structure of the disks is well resolved, and that some galaxies are a good match to observations of the Milky Way.\par
 
Overall, this is a strong indication that global stellar dynamics is adequately modelled in our simulations, in spite of a spatial resolution of only 150 pc. The global distribution of gas in the central regions also appears to be consistent with observations. In particular the absence of gas within SFDs is clear in the ALMA observations of molecular gas shown by \cite{George2019}.\par

However, a resolution of 150 pc does not allow us to properly track the movement of gas particles within the central regions, or to follow the formation of features like nuclear disks. The motion of gas particles along the bar is also not properly modelled, and for instance we do not see dense gas lanes along the leading edges of the bars.\par

Additionally, the Schmidt relation used to model star formation is based solely on the local  gas density, and does not account for dynamical heating from shocks halting the collapse of dense gas regions. Indeed, observations suggest that the star formation efficiency might be reduced in bars \citep{Momose2010}.\par

An imperfect modelling of star formation might be the reason why a majority of our simulated bars are star forming, which is not the case of bars generally in the Local Universe. However, star forming bars do exist \citep{Martin1997,Verley2007}, but a detailed comparison of the fraction of star-forming bars in simulations and observations (controlling for environment and mass) is beyond the scope of this paper.\par

With all of this in consideration, our simulations might overestimate star formation in bars, but probably model SFDs adequately in terms of the global dynamics of gas and stars.\par

\subsection{Potential bar dating method}

For all of the galaxies in our sample, the number of young stars (born after the bar formed) drops with time for the SFD compared to the bar. 
In five out of the six galaxies, the time of bar formation closely coincides with a change in the sign of the ``bar-SFD" residual age distribution (galaxy 82 is the exception, and with this case the residual changes sign long before the bar forms). This suggests the possibility to use the sign of the residual as an indicator of the epoch of bar formation. 
However, this signal appears to be very subtle, and consists in a gradual downturn in the age distribution instead of the sharp truncation assumed by \cite{James2016,James2018} to model star formation histories in their sample of observed SFDs. This is because young stars coming from the disk are migrating to the SFD, and are ``polluting" it with a young population that should not be present if only in-situ star formation happened.
In the following two subsections, we first explore the possible reasons for the strange behaviour of galaxy 82 and then discuss the usefulness of our method to date bar formation with observational data.\par

\subsubsection{The unusual behaviour of galaxy 82}\label{section:82}

Galaxy 82 is the only galaxy in which the change of sign of the Bar-SFD residual does not coincide with the epoch of bar formation. Within our full sample of 33 galaxies, galaxy 82 is unique in forming a bar as recently as 2 Gyr ago - all others formed their bars no later than 4 Gyr ago. To understand whether galaxy 82's strangeness could come from having a very young bar, we ran the simulation for a further 3 Gyr. We can confirm that even after 3 more Gyr, the age distributions still look different from the ones for the other simulated galaxies. Those differences are probably due to galaxy 82's very unique formation history that in turn could explain why it formed its bar so late.\par

At early times (10 Gyr) it consists of a central low density disk that persists throughout its evolution up until the time of bar formation. Additionally, at this time (from 10 to 9 Gyr) it undergoes the accretion of a satellite which leaves a gaseous ring surrounding the central disk. \par

The ring quickly undergoes fragmentation which is then followed by the formation of spiral arms. After the spiral arms have strengthened, the central regions become bar unstable leading to the formation of the bar. This varies drastically from the other evolutionary histories for the galaxies in our sample. Furthermore, there is a spatial segregation of the bar and SFD stars' birth positions well before the epoch of bar formation (with SFD stars being born at the edge of the low density disk, in the ring, and along the spiral arms while the bar stars are primarily born in the central disk), which is a feature we see only after the formation of the bar in the rest of our sample. This spatial segregation is most likely the cause of the early bar-SFD residual sign change, although what precisely leads to the segregation of the birth positions is not entirely clear.\par

\subsubsection{Application to observational data}

The method we propose to date bar formation in a galaxy relies on a very weak signal, which makes applying the SFD bar dating method more complex than previously suggested in \cite{James2016,James2018}. 
Indeed, the method we propose relies on the accurate recovery of SFH shapes for the bar and SFD. Spectra at old ages look very similar to one another and the effect of age and metallicity can be degenerate, which will make finding a bar formation signal for early bars more challenging. Bar and SFD average ages differ by approximately 2 Gyr, which makes comparisons between the SFHs of the components for early bars difficult given the constraints stated above. Additionally, if we have overestimated the star formation efficiency of the bar in the simulations then the signal could be even weaker than anticipated. \par

Should we find a signal in observational data, we face the additional problem that the bar formation time cannot be reliably determined for all simulated galaxies in our sample. Even considering that galaxy 82 may be an unusual case we can not assume that any signal we find is directly related to bar formation. However, we can use the SFD bar dating method in conjunction with several other methods. By measuring the vertical velocity dispersion \citep{Gadotti2005} or shape of the light profiles \citep{Kim2014} we can determine if the bars are old or young and so better constrain the region of the SFH where we would expect to see a signal. In cases where these age indicators disagree the studied galaxy could be flagged as having an unusual history.\par

We can also define a lower limit on the epoch of bar formation by looking at the ages of nuclear disks \citep{Gadotti2015}, which form after the formation of the bar. Additionally, we might also be able to date bar formation by comparing the metallicities of bar and SFD stars as a function of age, due to the spatial segregation in birth positions of bar and SFD stars younger than the bar. \par

\section{Summary}\label{section:conclusion}

\cite{James2015} first described the properties of star formation deserts, regions swept up by bars with very low levels of line emission and little recent star formation. \cite{James2016,James2018}  then proposed that the cessation of star formation in those regions was due to the formation of the bar. This would mean that finding a sharp truncation in star formation histories in SFDs could be a way to determine the epoch of bar formation.\par

In this paper, we investigated the validity of these conclusions by studying the properties of SFDs using zoom-in cosmological re-simulations. From the sample of \cite{Martig2012}, we chose 6 simulated disk galaxies with bar formation times ranging from 2 to 8 Gyr ago. We find that the formation of the bar does not appear to have an effect on the global star formation rate of the galaxies but affects the distribution of gas and star formation within the central regions. At $z=0$, we find on both sides of the bar regions that are dominated by old stars, and that resemble the observed SFDs. However, the SFDs in the simulated galaxies actually contain stars of all ages:

\begin{itemize}
\item SFD stars older than the bar are born in similar regions to similarly old stars that end up in the bar.
\item When the bar forms, it efficiently removes gas from the SFD on 1 Gyr timescales, which quickly truncates the local star formation. 
\item SFD stars younger than the bar are not formed in-situ but are born in the disk and migrate to the SFD (unlike bar stars of similar ages, which are mostly born in-situ).  
\end{itemize}

If there were no radial migration of young stars from the disk to the SFD, then the age distribution of SFD stars would show a truncation within $\sim 1$ Gyr after the time of bar formation. However, this is not the case, and the SFD age distributions show a gradual downturn instead of a truncation, which makes recovering the epoch of bar formation more complicated than \cite{James2016,James2018} anticipated. The different shapes of age distributions for SFD and bar stars can provide an indication of when the bar formed, but the signal is weak and potentially hard to detect. This might still be used to date bars, especially in conjunction with other methods.\par

SFDs could also be used to investigate radial migration. Indeed, they are unique regions with no in-situ star formation: stars younger than the bar all come from the disk (outside of the bar radius).
This can provide an uncontaminated sample of stars only affected by radial migration.
We plan to investigate this further to see if SFDs can be used to constrain migration efficiency and timescales.\par

We also plan to apply our bar dating method to observed galaxies using MUSE data from the TIMER consortium \citep{Gadotti2019} and supplementary long-slit spectroscopy. The signatures we expect in the SFD star formation histories are quite weak, but the comparison of bar, SFD and nuclear ring properties could provide better constraints on the epoch of bar formation in different types of galaxies.\par

\section*{Acknowledgements}
CEDK would like to thank the support and useful comments from B. Mummery and D. Gadotti which helped to improve the clarity of this paper. We thank the reviewer for insightful comments.  
CEDK would also like to acknowledge the receipt of a PhD studentship from STFC UK.




\bibliographystyle{mnras}
\bibliography{Biblip} 

\begin{thebibliography}{}
\makeatletter
\relax
\def\mn@urlcharsother{\let\do\@makeother \do\$\do\&\do\#\do\^\do\_\do\%\do\~}
\def\mn@doi{\begingroup\mn@urlcharsother \@ifnextchar [ {\mn@doi@}
  {\mn@doi@[]}}
\def\mn@doi@[#1]#2{\def\@tempa{#1}\ifx\@tempa\@empty \href
  {http://dx.doi.org/#2} {doi:#2}\else \href {http://dx.doi.org/#2} {#1}\fi
  \endgroup}
\def\mn@eprint#1#2{\mn@eprint@#1:#2::\@nil}
\def\mn@eprint@arXiv#1{\href {http://arxiv.org/abs/#1} {{\tt arXiv:#1}}}
\def\mn@eprint@dblp#1{\href {http://dblp.uni-trier.de/rec/bibtex/#1.xml}
  {dblp:#1}}
\def\mn@eprint@#1:#2:#3:#4\@nil{\def\@tempa {#1}\def\@tempb {#2}\def\@tempc
  {#3}\ifx \@tempc \@empty \let \@tempc \@tempb \let \@tempb \@tempa \fi \ifx
  \@tempb \@empty \def\@tempb {arXiv}\fi \@ifundefined
  {mn@eprint@\@tempb}{\@tempb:\@tempc}{\expandafter \expandafter \csname
  mn@eprint@\@tempb\endcsname \expandafter{\@tempc}}}

\bibitem[\protect\citeauthoryear{{Aguerri}, {Beckman}  \& {Prieto}}{{Aguerri}
  et~al.}{1998}]{AJAL1998}
{Aguerri} J.~A.~L.,  {Beckman} J.~E.,   {Prieto} M.,  1998, \aj, 116, 2136

\bibitem[\protect\citeauthoryear{{Athanassoula}}{{Athanassoula}}{1992}]{Athanassoula1992}
{Athanassoula} E.,  1992, \mnras, 259, 345

\bibitem[\protect\citeauthoryear{{Athanassoula}}{{Athanassoula}}{2003}]{Athanassoula2003}
{Athanassoula} E.,  2003, \mnras, 341, 1179

\bibitem[\protect\citeauthoryear{{Berentzen}, {Heller}, {Shlosman}  \&
  {Fricke}}{{Berentzen} et~al.}{1998}]{Berentzen1998}
{Berentzen} I.,  {Heller} C.~H.,  {Shlosman} I.,   {Fricke} K.~J.,  1998,
  \mnras, 300, 49

\bibitem[\protect\citeauthoryear{{Bournaud} \& {Combes}}{{Bournaud} \&
  {Combes}}{2002}]{Bournaud2002}
{Bournaud} F.,  {Combes} F.,  2002, \aap, 392, 83

\bibitem[\protect\citeauthoryear{{Bournaud} \& {Combes}}{{Bournaud} \&
  {Combes}}{2003}]{Bournaud2003}
{Bournaud} F.,  {Combes} F.,  2003, \aap, 401, 817

\bibitem[\protect\citeauthoryear{{Bournaud}, {Combes}  \& {Semelin}}{{Bournaud}
  et~al.}{2005}]{Bournaud2005}
{Bournaud} F.,  {Combes} F.,   {Semelin} B.,  2005, \mnras, 364, L18

\bibitem[\protect\citeauthoryear{{Buta}, {Vasylyev}, {Salo}  \&
  {Laurikainen}}{{Buta} et~al.}{2005}]{Buta2005}
{Buta} R.,  {Vasylyev} S.,  {Salo} H.,   {Laurikainen} E.,  2005, \aj, 130, 506

\bibitem[\protect\citeauthoryear{{Cacho}, {S{\'a}nchez-Bl{\'a}zquez}, {Gorgas}
  \& {P{\'e}rez}}{{Cacho} et~al.}{2014}]{Cacho2014}
{Cacho} R.,  {S{\'a}nchez-Bl{\'a}zquez} P.,  {Gorgas} J.,   {P{\'e}rez} I.,
  2014, \mnras, 442, 2496

\bibitem[\protect\citeauthoryear{{Carles}, {Martel}, {Ellison}  \&
  {Kawata}}{{Carles} et~al.}{2016}]{Carles2016}
{Carles} C.,  {Martel} H.,  {Ellison} S.~L.,   {Kawata} D.,  2016, \mnras, 463,
  1074

\bibitem[\protect\citeauthoryear{{Chapelon}, {Contini}  \&
  {Davoust}}{{Chapelon} et~al.}{1999}]{Chapelon1999}
{Chapelon} S.,  {Contini} T.,   {Davoust} E.,  1999, \aap, 345, 81

\bibitem[\protect\citeauthoryear{{Cheung} et~al.,}{{Cheung}
  et~al.}{2013}]{Cheung2013}
{Cheung} E.,  et~al., 2013, \apj, 779, 162

\bibitem[\protect\citeauthoryear{{Combes}}{{Combes}}{2001}]{Combes2001}
{Combes} F.,  2001, Astrophysics and Space Science Supplement, 277, 29

\bibitem[\protect\citeauthoryear{{Combes}}{{Combes}}{2008}]{Combes2008}
{Combes} F.,  2008, preprint, p. arXiv:0811.0153

\bibitem[\protect\citeauthoryear{{Consid{\`e}re}, {Coziol}, {Contini}  \&
  {Davoust}}{{Consid{\`e}re} et~al.}{2000}]{Considere2000}
{Consid{\`e}re} S.,  {Coziol} R.,  {Contini} T.,   {Davoust} E.,  2000, \aap,
  356, 89

\bibitem[\protect\citeauthoryear{{Devereux}}{{Devereux}}{1987}]{Devereux1987}
{Devereux} N.,  1987, \apj, 323, 91

\bibitem[\protect\citeauthoryear{{Ellison}, {Nair}, {Patton}, {Scudder},
  {Mendel}  \& {Simard}}{{Ellison} et~al.}{2011}]{Ellison2011}
{Ellison} S.~L.,  {Nair} P.,  {Patton} D.~R.,  {Scudder} J.~M.,  {Mendel}
  J.~T.,   {Simard} L.,  2011, \mnras, 416, 2182

\bibitem[\protect\citeauthoryear{{Elmegreen}, {Galliano}  \&
  {Alloin}}{{Elmegreen} et~al.}{2009}]{Elmegreen2009}
{Elmegreen} B.~G.,  {Galliano} E.,   {Alloin} D.,  2009, \apj, 703, 1297

\bibitem[\protect\citeauthoryear{{Fiacconi}, {Feldmann}  \& {Mayer}}{{Fiacconi}
  et~al.}{2015}]{Fiacconi2015}
{Fiacconi} D.,  {Feldmann} R.,   {Mayer} L.,  2015, \mn@doi [\mnras]
  {10.1093/mnras/stu2228}, \href
  {http://adsabs.harvard.edu/abs/2015MNRAS.446.1957F} {446, 1957}

\bibitem[\protect\citeauthoryear{{Fragkoudi}, {Athanassoula}  \&
  {Bosma}}{{Fragkoudi} et~al.}{2016}]{fragkoudi2016}
{Fragkoudi} F.,  {Athanassoula} E.,   {Bosma} A.,  2016, \mn@doi [\mnras]
  {10.1093/mnrasl/slw120}, \href
  {https://ui.adsabs.harvard.edu/\#abs/2016MNRAS.462L..41F} {462, L41}

\bibitem[\protect\citeauthoryear{{Gadotti}}{{Gadotti}}{2008}]{Gadotti2008}
{Gadotti} D.~A.,  2008, \mn@doi [\mnras] {10.1111/j.1365-2966.2007.12723.x},
  \href {https://ui.adsabs.harvard.edu/abs/2008MNRAS.384..420G} {384, 420}

\bibitem[\protect\citeauthoryear{{Gadotti} \& {de Souza}}{{Gadotti} \& {de
  Souza}}{2003}]{Gadotti2003}
{Gadotti} D.~A.,  {de Souza} R.~E.,  2003, \mn@doi [\apjl] {10.1086/368159},
  \href {https://ui.adsabs.harvard.edu/abs/2003ApJ...583L..75G} {583, L75}

\bibitem[\protect\citeauthoryear{{Gadotti} \& {de Souza}}{{Gadotti} \& {de
  Souza}}{2005}]{Gadotti2005}
{Gadotti} D.~A.,  {de Souza} R.~E.,  2005, \apj, 629, 797

\bibitem[\protect\citeauthoryear{{Gadotti} \& {dos Anjos}}{{Gadotti} \& {dos
  Anjos}}{2001}]{Gadotti2001}
{Gadotti} D.~A.,  {dos Anjos} S.,  2001, \aj, 122, 1298

\bibitem[\protect\citeauthoryear{{Gadotti}, {Seidel}, {S{\'a}nchez-
  Bl{\'a}zquez}, {Falc{\'o}n-Barroso}, {Husemann}, {Coelho}  \&
  {P{\'e}rez}}{{Gadotti} et~al.}{2015}]{Gadotti2015}
{Gadotti} D.~A.,  {Seidel} M.~K.,  {S{\'a}nchez- Bl{\'a}zquez} P.,
  {Falc{\'o}n-Barroso} J.,  {Husemann} B.,  {Coelho} P.,   {P{\'e}rez} I.,
  2015, \aap, 584, A90

\bibitem[\protect\citeauthoryear{{Gadotti} et~al.,}{{Gadotti}
  et~al.}{2019}]{Gadotti2019}
{Gadotti} D.~A.,  et~al., 2019, \mn@doi [\mnras] {10.1093/mnras/sty2666}, \href
  {https://ui.adsabs.harvard.edu/\#abs/2019MNRAS.482..506G} {482, 506}

\bibitem[\protect\citeauthoryear{{George}, {Joseph}, {Mondal}, {Subramanian},
  {Subramaniam}  \& {Paul}}{{George} et~al.}{2019}]{George2019}
{George} K.,  {Joseph} P.,  {Mondal} C.,  {Subramanian} S.,  {Subramaniam} A.,
   {Paul} K.~T.,  2019, \mn@doi [\aap] {10.1051/0004-6361/201834500}, \href
  {https://ui.adsabs.harvard.edu/\#abs/2019A&A...621L...4G} {621, L4}

\bibitem[\protect\citeauthoryear{{Hawarden}, {Mountain}, {Leggett}  \&
  {Puxley}}{{Hawarden} et~al.}{1986}]{Hawarden1986}
{Hawarden} T.~G.,  {Mountain} C.~M.,  {Leggett} S.~K.,   {Puxley} P.~J.,  1986,
  \mnras, 221, 41P

\bibitem[\protect\citeauthoryear{{Heckman}}{{Heckman}}{1980}]{Heckman1980}
{Heckman} T.~M.,  1980, \aap, 88, 365

\bibitem[\protect\citeauthoryear{{Heller} \& {Shlosman}}{{Heller} \&
  {Shlosman}}{1994}]{Heller1994}
{Heller} C.~H.,  {Shlosman} I.,  1994, \apj, 424, 84

\bibitem[\protect\citeauthoryear{{Henry} \& {Worthey}}{{Henry} \&
  {Worthey}}{1999}]{Henry1999}
{Henry} R.~B.~C.,  {Worthey} G.,  1999, Publications of the Astronomical
  Society of the Pacific, 111, 919

\bibitem[\protect\citeauthoryear{{Ho}, {Filippenko}  \& {Sargent}}{{Ho}
  et~al.}{1997}]{Ho1997}
{Ho} L.~C.,  {Filippenko} A.~V.,   {Sargent} W. L.~W.,  1997, \apj, 487, 591

\bibitem[\protect\citeauthoryear{{Hoyle} et~al.,}{{Hoyle}
  et~al.}{2011}]{Hoyle2011}
{Hoyle} B.,  et~al., 2011, \mnras, 415, 3627

\bibitem[\protect\citeauthoryear{{Hozumi}}{{Hozumi}}{2012}]{Hozumi2012}
{Hozumi} S.,  2012, Publications of the Astronomical Society of Japan, 64, 5

\bibitem[\protect\citeauthoryear{{Hozumi} \& {Hernquist}}{{Hozumi} \&
  {Hernquist}}{2005}]{Hozumi2005}
{Hozumi} S.,  {Hernquist} L.,  2005, Publications of the Astronomical Society
  of Japan, 57, 719

\bibitem[\protect\citeauthoryear{{Hummel}, {van der Hulst}, {Kennicutt}  \&
  {Keel}}{{Hummel} et~al.}{1990}]{Hummel1990}
{Hummel} E.,  {van der Hulst} J.~M.,  {Kennicutt} R.~C.,   {Keel} W.~C.,  1990,
  \aap, 236, 333

\bibitem[\protect\citeauthoryear{James \& Percival}{James \&
  Percival}{2015}]{James2015}
James P.~A.,  Percival S.~M.,  2015, Monthly Notices of the Royal Astronomical
  Society

\bibitem[\protect\citeauthoryear{{James} \& {Percival}}{{James} \&
  {Percival}}{2016}]{James2016}
{James} P.~A.,  {Percival} S.~M.,  2016, \mn@doi [\mnras]
  {10.1093/mnras/stv2978}, \href
  {https://ui.adsabs.harvard.edu/\#abs/2016MNRAS.457..917J} {457, 917}

\bibitem[\protect\citeauthoryear{{James} \& {Percival}}{{James} \&
  {Percival}}{2018}]{James2018}
{James} P.~A.,  {Percival} S.~M.,  2018, \mn@doi [\mnras]
  {10.1093/mnras/stx2990}, \href
  {https://ui.adsabs.harvard.edu/\#abs/2018MNRAS.474.3101J} {474, 3101}

\bibitem[\protect\citeauthoryear{{James}, {Bretherton}  \& {Knapen}}{{James}
  et~al.}{2009}]{James2009}
{James} P.~A.,  {Bretherton} C.~F.,   {Knapen} J.~H.,  2009, \aap, 501, 207

\bibitem[\protect\citeauthoryear{{Jogee}, {Scoville}  \& {Kenney}}{{Jogee}
  et~al.}{2005}]{Jogee2005}
{Jogee} S.,  {Scoville} N.,   {Kenney} J. D.~P.,  2005, \apj, 630, 837

\bibitem[\protect\citeauthoryear{{Kennicutt}}{{Kennicutt}}{1998}]{1998kennicutt}
{Kennicutt} Robert~C. J.,  1998, \apj, 498, 541

\bibitem[\protect\citeauthoryear{{Kim}, {Saitoh}, {Jeon}, {Figer}, {Merritt}
  \& {Wada}}{{Kim} et~al.}{2011}]{Kim2011}
{Kim} S.~S.,  {Saitoh} T.~R.,  {Jeon} M.,  {Figer} D.~F.,  {Merritt} D.,
  {Wada} K.,  2011, \apj, 735, L11

\bibitem[\protect\citeauthoryear{{Kim}, {Seo}, {Stone}, {Yoon}  \&
  {Teuben}}{{Kim} et~al.}{2012}]{Kim2012}
{Kim} W.-T.,  {Seo} W.-Y.,  {Stone} J.~M.,  {Yoon} D.,   {Teuben} P.~J.,  2012,
  \apj, 747, 60

\bibitem[\protect\citeauthoryear{{Kim} et~al.,}{{Kim} et~al.}{2014}]{Kim2014}
{Kim} T.,  et~al., 2014, \mn@doi [\apj] {10.1088/0004-637X/782/2/64}, \href
  {https://ui.adsabs.harvard.edu/abs/2014ApJ...782...64K} {782, 64}

\bibitem[\protect\citeauthoryear{{Kim}, {Gadotti}, {Athanassoula}, {Bosma},
  {Sheth}  \& {Lee}}{{Kim} et~al.}{2016}]{Kim2016}
{Kim} T.,  {Gadotti} D.~A.,  {Athanassoula} E.,  {Bosma} A.,  {Sheth} K.,
  {Lee} M.~G.,  2016, \mn@doi [\mnras] {10.1093/mnras/stw1899}, \href
  {https://ui.adsabs.harvard.edu/abs/2016MNRAS.462.3430K} {462, 3430}

\bibitem[\protect\citeauthoryear{Kim, {Seong Hwang}, Chung, Lee, Park,
  {Cervantes Sodi}  \& Kim}{Kim et~al.}{2017}]{Kim2017}
Kim E.,  {Seong Hwang} H.,  Chung H.,  Lee G.-H.,  Park C.,  {Cervantes Sodi}
  B.,   Kim S.~S.,  2017, The Astrophysical Journal, 845, 93

\bibitem[\protect\citeauthoryear{{Knapen}, {Beckman}, {Heller}, {Shlosman}  \&
  {de Jong}}{{Knapen} et~al.}{1995}]{Knapen1995}
{Knapen} J.~H.,  {Beckman} J.~E.,  {Heller} C.~H.,  {Shlosman} I.,   {de Jong}
  R.~S.,  1995, \apj, 454, 623

\bibitem[\protect\citeauthoryear{{Knapen}, {Shlosman}  \& {Peletier}}{{Knapen}
  et~al.}{2000}]{Knapen2000}
{Knapen} J.~H.,  {Shlosman} I.,   {Peletier} R.~F.,  2000, \apj, 529, 93

\bibitem[\protect\citeauthoryear{{Kraljic}, {Bournaud}  \& {Martig}}{{Kraljic}
  et~al.}{2012}]{KK2012}
{Kraljic} K.,  {Bournaud} F.,   {Martig} M.,  2012, \apj, 757, 60

\bibitem[\protect\citeauthoryear{{Laurikainen}, {Salo}, {Buta}  \&
  {Vasylyev}}{{Laurikainen} et~al.}{2004}]{Laurikainen2004}
{Laurikainen} E.,  {Salo} H.,  {Buta} R.,   {Vasylyev} S.,  2004, \mnras, 355,
  1251

\bibitem[\protect\citeauthoryear{{Lin}, {Li}, {He}, {Xiao}  \& {Wang}}{{Lin}
  et~al.}{2017}]{Lin2017}
{Lin} L.,  {Li} C.,  {He} Y.,  {Xiao} T.,   {Wang} E.,  2017, \apj, 838, 105

\bibitem[\protect\citeauthoryear{{Lynden-Bell}}{{Lynden-Bell}}{1979}]{LyndenBell1979}
{Lynden-Bell} D.,  1979, \mnras, 187, 101

\bibitem[\protect\citeauthoryear{{Marinova} \& {Jogee}}{{Marinova} \&
  {Jogee}}{2007}]{Marinova2007}
{Marinova} I.,  {Jogee} S.,  2007, \apj, 659, 1176

\bibitem[\protect\citeauthoryear{Martig, Bournaud, Teyssier  \& Dekel}{Martig
  et~al.}{2009}]{Martig2009}
Martig M.,  Bournaud F.,  Teyssier R.,   Dekel A.,  2009, The Astrophysical
  Journal, 707, 250

\bibitem[\protect\citeauthoryear{{Martig}, {Bournaud}, {Croton}, {Dekel}  \&
  {Teyssier}}{{Martig} et~al.}{2012}]{Martig2012}
{Martig} M.,  {Bournaud} F.,  {Croton} D.~J.,  {Dekel} A.,   {Teyssier} R.,
  2012, \apj, 756, 26

\bibitem[\protect\citeauthoryear{{Martig}, {Minchev}  \& {Flynn}}{{Martig}
  et~al.}{2014a}]{Martig2014a}
{Martig} M.,  {Minchev} I.,   {Flynn} C.,  2014a, \mnras, 442, 2474

\bibitem[\protect\citeauthoryear{{Martig}, {Minchev}  \& {Flynn}}{{Martig}
  et~al.}{2014b}]{Martig2014b}
{Martig} M.,  {Minchev} I.,   {Flynn} C.,  2014b, \mnras, 443, 2452

\bibitem[\protect\citeauthoryear{{Martin}}{{Martin}}{1995}]{Martin1995}
{Martin} P.,  1995, \aj, 109, 2428

\bibitem[\protect\citeauthoryear{{Martin} \& {Friedli}}{{Martin} \&
  {Friedli}}{1997}]{Martin1997}
{Martin} P.,  {Friedli} D.,  1997, \aap, 326, 449

\bibitem[\protect\citeauthoryear{{Martinet} \& {Friedli}}{{Martinet} \&
  {Friedli}}{1997}]{Martinet1997}
{Martinet} L.,  {Friedli} D.,  1997, \aap, 323, 363

\bibitem[\protect\citeauthoryear{{Momose}, {Okumura}, {Koda}  \&
  {Sawada}}{{Momose} et~al.}{2010}]{Momose2010}
{Momose} R.,  {Okumura} S.~K.,  {Koda} J.,   {Sawada} T.,  2010, \apj, 721, 383

\bibitem[\protect\citeauthoryear{{Nair} \& {Abraham}}{{Nair} \&
  {Abraham}}{2010}]{Nair2010}
{Nair} P.~B.,  {Abraham} R.~G.,  2010, \apj, 714, L260

\bibitem[\protect\citeauthoryear{{Oh}, {Oh}  \& {Yi}}{{Oh}
  et~al.}{2012}]{Oh2012}
{Oh} S.,  {Oh} K.,   {Yi} S.~K.,  2012, The Astrophysical Journal Supplement
  Series, 198, 4

\bibitem[\protect\citeauthoryear{{P{\'e}rez} \&
  {S{\'a}nchez-Bl{\'a}zquez}}{{P{\'e}rez} \&
  {S{\'a}nchez-Bl{\'a}zquez}}{2011}]{Perez2011}
{P{\'e}rez} I.,  {S{\'a}nchez-Bl{\'a}zquez} P.,  2011, \aap, 529, A64

\bibitem[\protect\citeauthoryear{{P{\'e}rez}, {S{\'a}nchez-Bl{\'a}zquez}  \&
  {Zurita}}{{P{\'e}rez} et~al.}{2009}]{Perez2009}
{P{\'e}rez} I.,  {S{\'a}nchez-Bl{\'a}zquez} P.,   {Zurita} A.,  2009, \aap,
  495, 775

\bibitem[\protect\citeauthoryear{{Pompea} \& {Rieke}}{{Pompea} \&
  {Rieke}}{1990}]{Pompea1990}
{Pompea} S.~M.,  {Rieke} G.~H.,  1990, \apj, 356, 416

\bibitem[\protect\citeauthoryear{{Regan} et~al.,}{{Regan}
  et~al.}{2006}]{Regan2006}
{Regan} M.~W.,  et~al., 2006, \apj, 652, 1112

\bibitem[\protect\citeauthoryear{{Sakamoto}, {Okumura}, {Ishizuki}  \&
  {Scoville}}{{Sakamoto} et~al.}{1999}]{Sakamoto1999}
{Sakamoto} K.,  {Okumura} S.~K.,  {Ishizuki} S.,   {Scoville} N.~Z.,  1999,
  \apj, 525, 691

\bibitem[\protect\citeauthoryear{{Sellwood}}{{Sellwood}}{2014}]{Sellwood2014}
{Sellwood} J.~A.,  2014, Reviews of Modern Physics, 86, 1

\bibitem[\protect\citeauthoryear{{Seo} \& {Kim}}{{Seo} \&
  {Kim}}{2013}]{Seo2013}
{Seo} W.-Y.,  {Kim} W.-T.,  2013, \apj, 769, 100

\bibitem[\protect\citeauthoryear{{Sheth}, {Vogel}, {Regan}, {Thornley}  \&
  {Teuben}}{{Sheth} et~al.}{2005}]{Sheth2005}
{Sheth} K.,  {Vogel} S.~N.,  {Regan} M.~W.,  {Thornley} M.~D.,   {Teuben}
  P.~J.,  2005, \apj, 632, 217

\bibitem[\protect\citeauthoryear{{Sheth} et~al.,}{{Sheth}
  et~al.}{2008}]{Sheth2008}
{Sheth} K.,  et~al., 2008, \apj, 675, 1141

\bibitem[\protect\citeauthoryear{{Shin}, {Kim}, {Baba}, {Saitoh}, {Hwang},
  {Chun}  \& {Hozumi}}{{Shin} et~al.}{2017}]{Shin2017}
{Shin} J.,  {Kim} S.~S.,  {Baba} J.,  {Saitoh} T.~R.,  {Hwang} J.-S.,  {Chun}
  K.,   {Hozumi} S.,  2017, \apj, 841, 74

\bibitem[\protect\citeauthoryear{{Shlosman}, {Frank}  \& {Begelman}}{{Shlosman}
  et~al.}{1989}]{Shlosman1989}
{Shlosman} I.,  {Frank} J.,   {Begelman} M.~C.,  1989, \nat, 338, 45

\bibitem[\protect\citeauthoryear{{Simmons} et~al.,}{{Simmons}
  et~al.}{2014}]{Simmons2014}
{Simmons} B.~D.,  et~al., 2014, \mn@doi [\mnras] {10.1093/mnras/stu1817}, \href
  {http://adsabs.harvard.edu/abs/2014MNRAS.445.3466S} {445, 3466}

\bibitem[\protect\citeauthoryear{{Teyssier}}{{Teyssier}}{2002}]{teyssier2002}
{Teyssier} R.,  2002, \aap, 385, 337

\bibitem[\protect\citeauthoryear{{Vera}, {Alonso}  \& {Coldwell}}{{Vera}
  et~al.}{2016}]{Vera2016}
{Vera} M.,  {Alonso} S.,   {Coldwell} G.,  2016, \aap, 595, A63

\bibitem[\protect\citeauthoryear{{Verley}, {Combes}, {Verdes-Montenegro},
  {Bergond}  \& {Leon}}{{Verley} et~al.}{2007}]{Verley2007}
{Verley} S.,  {Combes} F.,  {Verdes-Montenegro} L.,  {Bergond} G.,   {Leon} S.,
   2007, \aap, 474, 43

\bibitem[\protect\citeauthoryear{{Wang} et~al.,}{{Wang}
  et~al.}{2012}]{Wang2012}
{Wang} J.,  et~al., 2012, \mnras, 423, 3486

\bibitem[\protect\citeauthoryear{{Willett} et~al.,}{{Willett}
  et~al.}{2015}]{Willett2015}
{Willett} K.~W.,  et~al., 2015, \mnras, 449, 820

\bibitem[\protect\citeauthoryear{{Wozniak}}{{Wozniak}}{2007}]{Wozniak2007}
{Wozniak} H.,  2007, \aap, 465, L1

\makeatother
\end{thebibliography}





\bsp	
\label{lastpage}
\end{document}